\documentclass[useAMS,usenatbib]{mn2e}
\usepackage{epsfig}
\def \ie  {{\it i.e.\ }}
\def \eg  {{\it e.g.\ }}
\def \mnras {{MNRAS}}
\def \apj {{ApJ}}
\def \apjs {{ApJS}}
\def \aap {{A\&A}}
\def \aaps {{A\&AS}}
\def \aj {{AJ}}
\def \apjl {{ApJL}}
\def \araa {{ARA\&A}}
\def \nat {{Nature}}
\newcommand\ion[2]{#1$\;${\scshape{#2}}}%     

\title[An N-body/SPH Study of Isolated Galaxy Mass Density Profiles]{An N-body/SPH Study of Isolated Galaxy Mass Density Profiles}

\author[Foyle et al.]{Kelly Foyle$^{1}$, St\'{e}phane Courteau$^{2}$ and Robert J. Thacker$^{3}$\thanks{Canada Research Chair}\\
$^{1}$Max-Planck-Institut f\"ur Astronomie, K\"onigstuhl 17, Heidelberg, Germany\\
$^{2}$Department of Physics, Engineering Physics and Astronomy, Queen's University, Kingston, Ontario, Canada\\
$^{3}$Department of Astronomy and Physics, Saint Mary's University, Halifax, Nova Scotia, Canada\\}
\date{Accepted 2008 February 29.}
\pagerange{\pageref{firstpage}--\pageref{lastpage}}
\pubyear{2008}
\begin{document}
\label{firstpage}

\maketitle

\begin{abstract} 
We investigate the evolution of mass
density profiles in secular disk galaxy models, paying special attention
to the development of a two-component profile from a single initial exponential disk free of cosmological evolution (i.e., no accretion or interactions).  As the source of density
profile variations, we examine the parameter space of the spin parameter,
halo concentration, virial mass, disk mass and bulge mass, for a total of
162 simulations in the context of a plausible model of star formation and
feedback. The isolated galaxy models are based on the method of Springel \& White (1999) and were evolved using the N-body/SPH
code GADGET-2.  The initially pure exponential disks have a minimum of
1.4 million particles and most models were evolved over a
period of 10 Gyr. We find that the slope of the outer density profile is
in close agreement with that of the initial profile and remains stable
over time, whereas the inner density profile slope evolves considerably as
a result of angular momentum redistribution.  The evolution of the galaxy
mass density profile, including the development of a two-component profile with an inner and outer
segment, is controlled by the ratio of the disk mass fraction, $m_{d}$,
to the halo spin parameter, $\lambda$.  The location of the break between the two components and
speed at which it develops is directly proportional to $m_{d}/\lambda$;
the amplitude of the transition between the inner and outer regions is however controlled by the ratio of halo
concentration to virial velocity.  The location of the divide between the inner and outer profile does not change with time.  The condition for a two-component profile is roughly
$m_{d}/\lambda \geq 1$. While the development of a two-component density
profile is coupled to bar formation, not all barred galaxies develop a
two-component profile.  A galaxy model showing a clear minimum
Toomre $Q$, normally linked to a double exponential in the stellar profile, may never
exhibit any two-component feature thus yielding a fully evolved pure exponential disk.

\end{abstract}
%Section heading
\section{Introduction}

The evolution of galaxies is a complex process involving angular 
momentum redistribution, star formation, chemical evolution, feedback 
and numerous merging events.  Understanding how and why galaxies have the 
properties we observe is further complicated by the fact that 
observations are limited in a number of complicated ways, including dust 
obscuration and our inability to view a galaxy in all its projections.

One of the key observations of disk galaxy properties is the ubiquitous 
exponential surface brightness profile (de Vaucouleurs 1959; Freeman 
1970).  Freeman, and many others since, modeled galaxies as two component systems with: 
an inner spheroid where $\log I(R) \approx R^{1/4}$ and an outer 
exponential disk component with $I(R) = I_{o}e^{-\alpha R}$.  It is now believed that the inner galaxy components are best modeled by a S\'{e}rsic function, $I(R)\simeq I_{e}e^{-(R/R_{e})^{1/n}}$, with $n<4$ (MacArthur et al. 2003; Pohlen \& Trujillo 2006, hereafter PT06).  Exponential surface brightness profiles 
have largely been explained as a relic of the initial angular momentum 
distribution of the gas which collapsed to form the observed stellar 
disk population (Fall \& Efstathiou 1980; Ferguson \& Clarke 2001).  While the exponential profile is certainly common among disk 
galaxies, many galaxies deviate significantly from it.  A 
large percentage of galaxies show a two-component exponential profile 
with a steep outer region  (PT06).   The transition between the two regimes as sometimes been referred to as the truncation or break radius.  In order to avoid confusion in notation we will refer to these profiles as two-component or broken exponentials.  The issue of broken profiles 
was first addressed by van der Kruit (1979) and while improved 
observations have led to subsequent refinements in the analysis of these 
profiles, confusion still remains as to the nature of these 
profile excursions.
 
Cosmological simulations coupled with semi-analytical models have been able to reproduce many of the observed features of galaxies.  Recent high resolution simulations by Governato et al. (2007) with gas cooling and feedback have shown a rotationally supported disk with an exponential density profile can be generated.  However, current models still exhibit striking failures.  For instance, the well-known angular momentum 
catastrophe may be caused by dynamical friction of infalling gas 
leading to the loss of angular momentum (Navarro \& Benz 1991; Navarro 
\& White 1994; Navarro \& Steinmetz 1999).  Smooth mass accretion, additional resolution and angular momentum conservation 
through feedback processes also yield disks that are too centrally concentrated.  
Producing a realistic disk galaxy still remains a challenging task for 
all galaxy modelers (Lake \& Carlberg 1988; Katz 1992; Navarro \& White 
1994; Navarro \& Steinmetz 1997; Thacker \& Couchman 2000; Sommer-Larsen 
et al. 2003; Governato et al. 2004; Kaufmann et al. 2007; Governato et al. 
2007).  To date, no single galaxy model has ever matched simultaneously the size-velocity-luminosity and color luminosity of galaxies (Courteau et al. 2007; Dutton et al. 2007).  Cosmological simulations of galaxies have yet to produce pure exponential disk galaxies as seen in nature (\eg NGC 300 (Bland-Hawthorne et al. 2005).

Due to measurement limitations and the fact that the observed
outer disks blend into the sky noise especially for face-on orientations, it is difficult to address the origin of double exponential profiles using observations alone.  Simulations allow us to examine the evolution of the galaxy in time free of any observational constraints.  However, careful comparisons with observations are still required. 
While semi-analytical studies and N-body simulations have been performed
in this context (\eg Dalcanton, Summers \& Spergel 1997, hereafter DSS97;  
Mo, Mao \& White 1998, hereafter MMW98; van den Bosch 2001, hereafter
vdB01; Debattista et al. 2006;  hereafter D06; Kauffman et al. 2007;
Governato et al. 2007), an exhaustive exploration of the full parameter
space of galaxy structure models has not yet been attempted.

vdB01 conducted a semi-analytical study of the nature of galaxy
density distributions, and focused specifically on the angular momentum 
catastrophe.  
Like most treatments that ignore angular momentum exchange between the
bulge, disk and halo components, the disks produced by vdB01 were too
centrally concentrated in both the stars and gas.  However, the stellar
breaks in the exponential profiles were in close agreement with observations due to the use of
a star formation threshold.  Because the gas distribution was directly
related to angular momentum, the subsequent development of a high central
density in the disks drove the gas to the inner regions producing outer
breaks interior to those observed.  Furthermore the models of vdB01 did not produce the observed double exponential profile in the gas and stars, but were rather sharply truncated in the outer regions.

This work addresses the nature of physical two-component profiles in isolated galaxies free of cosmological evolution (such as merging and accretion) and whether
these profiles can be attributed to dynamical processes in a disk or alternatively
to star formation, or indeed both.  As noted by D06, a semi-analytical
approach like that of vdB01, cannot account for the fundamental role of
bars and secular evolution on the profiles.  The pioneering numerical
study of D06 was however restricted to a small range of galaxy parameters.  
D06 used an N-body approach to examine whether secular evolution could
reproduce observed stellar density profiles.  
 D06 found that the density profiles may evolve considerably over
time, particularly under the action of a bar.  Furthermore, their study of broken exponentials relied on rigid halo models and also lacked a representation of 
star formation and 
feedback.  For our study of galaxy density profiles, we have explored a
more extensive parameter space of the dark and baryonic matter in
galaxies, with combinations of the structural parameters that control the
various observed profile types and how those profiles evolve over time.  
Our study involves a sensitivity analysis of galaxy models based
on five fundamental parameters, the spin parameter, $\lambda$, halo
concentration, $c$, mass or virial velocity, $V_{200}$, disk mass
fraction, $m_{d}$, and the presence of a bulge with a mass fraction,
$m_{b}$.  Our study considers strictly isolated galaxies free of gas infall or interactions.  While such models do not accurately represent nature, they enable a clear focus on the physical processes associated with secular evolution, that would otherwise be masked by the addition of merging histories and gas accretion.  Upon identifying the most significant parameters that control break development, one can then begin to consider a more complete environment (\eg Roskar et al. 2007, Curir et al. 2006 ).

This research makes use of an N-body/SPH code, GADGET-2,
developed and made available to us by Volker Springel (see Springel
2005 for a description of this code).  The star formation model uses the parameters of Springel \& Hernquist (2003) and was held fixed to 
keep the exploration of the dynamical parameter space manageable.

The layout of this paper is as follows: in \S 2 we review recent works and discuss our motivations.  In \S 3, we describe our simulations and the choice of parameter space.  We present our findings in \S 4 and compare our results with observations  in \S 5. A discussion and a summary of our main findings are presented in \S 6.

\section{Background \& Motivation}
\subsection{Angular Momentum}

Recent theoretical models of galaxy formation have shown that the angular
momentum, $J$, is the principal determinant for the observable features of
a galaxy (DSS97; vdB01; Hernandez \& Cervantes-Sodi 2006; hereafter HS06;
Dutton et al. 2007).  DSS97 showed that the angular momentum distribution
of a galaxy controls the shape of its rotation curve and its overall
surface brightness.  Low angular momentum galaxies are more centrally
concentrated, have high surface brightness (HSB) and are globally unstable
to the formation of bars and bulges.  Conversely, high angular momentum
galaxies have greater disk scale lengths, have lower surface brightnesses
(LSB) and a higher dynamical mass-to-light ratio.

Angular momentum is also central to determining if a system is dynamically
stable to bar formation (MMW98).  The early simulations of Efstathiou, Lake
\& Negroponte (1982) showed that disks are unstable to bar formation if:\\
\begin{equation}
\frac{V_{max}}{(GM_{d}/h)^{1/2}}\leq 1.1
\end{equation}

where $V_{max}$ is the maximum rotation velocity of the disk, $h$ is the
disk scale length and $M_{d}$ is the total disk mass.  For a given disk
mass, only a range of spin parameters would produce a stable disk.  In the
model of MMW98 the stability condition is recast as:\\
\begin{equation}
\lambda \geq \lambda_{crit}=\sqrt{2}\epsilon_{m,crit}^{2}m_{d}f_{c}^{1/2}f_{R}^{-1}f_{V}^{-2}
\end{equation}
where $\epsilon_{m,crit}\approx 1$ and $f_{c}$, $f_{R}$ and $f_{V}$ are functions introduced for a self-gravitating disk (see Eqs. 23, 29, 34 and 37 of MMW98).  Since the effect of disk self-gravity on $h$ and $V_{max}$ is weak, the condition for stability can be approximated by $\lambda_{crit}\sim m_{d}$.  Thus, a galaxy with a high disk mass fraction and low angular momentum will be dynamically unstable to bar formation. We contrast this global dynamical stability estimator based on $m_{d}/\lambda$ with the local Toomre stability criterion in \S 4.3.

The maximum specific angular momentum of the baryons in the disk may
potentially cause disk breaks.  If angular momentum redistribution does
not occur in the disk, then the outermost stellar radius will reflect the
maximum value of the angular momentum of the baryonic material.  Whether or not a galaxy develops a broken exponential could be an important indicator of the angular momentum
properties of the original halo.  Thus, the outer disk could serve as an
important diagnostic of properties of the proto-galaxy (Freeman \&
Bland-Hawthorne 2002).

Several ongoing problems have plagued our understanding of the role of 
angular momentum in disk formation.  The assumption of angular momentum 
conservation for the baryonic component is essential to avoid the 
so-called {\it angular momentum catastrophe}.  A simple expression of
the disk angular momentum is given by $J_{d}=2M_{d}hV_{max}$ (MMW98).  
Angular momentum conservation implies that if the disk mass increases, the 
disk scale length must also decrease (assuming constant $V_{max}$), 
producing small compact galaxies. Numerical simulations have shown that 
angular momentum loss yields disk galaxies that are an order of magnitude 
too small/compact (Navarro \& Benz 1991).  To prevent angular momentum 
loss and small disks, stellar feedback and ionizing UV background 
radiation definitely play some role (\eg Thacker \& Couchman 2001; Somerville 
2002; Governato et al. 2007). Recent work by Kauffman et al. (2007) and Naab et al. (2007), which extends the work of Okamoto et al. (2003), have shown that increased numerical resolution also lessens the angular momentum catastrophe.  Nonetheless, 
current galaxy models simulated from cosmological initial conditions still 
fail to reproduce the observed 
galaxy scaling 
relations (\eg Gnedin et al. 2006; Dutton et al. 2007; Governato et al. 
2007; Dutton \& Courteau 2008) and more research is needed.

\subsection{Disk Types}
While the light profile of disk galaxies may generally be described by an exponential function, observed profiles are in fact much more complex due to the presence of a bulge, bar, spiral arms and transient features.  Freeman (1970) already distinguished two types of disk profiles: the Type I's that are closely exponential over all radii except for the possible presence of a bulge component, and Type II's that display a depression of the surface brightness between the outer bulge and the inner disk.

Van der Kruit (1979) and later van der Kruit \& Searle (1981, 1982), found 
departures from a pure exponential profile at roughly 4-5 disk scale 
lengths. This range of break radii was later systematically revised 
to lower values 
(2.2-3.6 disk scale lengths) (PT06), thanks to a better understanding of observational 
issues.  
Specifically, these early studies used edge-on systems since the 
line of sight integration enhances the surface brightness of a break.  
However, line of sight integration distributes the light along the 
exponential decline meaning that the measured scale lengths in projection 
will systematically be smaller than in the face-on case (Pohlen 
et al. 2004). For face-on galaxies alone, van der Kruit (1988) and Pohlen 
et al. (2002) found mean break radii at $4.5 \pm 1.0$ and $3.9 \pm 
0.7$ disk scale lengths respectively. There is considerable controversy 
surrounding what fraction of galaxies exhibit breaks, due in part to 
conflicting definitions for breaks and galaxy types.  For instance, 
unlike Freeman (1970), PT06 label 
Type II galaxies as having a break in the outer region (but those two are 
not inconsistent).

Erwin, Beckman and Pohlen (2005) also defined Type III galaxies as anti-truncated or upbending systems.  While sky subtraction errors (underestimated sky level) could explain away these profiles, bona fide Type III's are associated with galaxy mergers.  Since our study focuses on the secular evolution of isolated galaxies, this galaxy type will not be investigated.  Proper statistics of truncated systems will be provided elsewhere but suffice to say here that roughly two thirds of observed disk systems exhibit a two-component profile.  It is not surprising that this fraction should match that of barred systems since the bars and profile breaks are intimately connected (as we shall see in \S4.2). Menendez-Delmestre et al. (2007) has shown that roughly $2/3$ of all nearby bright galaxies in the IR are barred.

Broken exponentials are also ubiquitous at higher redshift.  P\'{e}rez (2004) 
found evidence for ``early truncation'' in a sample of 16 galaxies at $z \approx 1$ with an average break radius $\leq$ 3.5 disk scale lengths.  Six of these distant galaxies had an average break radius $<$1.8 disk scale lengths.  Trujillo \& Pohlen (2005) studied 36 face-on galaxies in the HST Ultra Deep Field galaxies of which 21 exhibited a two-component profile with break radii smaller than 1-3 kpc from observed local values.  These findings would suggest an increase in galaxy disk sizes by 25\% since z $\approx$1 (\eg Somerville et al. 2008).

In summary, while there is considerable scatter in the frequency of galaxy types and the location of breaks, the fact that so many groups have identified broken exponentials using different samples and different methods, means that breaks are certainly {\it real}.  Any complete theory of galaxy formation and evolution must account for the origin and range of observed surface brightness profiles.

\subsection{Bars, Spiral Structure \& Secular Evolution}

Both the early processes of disk assembly, which are dominated by the 
results of major and minor mergers, and the long term effects of secular 
evolution play central roles in defining many of the observed features of 
galaxy surface brightness profile.  Non-axisymmetric perturbations, such 
as those generated by a bar, may play a significant role in shaping the 
observed light distributions.  As they evolve, bars drive spiral 
structure producing rings and driving gas to the center of galaxies 
triggering starbursts (Hohl 1971).

Bars are also thought to trigger bulge formation (Kormendy \& Kennicutt 
2004) . The action of the bar induces resonances in the disk gas (\eg inner Lindblad Resonance at the end of the bar and outer Lindblad resonance at $\approx$ 2.2 bar radii), and funnels gas to the inner regions of the galaxy ultimately yielding the formation of a ``pseudo-bulge'' (Kormendy 1993; Courteau et al. 1996; Kormendy \& Kennicutt 2004).
 
Bars have also been linked to broken exponential profiles.  Anderson, Bagget \&
Bagget (2004) found from the study of surface brightness profiles for 218
spiral and lenticular galaxies that barred galaxies are four times more
likely to exhibit breaks than unbarred ones.  They postulated that
unbarred truncated galaxies might be produced by resonances driven by other
asymmetries.  They proposed that lopsided features which were identified in
30\% of the galaxies by Zaritsky \& Rix (1997) might produce such
resonances. Some authors have suggested that broken exponentials may reflect
variations in disk intensity due to clumpy star formation, spiral arms or
ring-like features (Narayan \& Jog 2003).  Erwin et al. (2008) found that 42\% of a sample of  66 barred galaxies showed breaks.  They showed that these breaks may indeed be associated with the bar or resonances in the disk.

Simulations have also shown that bars can induce a substantial amount of matter redistribution in the outer disk (Athanassoula \& Misiriotis 2002; Valenzuela \& Klypin 2003; D06).  D06 explored the evolution of disks under the presence of a bar and showed that breaks are produced interior to where the bar sheds its angular momentum and carried away by resonantly-coupled spiral arms.  While their work highlighted the importance of secular evolution in determining the morphology of galaxies, it only covered a limited parameter space.  Our study follows in their footsteps and explores secular evolution in general, not just that induced by bar formation.  

\subsection{Physical Origin of Broken Exponentials}
There are two favored hypotheses for the physical origin of broken exponentials.  The {\it collapse model} of van der Kruit (1987) links the break radius to the angular momentum of the protogalaxy.  Kennicutt (1989), on the other hand, suggested a {\it threshold model} whereby a dynamical critical threshold for star formation yields a visible break in the stellar luminosity profile.  The break radius would be located where the gas density is less than the critical value defined by the Toomre criterion (Toomre 1964):
\begin{equation}
\label{crit_gas}
\sum_{crit}\approx \frac{\sigma_{gas} \kappa}{\pi G}.
\end{equation}
where $\sigma_{gas}$ is the radial velocity dispersion of the gas and $\kappa$ is the epicyclic frequency of the stars and gas. The origin of a gas cloud that exceeds this critical density may be a product of swing-amplified or shear instabilities, ring instabilities or giant expanding shells in turbulence-compressed regions (Elmegreen \& Hunter 2006).

Stellar disks behave similarly to gaseous disks at long wavelengths (Rafikov 2001).  The description of a two-component disk with a hot stellar fluid and a cold ISM fluid is however more elaborate.  Wang \& Silk (1994) added a contribution of the stellar component to the critical density and expressed the Toomre-Q parameter for the gas $+$ star model as:
\begin{equation}
\label{Wang}
Q=\gamma\frac{\sigma_{gas} \kappa}{\pi G \Sigma_{gas}} 
\end{equation}
with $\gamma=(1+\frac{\sum_{\star}\sigma_{gas}}{\sum_{gas}\sigma_{\star}})^{-1}$.  Thus, star formation can only occur in {\it locally} unstable regions of the disk\footnote{This stability criterion should not be confused with the bar instability criterion which is determined by the global parameters of the galaxy (see \S 4.3).}, $Q < 1$ and the timescale of star star formation is related to the local dynamical time, $1/\kappa \sim 1/\Omega$, where $\Omega$ is the orbital frequency.  $Q$ also correlates with the star formation rate, with small values implying more rapid star formation (Wang \& Silk 1994).

vdB01 considered a combination of the collapse and the threshold models in 
his semi-analytical approach to understand the density distribution of 
galaxies and the angular momentum catastrophe.  In his model, the 
break in the gas density profile is directly related to the angular 
momentum of the protogalaxy whereas the break radius of the stars is 
determined by a threshold density predicted by the Toomre criterion.  vdB01 inferred that stellar break radii were in good agreement 
with observations, due to the use of star formation threshold, but his gas 
breaks had too high a surface density in conflict with observations.  
His models predicted a sharp cut-off of the gas close to the radii 
$R_{HI}$, where the \ion{H}{I} column density falls to $\sim$$10^{20}$ 
cm$^{-2}$.  In real galaxies, however, \ion{H}{I} gas is found far beyond 
$R_{HI}$.  Like most treatments that ignore angular momentum exchange 
between the bulge, disk and halo components, the disks produced by vdB01 
were too centrally concentrated in both the stars and gas.  Because the 
gas distribution was directly related to angular momentum, the subsequent 
development of a high central density in the disks would drive the gas to the 
inner regions producing an outer break interior to those observed.

Another possible explanation for the origin of the exponential profile lies in a mechanism that could drive the stellar profile towards an exponential regardless of the initial gas profile.  The fact that many observed gas profiles depart from an exponential profile at large radii further supports this idea (Ferguson \& Clarke 2001).  Lin \& Pringle (1987) first introduced the idea of viscosity to regulate the angular momentum redistribution and radial flows.  Viscous models lead to matter funneling into a central concentration and the outer regions, subsequently moving outward, and forming an exponential profile.  Ferguson \& Clarke (2001) did not acknowledge the development of two-component profiles but, as noted by Pohlen (2002), their Figs. 2 and 5 do show break-like features at 4-5 disk scale lengths in agreement with observed values.  The timescale for development of the outer exponential profile can be long, comparable to the star formation timescale (Slyz et al. 2002), especially in cases where the initial disk is highly irregular.  Furthermore, irregular galaxies also exhibit exponential profiles, where there is little shear from viscous effects (Hunter \& Elmegreen 2006).

Elmegreen \& Hunter (2006) showed that a double exponential profile, in contrast to a sharp cut-off in a single exponential disk profile, could be produced by invoking gas turbulence in the outer regions of the disk. In the inner regions the star formation threshold is responsible for the inner disk component.  In the outer regions turbulent compression, which allows for cloud formation and star formation despite sub-critical threshold values, can reproduce observed density profiles of disk galaxies with great fidelity.  Their model thus explains the notion that HSB galaxies would have an increased ratio of the break radius to the inner exponential scale length, since unstable regions of the disk would extend further out, as compared to LSB galaxies.

In their N-body/SPH study D06 found that galaxy density profiles may evolve 
considerably over time, particularly under the action of a bar.  They surmised that previous investigations based on semi-analytic models (\eg vdB01) had failed to produce breaks since they lacked the necessary development of bars.  Angular momentum redistribution was shown to produce both pure exponential profiles and profiles with realistic breaks. Purely collisionless systems matched observations poorly, but the addition of gas and star formation produced much more realistic systems with breaks and mass density excursions from the pure exponential initial disk profile similar to those observed in real galaxies.  

D06 stressed that angular momentum cannot be trivially related to a distribution of scale lengths, which is often assumed (DSS97; MMW98).  Rather, the evolution of the density profile after bar formation should depend primarily on the Toomre-Q parameter.  If Q is small, a bar develops quickly.  If the disk is hotter, bar formation is suppressed and the disk scale length remains relatively unchanged.  The process of bar formation is thus directly related to the formation of breaks: the hotter disks that suppress bar formation will also preserve their pure exponential profile.  However, D06 acknowledge that breaks have been observed in unbarred galaxies and postulated that these galaxies may have been excited through some form of interaction (although typically Type III or upbending profiles are associated with such events).  They found that bars could not be destroyed, thus preventing the subsequent development of unbarred truncated galaxy as well.  

While their simulations achieved significant resolution, the approximation of purely collisionless dynamics in their break simulations may lack important physics such as the effects of star formation, star formation threshold and feedback (see \S 3.4). They also explored a limited section of the available parameter space, completing only 12 simulations, and focusing on variations of the ratio of the disk scale height to the scale length ($z_{d}/h=$0.025-0.2), the Toomre-Q parameter ($Q$=1-1.6), the halo type (logarithmic or Hernquist potential), the ratio of halo to disk scale length ($r_{h}/h=$ 3.3 and 20.8) and the S\'{e}rsic index of the disk ($n=$1.0, 1.5, 2.0).  Our study explores a more complete parameter space as well as the time evolution of the different features in the disk (see \S 3).  D06's study of breaks also relied on rigid halo models and lacked star formation.  We improve upon these limitations in our models.

Building on this body of previous work, we thus investigate the origin 
of two-component profiles in isolated galaxy disks using N-body/SPH methods.  We use 
an exhaustive grid search to compare the relative importance of the halo 
parameters including the mass, concentration and angular momentum with 
disk and bulge parameters.  Important physical processes such as star 
formation and feedback are included in our simulations using the  N-body/SPH models of SH03 for the GADGET-2 code (Springel 2005).   
GADGET-2 
includes a star formation threshold which, unlike D06, enables us to 
compare breaks produced by such a threshold with those resulting from 
matter redistribution in the disk.  Over half of our simulations were evolved up to 10 Gyr.  

Our study also addresses the time evolution of mass density profiles and their features.  By comparing with observations, we can test the validity of our findings and constrain the parameters that control the observed features in galaxy density profiles.

\section{Method \& Explored Parameters}
\subsection{Simulation Code}
Our simulations rely on the GADGET-2 code (Springel 2005) 
with star formation and feedback as described in Springel \& Hernquist 
(2003). The initial conditions of each galaxy model were generated with the 
initial
conditions generator of Springel \& White (1999). The simulations were computed on 
the Canadian SHARCNET facilities at the University of Waterloo.
To reduce running times four processors were used in parallel in each simulation and, on average, each simulation required 10 wallclock days to complete.  The softening length was chosen as 0.08 kpc for the star and gas particles and 0.18 kpc for the dark matter particles for all models. These particular values are sufficiently small so as not to suppress bar formation (Kaufmann et al. 2007).

\subsection{SF Model}

GADGET-2 uses a multiphase model for star formation which incorporates cooling and supernovae energy 
injection.  Star formation is assumed to convert cold clouds into stars on a characteristic timescale 
$t_{\star}$; a mass fraction, $\beta$, of these stars are short-lived and die instantly as SNe (SH03).  
The star formation rate parameterization in GADGET-2 is given as:
$\dot{\rho_{\star}}=(1-\beta)\frac{\rho_{c}}{t_{\star}}$, where $\rho_{c}$ is the density of cold star 
forming gas clouds.
\begin{figure}
\centering

     \epsfxsize=8cm\epsfbox{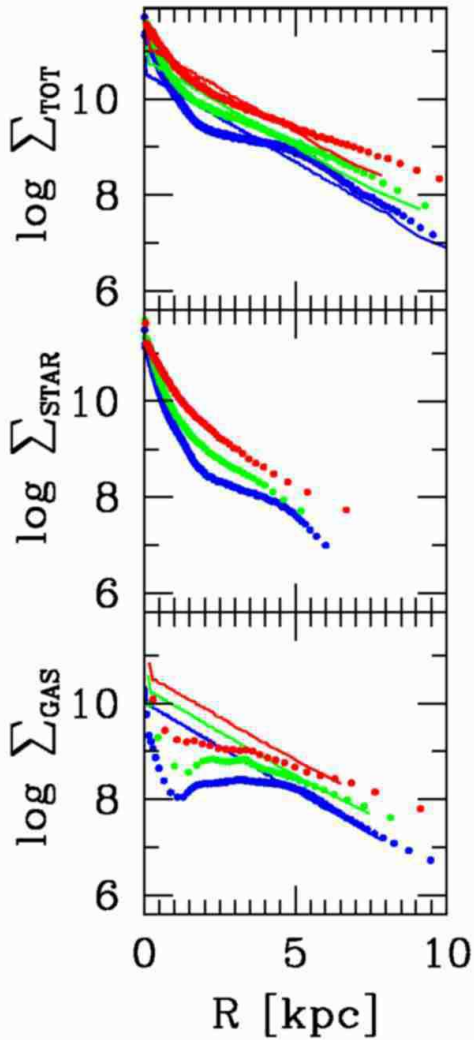}

\rm
\caption{Comparison of density profiles for the gas (bottom), stars
(middle) and total (stars + gas: top) for three particle number
resolutions for the same galaxy model evolved to 10 Gyr.  Blue
represents $ 1.4 \times 10^6$ particles, green $0.35 \times 10^6$
particles, and red $0.09 \times 10^6$ particles.}

\label{res}
\end{figure}

The star formation timescale is proportional to the dynamical time of the gas:
\begin{equation}
\label{timescale}
t_{\star}=t_{0}^{\star}\left(\frac{\rho}{\rho_{th}}\right)^{-1/2}
\end{equation}
where $t_{0}^{\star}$ is a model parameter and $\rho_{th}$ is the threshold density that depends on $t_{0}^{\star}$.  In the models of SH03, a value of $t_{0}^{\star}=2.1$ Gyr was adopted to reproduce the Schmidt law.

The star formation rate of an SPH particle of current mass $m$ is given by $\dot{M}_{\star}=(1-\beta)xm/t_{\star}$, where $x$ is the density fraction of cold clouds. In a given time step, $\Delta t$, a new star particle of mass $m_{\star}=m_{o}/N_{g}$ is born provided that a random number between [0,1] is below
\begin{equation}
p_{\star}=\frac{m}{m_{\star}}\left(1-exp\left[-1\frac{(1-\beta)x\Delta t}{t_{\star}}\right]\right)
\end{equation}
$m_{o}$ is the initial mass of a gas particle and $N_{g}$ is an integer which expresses the potential number of star generations that each gas particle might spawn.  GADGET-2 allows a gas particle to spawn multiple stars.  The total number of particles in the simulation will increase as stars are formed.

The temperature of the cold clouds, $T_{c}$, and the supernovae mass fraction factor, $\beta$, are adopted as 1000 K and 0.1 respectively.  As SH03 note, the results are not sensitive to these values provided $T_{c} < 10^{4}$ K and $\beta \approx 0.1$.  We do not assume a fixed gas mass density, $\rho_{crit}$, below which star formation is impeded; rather, GADGET-2 allows for a threshold density, $\rho_{th}$, which is determined via the star formation timescale, $t_{\star}$, which sets the overall gas consumption timescale.  The star formation timescale, $t_{o}^{\star}$, is chosen by fine-tuning to match the global Schmidt law of Kennicutt (1998).  A detailed overview of the resulting phase evolution between hot and cold gad can be found in SH03.

The power index on $\Sigma_{gas}$ was chosen to fit a value of 1.5, in order for the star formation to be proportional to the local dynamical timescale.   This adjustment prompts the normalization coefficient to be lowered by a factor of 2 in order for the star formation rate to be unchanged at an intermediate density of $10^{3}$ M$_{\odot}$ pc$^{-2}$.  Springel \& Hernquist (2005) showed that the best fitting timescale for star formation was 2.1 Gyr.  This led not only to a good fit to the Kennicutt slope, but also to the observed gas threshold density of $10$ M$_{\odot}$ pc$^{-2}$ (Kennicutt 1998).  Springel \& Hernquist (2005) could not explain why this choice of $t_{o}^{\star}$ reproduces both the Kennicutt law and the gas threshold simultaneously.   The significance of this, perhaps fortuitous, coincidence remains unclear.  We also note our choice of a 30\% gas fraction to allow for sufficient star formation in the disk over a 10 Gyr evolution time.

The use of a star formation threshold will only affect the newly formed stars and not the initial 
stars in the model.  Thus, when drawing comparisons with observations on the nature of a star 
formation threshold, we focus strictly on newly formed stars.

\subsection{Resolution Dependence}
Because density profile breaks typically occur at several scale lengths, a
sufficiently high number of particles is required to properly sample
the outer disk.  However, the particle number should also allow the
numerous simulations to be completed in a reasonable amount of time and
yet not compromise any scientific value.  Time constraints from our 
resource allocation lead us to choose a minimum of
1.4 million initial particles in the no-bulge case and 1.6 million initial
particles in the presence of a bulge. To assess the impact of 
resolution on our simulations we also conducted lower resolution 
simulations, one simulation using 350 000 particles and another using 90 
000 particles. Fig.~\ref{res} shows the impact of resolution on the 
simulated radial profiles for these three different
resolutions, on a galaxy model with parameters $\lambda=0.2$, $V_{200}=80$ km s$^{-1}$, $c=5$, $m_{d}=0.025$ and $m_{b}=0$. Profile features are essentially eliminated in
the low resolution case of 90 000 particles (red).  In the intermediate
range of 350 000 particles (green), the break feature is evident, but the
outer disk is poorly sampled. The highest achieved resolution can 
reproduce the break feature and provides adequate sampling in the outer region.
Naturally, we have used as many particles as
possible, as prescribed by the number of CPU hours 
available for the project.
\begin{figure*}
\centering

     \epsfxsize=8cm\epsfbox{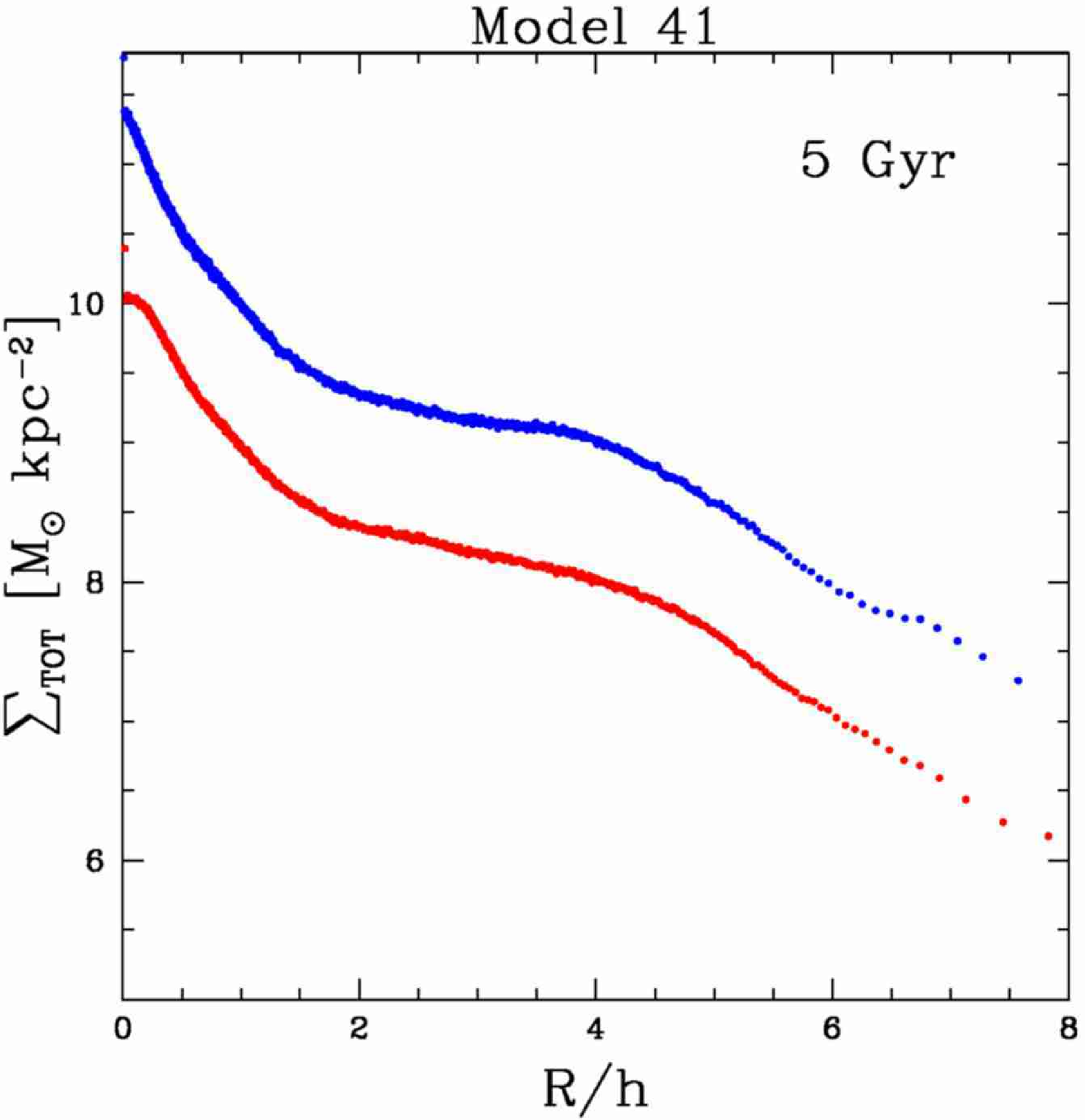}
     \epsfxsize=8cm\epsfbox{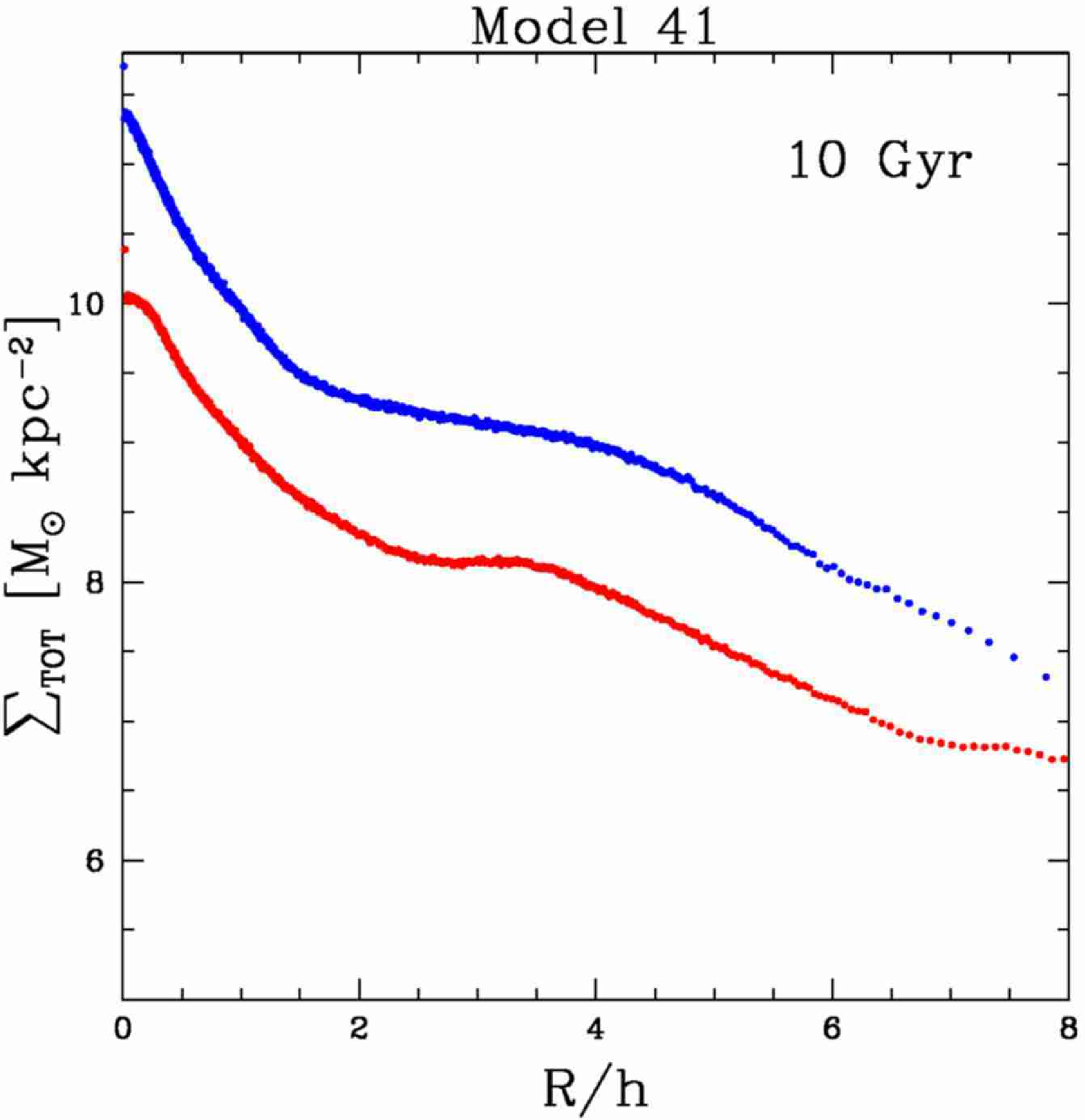}
     \epsfxsize=8cm\epsfbox{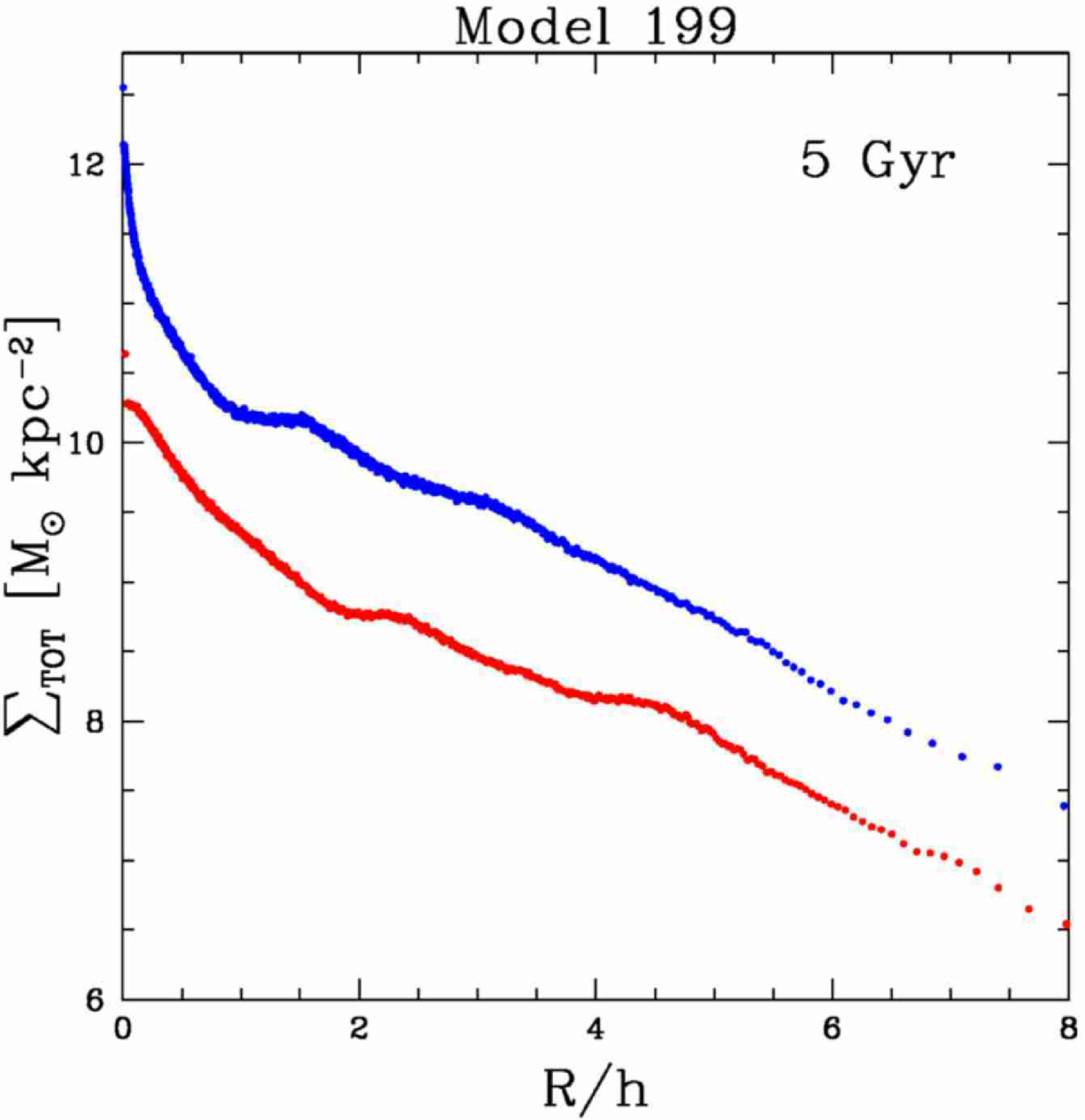}
     \epsfxsize=8cm\epsfbox{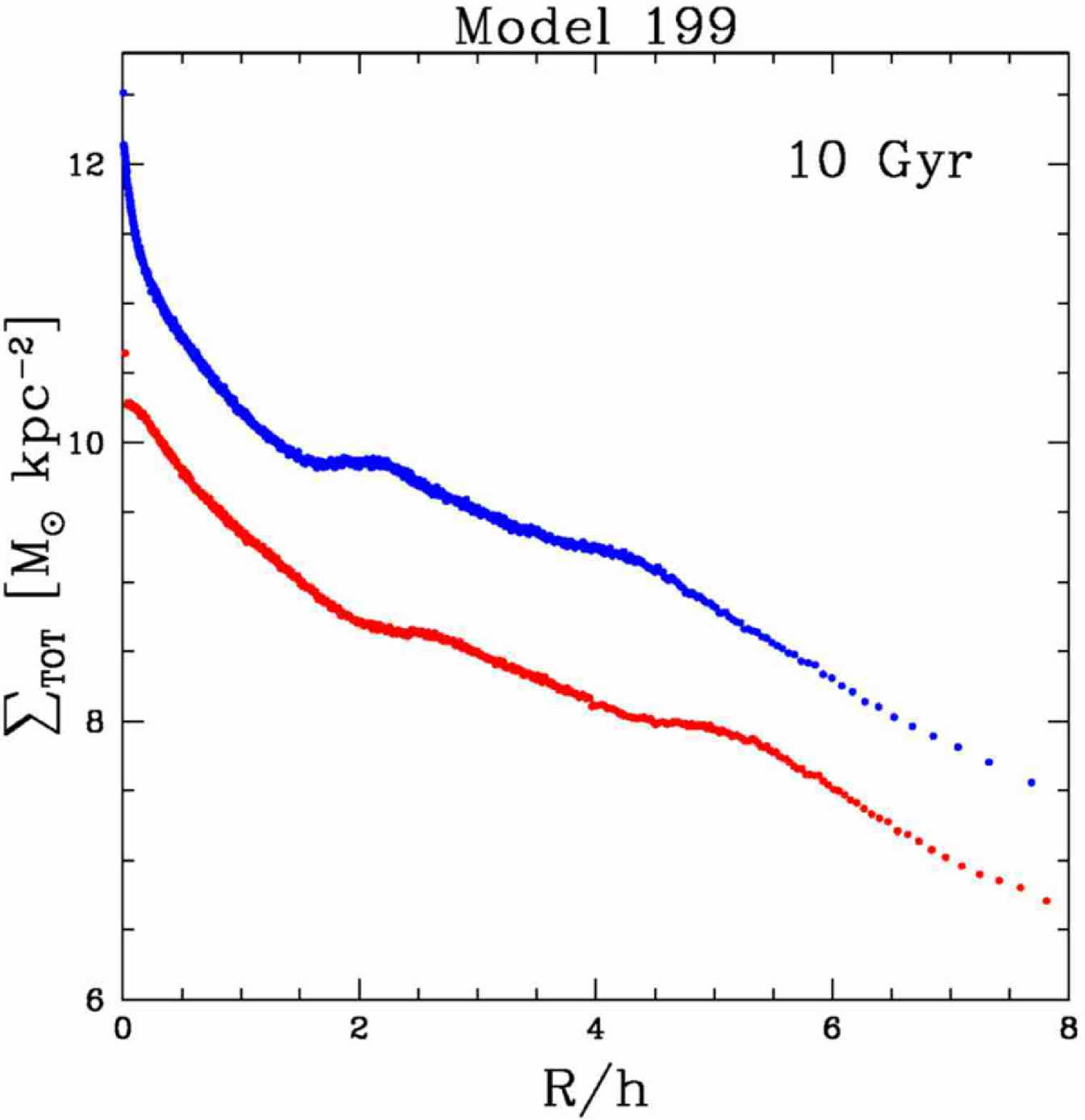}
\rm
\caption{Comparison of collisionless runs (red) versus runs including 
gas and star formation (blue) for two different models and two different
times (5 and 10 Gyr).  The vertical offset is a result of the additional
gas mass in the blue models. Note the significant differences between
the profile shapes and break locations.}
\label{col}
\end{figure*}

\subsection{Comparison with Collisionless Models}

D06 used collisionless particle models to investigate the development of a two-component profile.  Our 
study extends this work, not only by considering a broader parameter space but also by including gas 
particles and star formation.  In order to ascertain the effect, if any, of dissipational processes on 
the radial profiles, we ran several pure collisionless models and models with gas and star formation.  
Fig.~\ref{col} shows a comparison of two such models.  The collisionless runs are shown in red, while 
those with gas and star formation are shown in blue for models after 5 and 10 Gyr.  The radial 
profiles differ in several important ways: (1) the breaks in the collisionless runs tend to develop 
further out into the disk; (2) more matter is distributed to the inner regions of the disk in the 
collisionless models; and (3) the profiles evolve more rapidly in the collisionless case.  Thus, gas 
and star formation does affect significantly the radial profiles and the development of a 
two-component profile.

\subsection{Parameter Space} 
\begin{figure}
\centering

     \epsfxsize=8cm\epsfbox{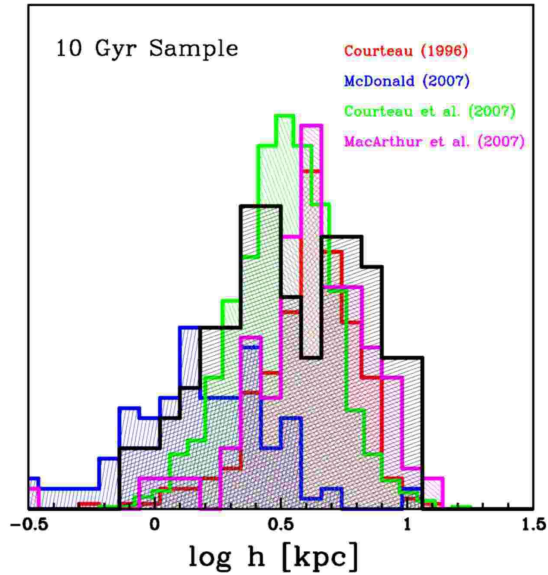}
\rm
\caption{Comparison of disk scale length distributions for existing surveys.  The current models
(black histogram, 10 Gyr sample) encompass nicely the range of observed scale lengths in various
galaxy imaging surveys.}
\label{mainhist}
\end{figure}
Our grid search involved permutations of the basic galaxy parameters 
listed in Table 1, resulting in a total of 162 galaxy models with gas and star formation (see Appendix 
A).  We focused on the five galaxy parameters, $\lambda$, $c$, $V_{200}$, 
$m_{d}$ and $m_{b}$ as a fundamental base to determine the rotation curve 
and the mass density profile of a galaxy.  For instance, it is known that 
the spin parameter, concentration of the halo, and its mass are 
fundamental to the development of bars within a galaxy (Athanassoula 
2003).  Furthermore, the relationship between disk mass fraction and spin 
parameter determines whether a galaxy is bar unstable (MMW98).  Finally, a 
significant concentration of mass in the center of a galaxy, such as 
classical bulge, can slow the formation of a bar or possibly impede its 
formation (van den Bosch 1998).

All successful simulations were computed for a minimum of 5 Gyr, and extended to 10 Gyr, computer time and resources permitting.  
The minimum evolution of 5 Gyr allowed the disk to stabilize and form a
substantial amount of stars.  Following an initial period of transient
evolution, we found that most simulated profiles had settled into a near
final configuration after 3-4 Gyr.

Upon completing the full permutations of model parameters, we also ran a
new suite of simulations to examine a finer grid of the parameter space.  
This enabled the confirmation and refinement of various trends found
with our coarser grid.  We varied $\lambda$ from 0.005 to 0.1 in
increments of 0.005 for those simulations (see Table 1 and Appendix A for details).
\begin{table}
\caption{Grid Search Parameters}
\begin{tabular}{cccc}
Parameter & &{Values}\\
\hline\hline

$\lambda$ & 0.02 & 0.03 & 0.08\\
$c$ & 5 & 10 & 15\\
$V_{200}$ & 80 & 160 & 180\\
$m_{d}$ & 0.025 & 0.05 & 0.1\\
$m_{b}$ & 0 & $0.2m_{d}$ &  \\
\hline
\end{tabular}
\end{table}

\begin{table}
\caption{Particle Number}
\begin{tabular}{cc}
\hline\hline
$N_{halo}$ & 600000 \\
$N_{disk}$ & 400000 \\
$N_{gas}$ & 400000 \\
$N_{bulge}$ & 200000 \\
\hline
\end{tabular}
\end{table}

Having considered a wide range of model parameters, it is only natural that
some models should depart significantly from the expected norm.  Some of
our simulations could not stabilize due to high star formation rates; this
either caused the simulation to stop due to too many star particles being 
created at one time, or alternatively the simulations became too heavily 
populated with star particles making any integration to 10 Gyr unrealistic.
These galaxies all
had $m_{d} \gg \lambda$.  All of the models with $m_{d}=0.1$ and $\lambda=0.02$
did not evolve to 5 Gyr (see Appendix A).

Out of our initial set of 162 model galaxies, 112 models evolved up to 5
Gyr of which 96 reached 10 Gyr.  The 10 Gyr sample proved to match
well with many of the observable properties of typical galaxy samples.  
For instance, we show in Fig.~\ref{mainhist} a histogram of the initial disk scale
lengths of the 10 Gyr models versus the observed distribution of disk scale
lengths for a variety of observed field and cluster spirals (Courteau et
al. 1996; MacArthur et al. 2003; Courteau et al. 2007; McDonald et al
2007).

\begin{figure*}
\centering

     \epsfxsize=8cm\epsfbox{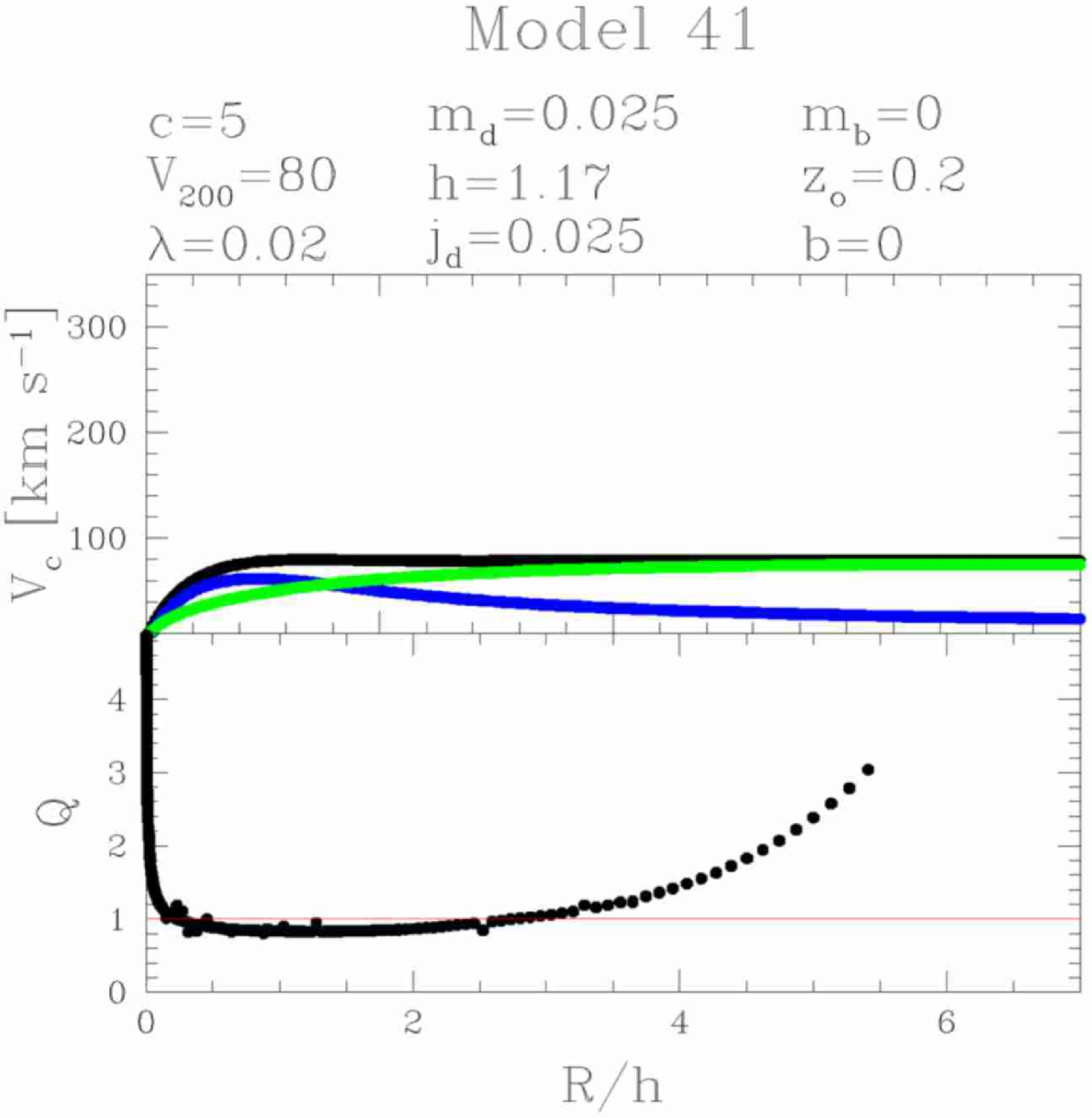}
     \epsfxsize=8cm\epsfbox{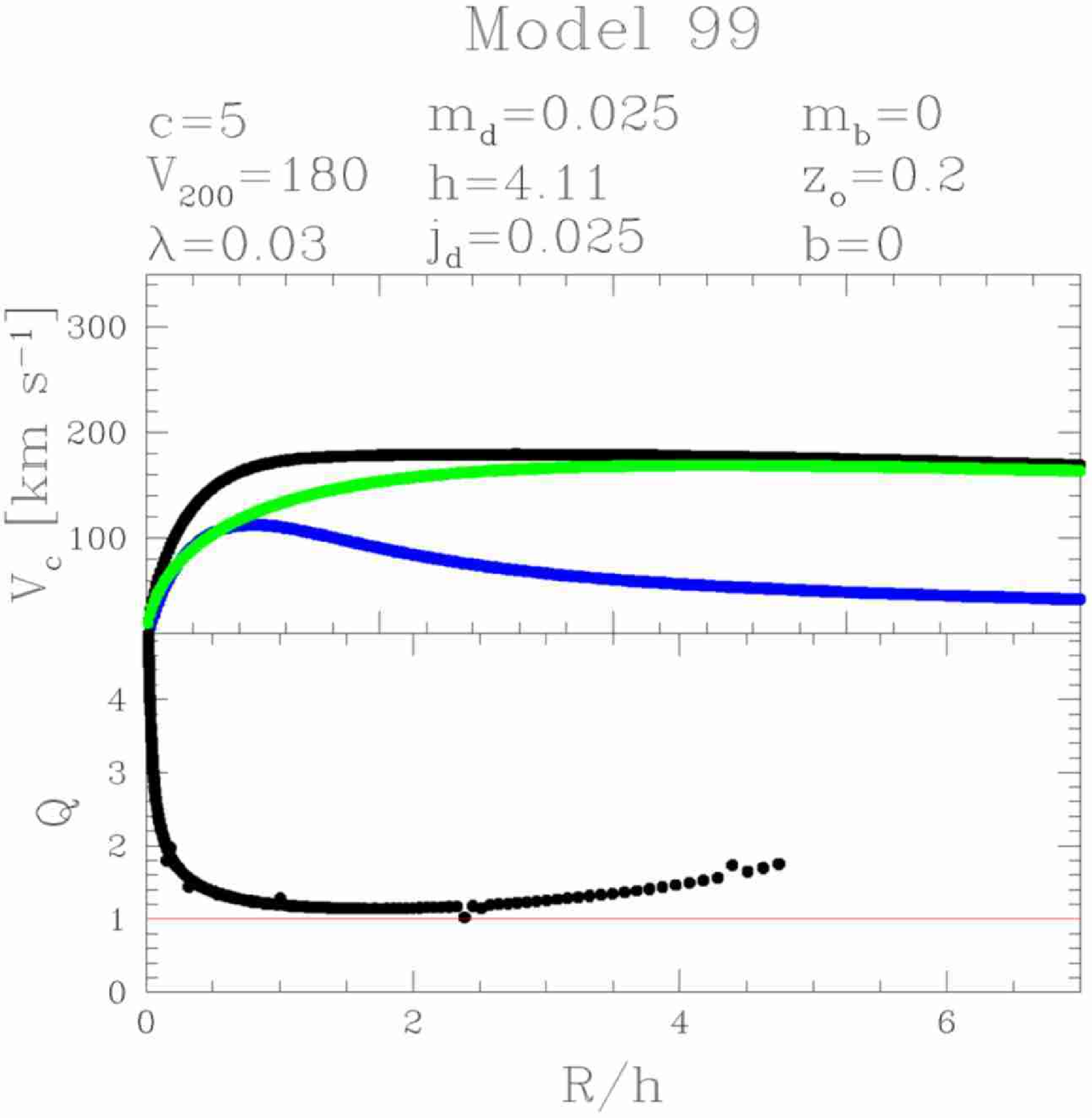}
     \epsfxsize=8cm\epsfbox{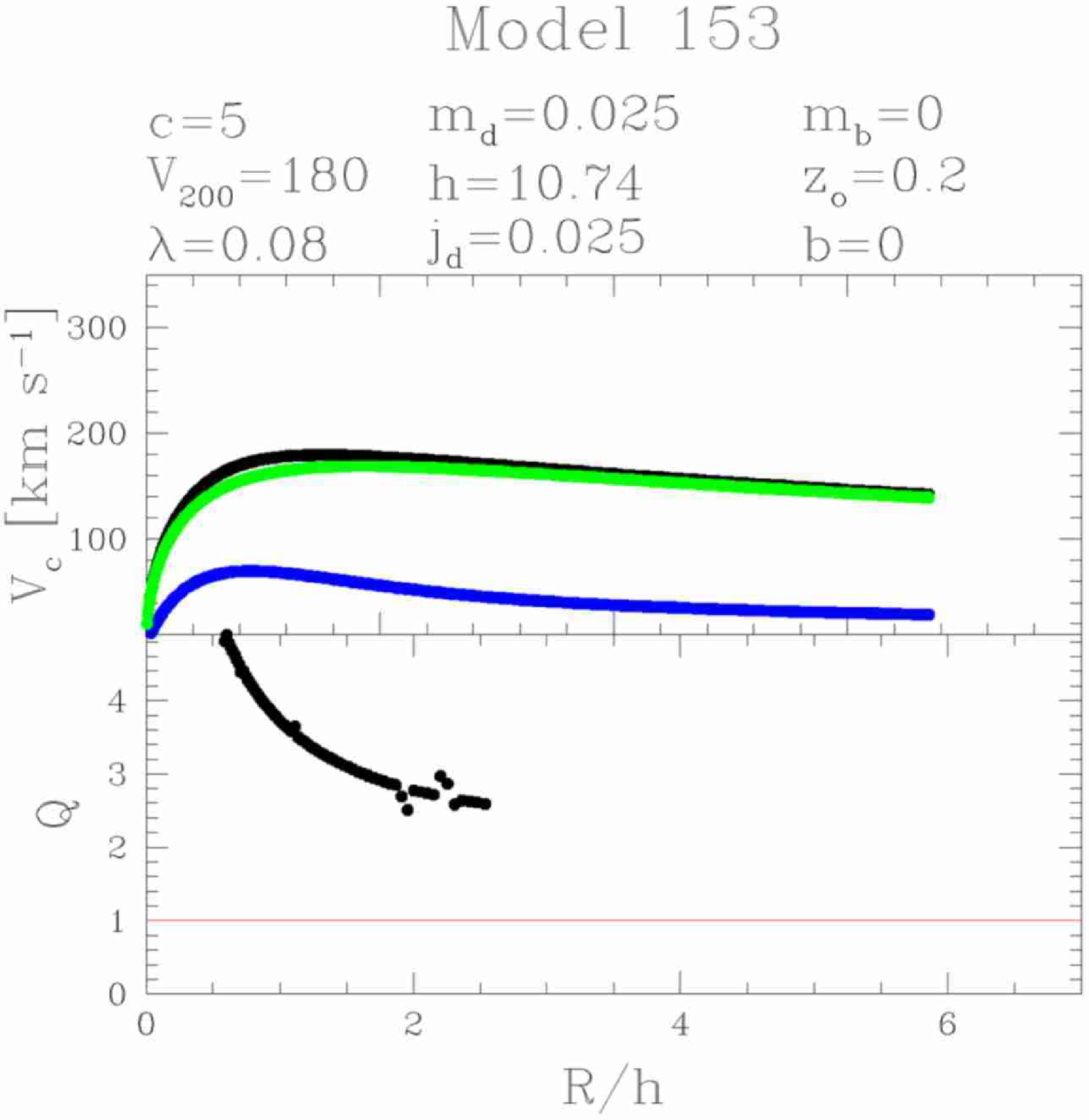}
     \epsfxsize=8cm\epsfbox{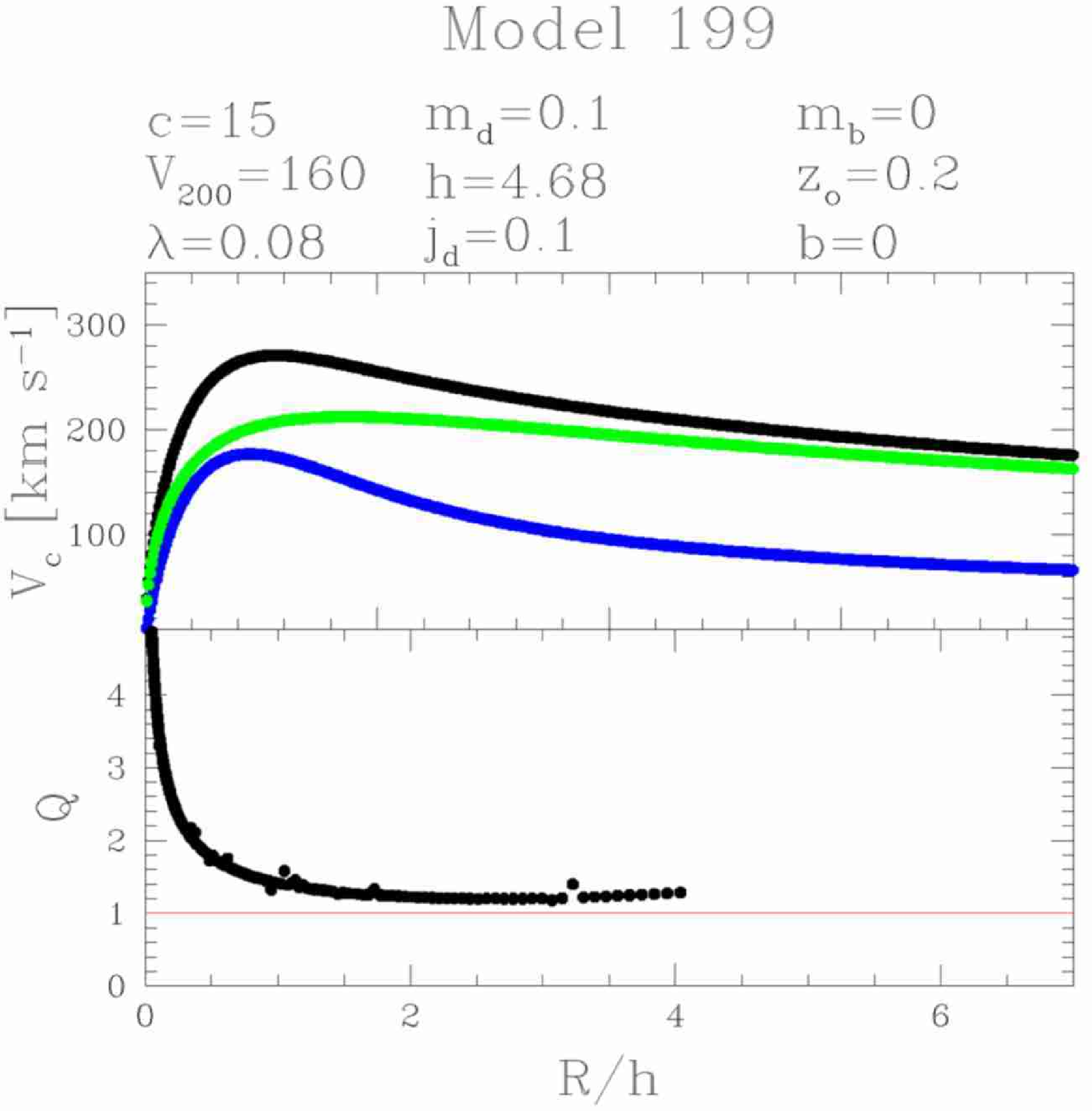}
\rm

\caption{Four model initial rotation curves and Q-curves.  The halo, disk and total are shown in 
green, blue and black respectively.  These curves are determined prior to running the simulations.}

\label{initQrot}
\end{figure*}

\begin{figure*}
\centering

     \epsfxsize=8cm\epsfbox{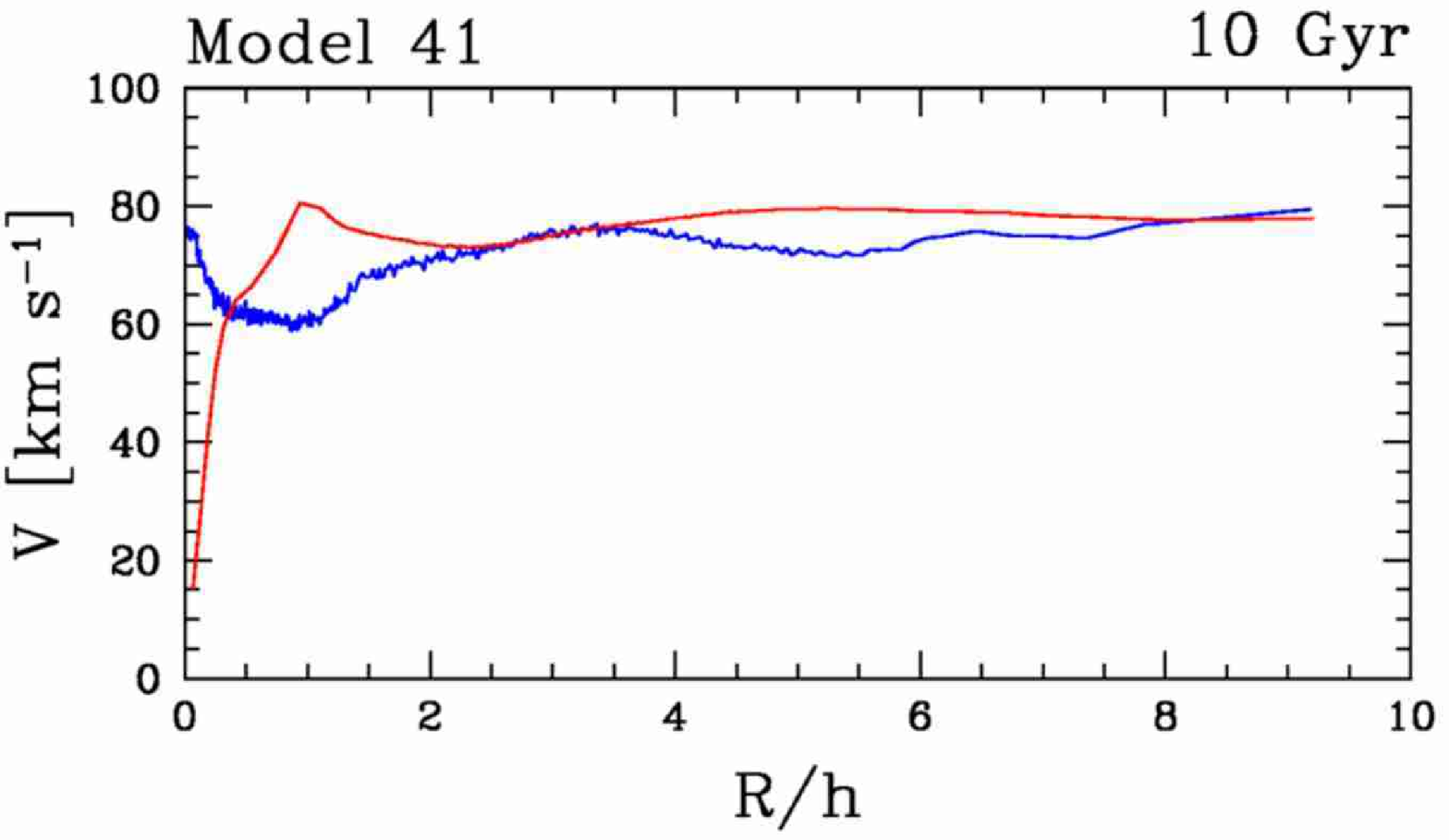}
     \epsfxsize=8cm\epsfbox{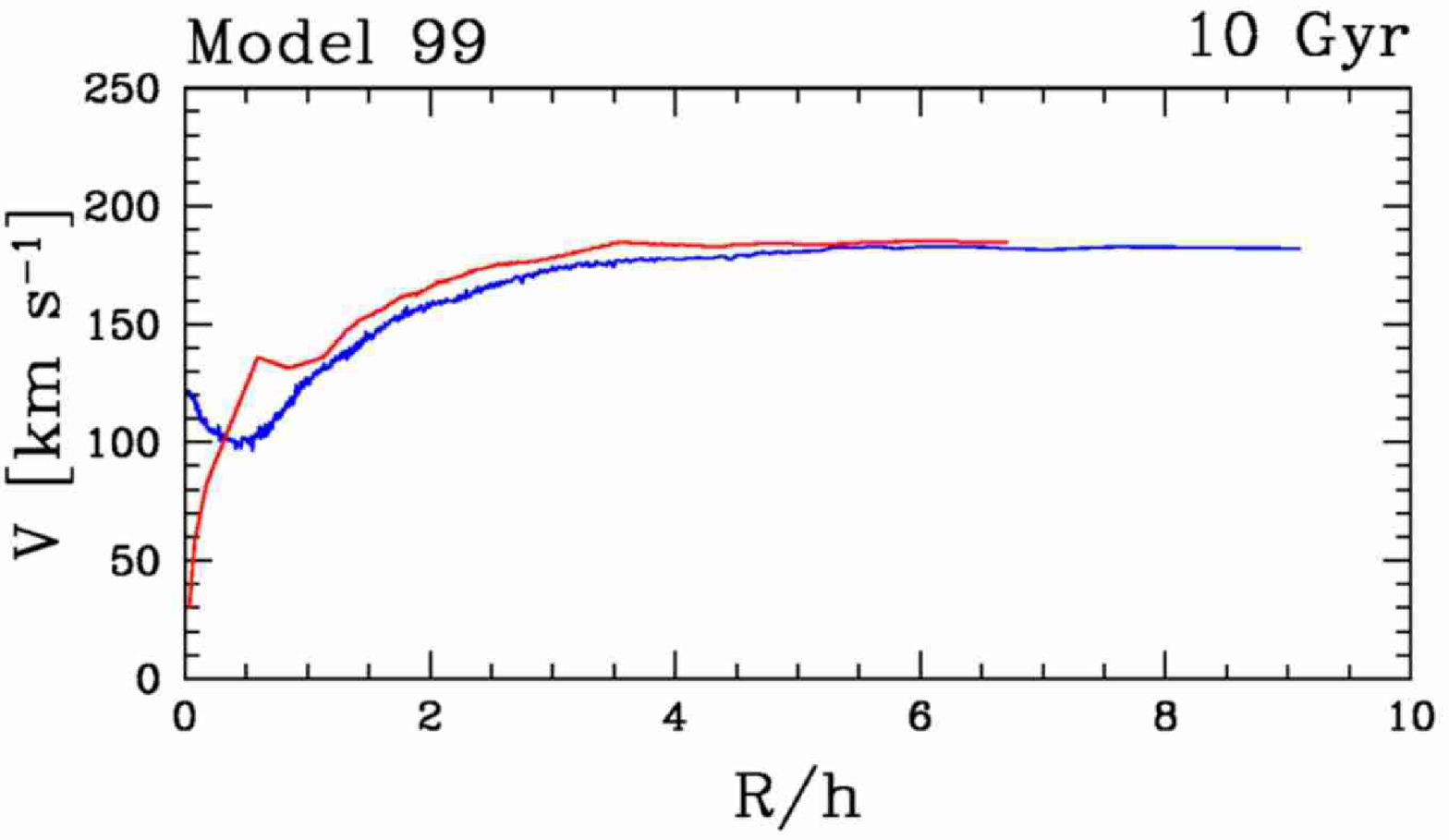}
     \epsfxsize=8cm\epsfbox{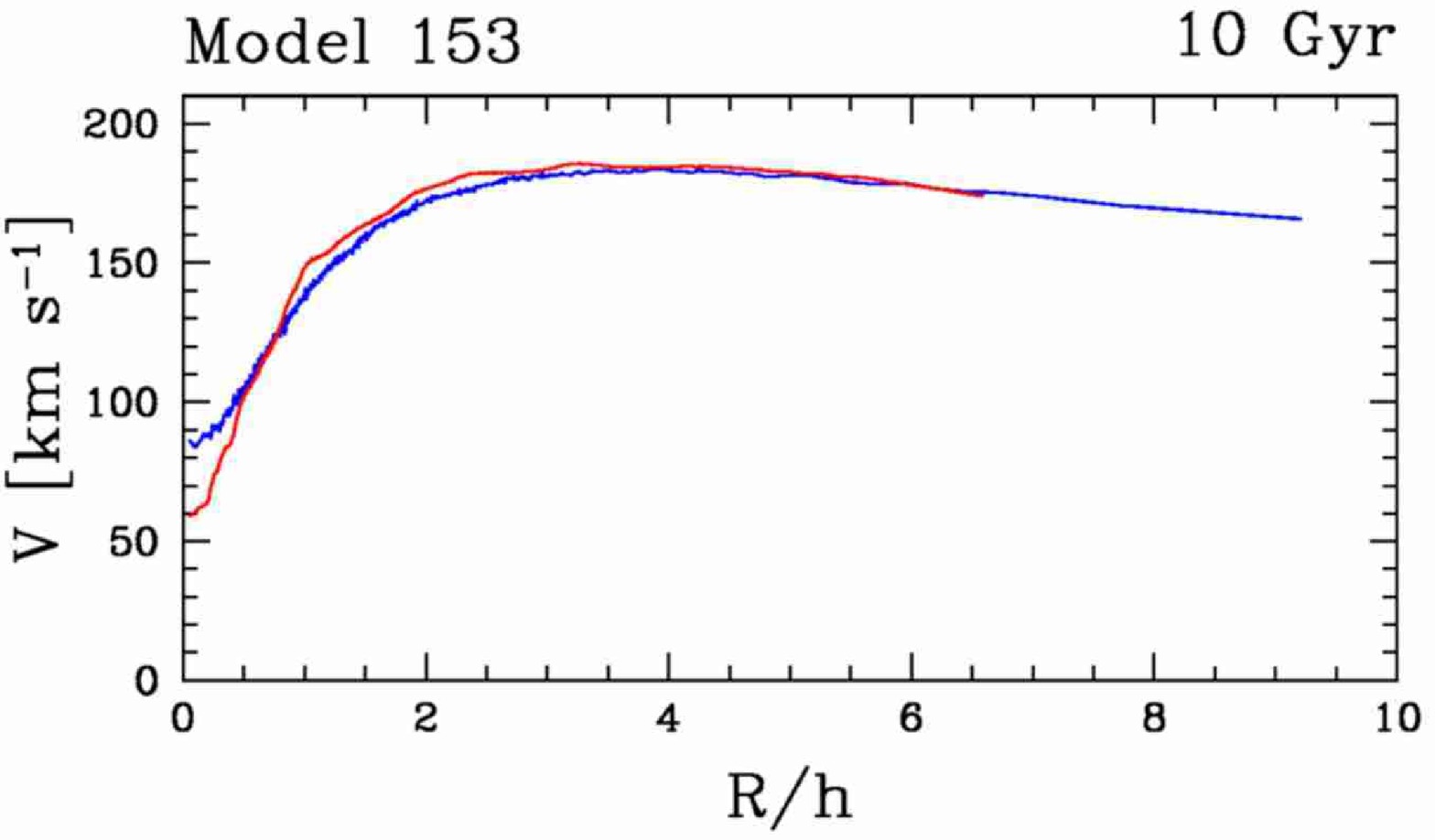}
     \epsfxsize=8cm\epsfbox{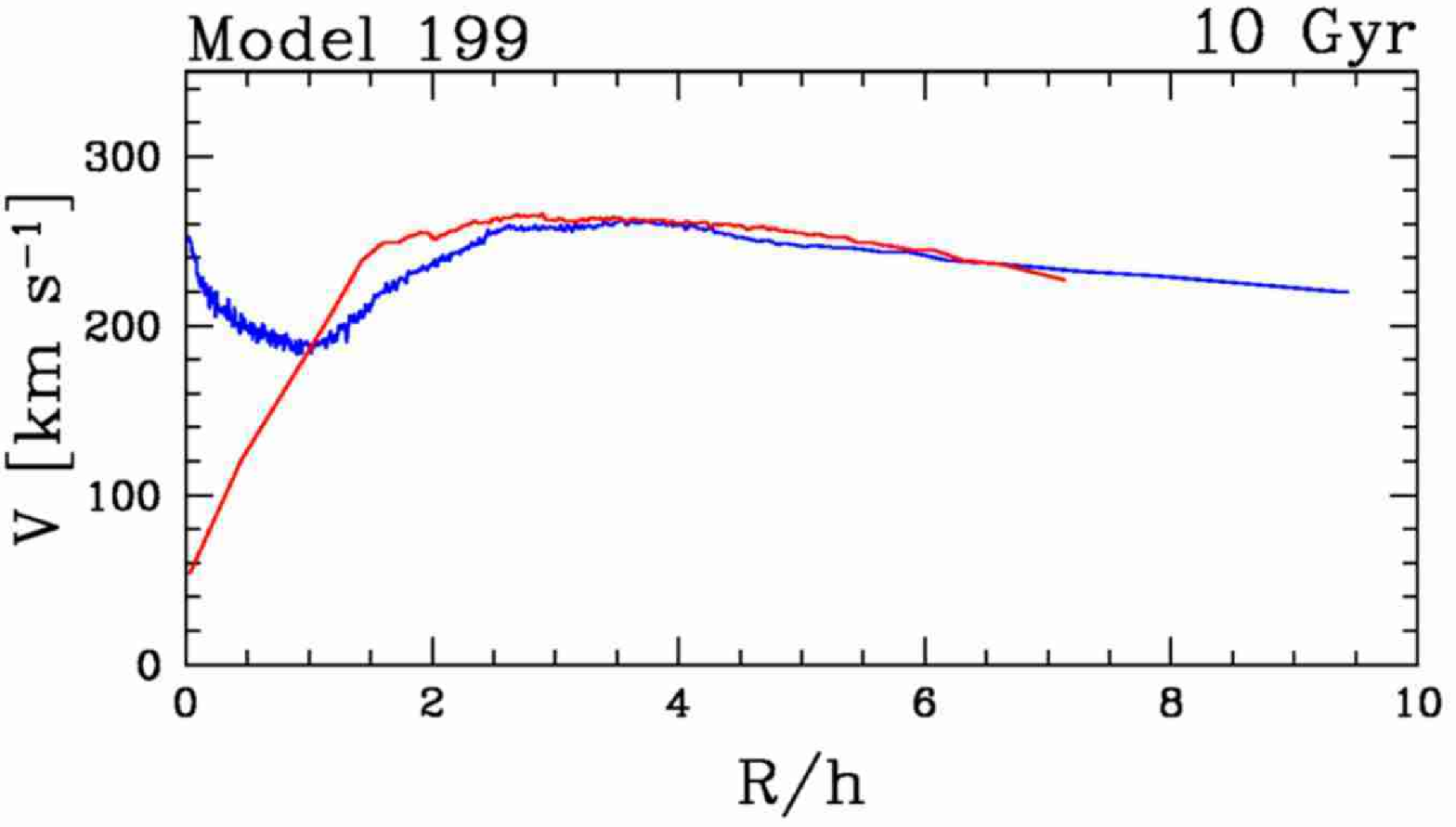}
\rm
\caption{Four model rotation curves after 10 Gyr of evolution.  The rotational velocity is plotted 
versus the number of disk scale lengths for the disk stars (blue) and gas (red).}
\label{rot_curves}
\end{figure*}

Fig.~\ref{initQrot} shows the initial rotation curves and the initial Q-curves for four galaxies.  
Fig.~\ref{rot_curves} shows four sample rotation curves for the gas and stars for the same galaxies 
evolved over 10 Gyr.  These curves all have realistic shapes.

%%%%%%%%%%%%%%%%%%%%%%%%%%%%%%%%%%%%%%%%%%%%%%%%%%%%%%%%%%%%%%%%%%%
%%%%%%%%%%%%%%%%%%%%%%%%%%%%%%%%%%%%%%%%%%%%%%%%%%%%%%%%%%%%%%%%%
%%%%%%%%%%%%%%%%%%%%%%%%%%%%%%%%%%%%%%%%%%%%%%%%%%%%%%%%%%%%%%%%%%%
%%%%%%%%%%%%%%%%%%%%%%%%%%%%%%%%%%%%%%%%%%%%%%%%%%%%%%%%%%%%%%%%%

\subsection{Analysis}
Lagrangian density profiles were derived from the simulation outputs 
with equal bin sizes of 1000 particles. The mean radius and surface density for each 
bin was
determined by calculating the mean radius of the particles and the total
mass within the circular area determined by the outer radius of that
bin.  Gas, newly formed star, or pre-existing disk or bulge 
star particles were all analyzed as individual species.  

To assess the locations of breaks we developed two 
separate numerical methods and compared these to 
eye-ball estimates. We summarize these methods in Appendix B.  In the end, we adopted the eye-ball estimates for our analysis.

\section{Results}
\subsection{Profile Breaks}
\begin{figure}
\centering

     \epsfxsize=8cm\epsfbox{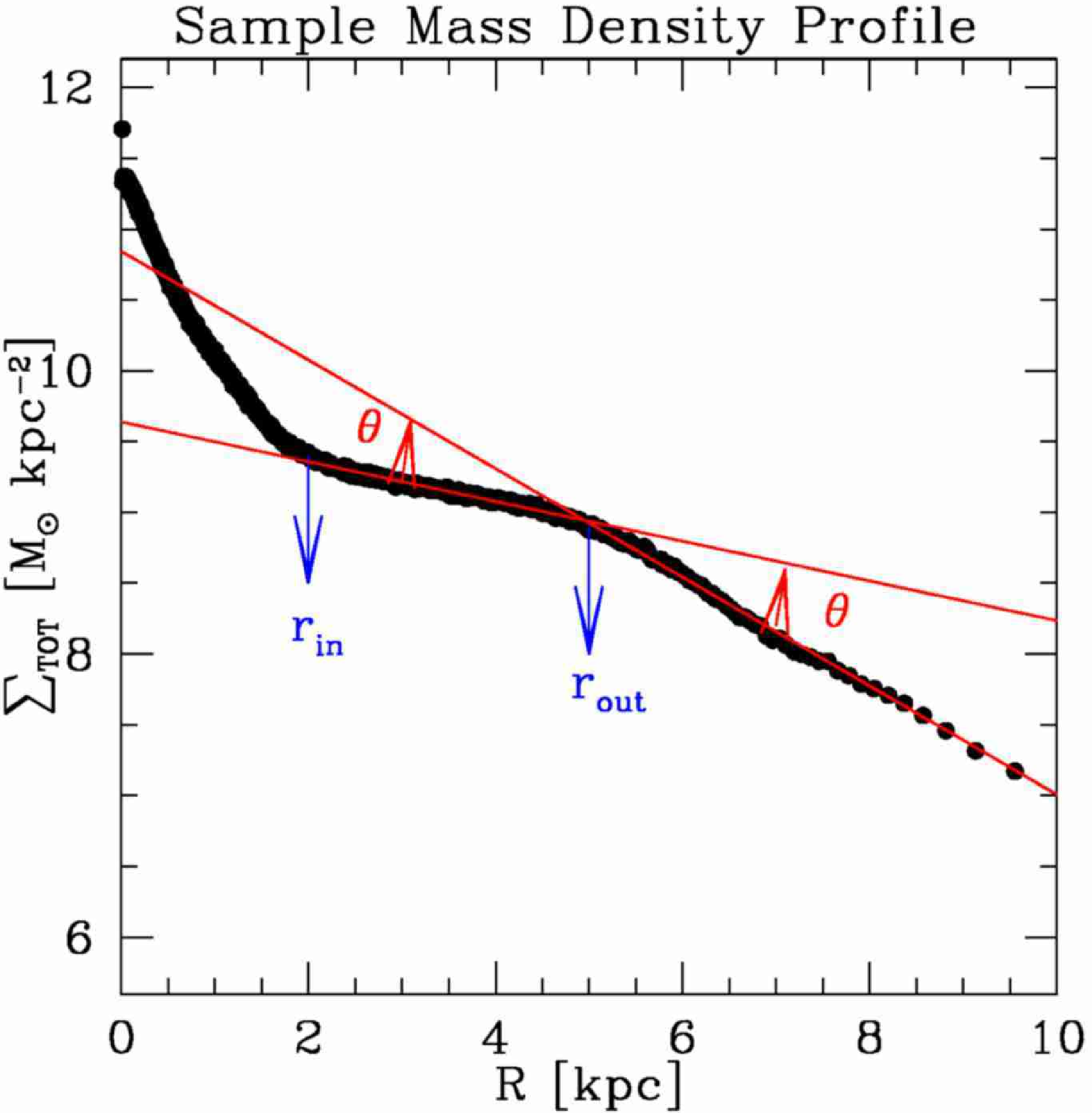}
\rm

\caption{A sample total (stars $+$ gas) density profile, showing the locations of inner and outer 
break radii, $r_{in}$ and $r_{out}$.  The angle $\theta$ measures the sharpness of the break from the
inner to the outer profile.}

\label{mark}
\end{figure}

Both the mass density profiles for the gas and newly formed stars
will be controlled primarily by the star formation gas density threshold,
$\rho_{th}$. Thus, we expect to see breaks in the
density profiles for the gas and newly formed stars.  By construction, the
gas will be depleted to a value of $\sim$10 M$_{\odot}$ pc$^{-2}$,
producing an inner plateau in the density profile. Stars will form in the
interior regions, but we expect little or no star formation where the
average gas density is below $\rho_{th}$.  While the dynamics of the disk
will cause matter redistribution in both profiles, the star formation
threshold will always produce a break feature.  The total baryonic component, on the other hand, consists predominantly of
pre-assembled stars in the initial models.  Their initial density profile
is described by a straight exponential function.  Deviations from a
straight exponential profile implies that matter has been redistributed as
a result of the dynamical properties of the disk.  Thus, we can distinguish
between two types of breaks: those formed by a SF threshold\footnote{This study is not centered with the specifics of breaks caused by SF since we do not vary the SF parameters.  We focus solely on the dependences of breaks and dynamical processes.} (in the gas and
newly formed stars) and those formed by dynamical properties (in the total
baryonic component).

\begin{figure}
\centering

     \epsfxsize=8cm\epsfbox{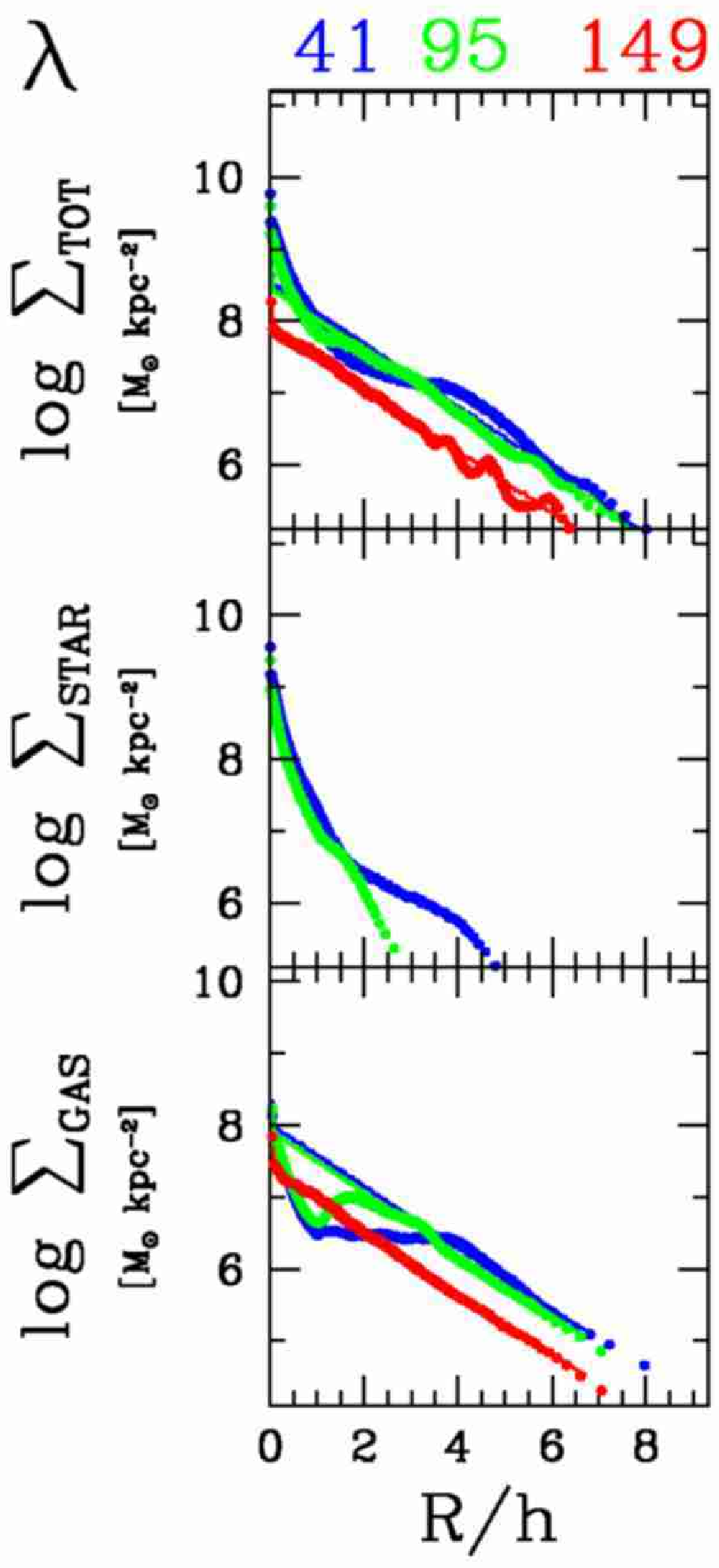}

\rm

\caption{Surface mass densities vs. $R/h$ for different values of $\lambda=$ 0.02 (blue), 0.03 (green) 
and 0.08 (red), respectively after 5 Gyr of evolution.  The coloured numbers at the top correspond to 
specific galaxy models and $h$ is the initial disk scale length of the model.  The other model 
parameters were held constant at $m_{d}$=0.025, $c$=5, $V_{200}$=80 and $m_{b}$=0.  The straight lines 
show the initial density profiles for the gas and total components.}

\label{profspin}
\end{figure}

\begin{figure}
\centering

     \epsfxsize=8cm\epsfbox{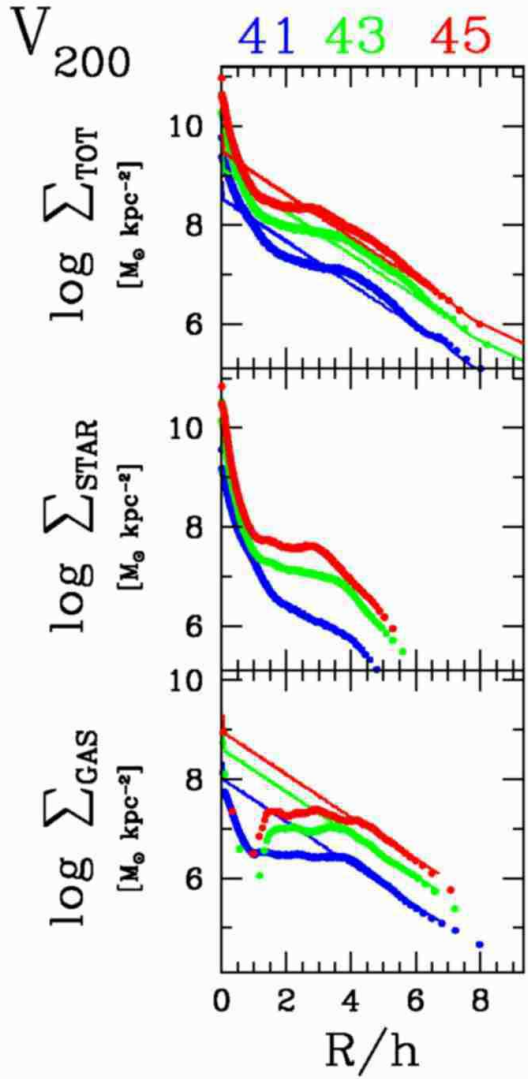}

\rm

\caption{Surface mass densities vs. $R/h$ for different values of $V_{200}$=80 (blue), 160 (green) and 
180 km s$^{-1}$ (red), respectively after 5 Gyr of evolution.  The coloured numbers correspond to 
specific galaxy models.  The other model parameters were held constant at $m_{d}$=0.025, $c$=5, 
$\lambda$=0.02 and $m_{b}$=0. The straight lines show the initial density profiles.}

\label{profvel}
\end{figure}

\begin{figure}
\centering

     \epsfxsize=8cm\epsfbox{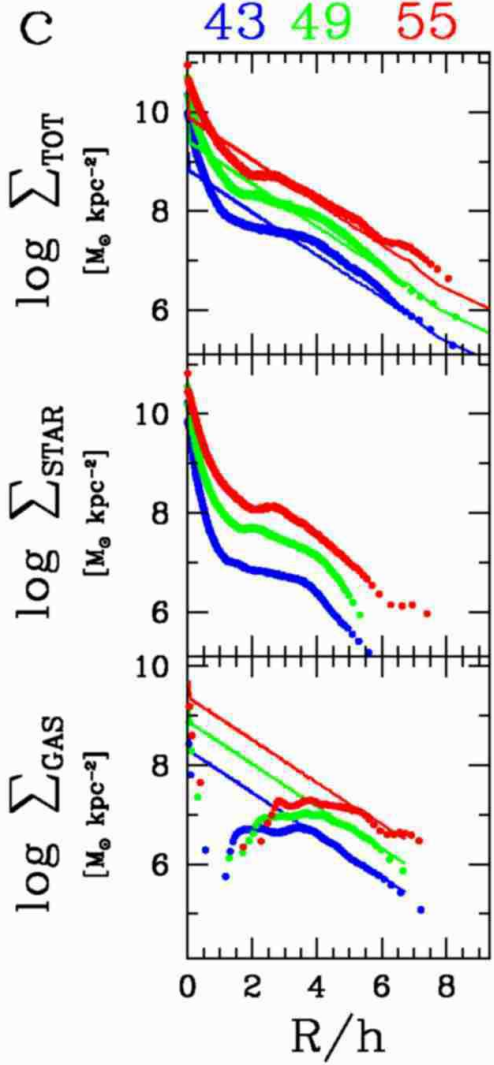}

\rm

\caption{Surface mass densities vs. $R/h$ for different values of $c=$5 (blue), 10 (green) and 15 
(red), respectively after 5 Gyr of evolution.  The other model parameters were held constant at 
$m_{d}$= 0.025, $V_{200}$=160, $\lambda$=0.02 and $m_{b}$=0. Other lines and labels are as in 
Fig.~\ref{profvel}.}

\label{profconc}
\end{figure}

\begin{figure}
\centering

     \epsfxsize=8cm\epsfbox{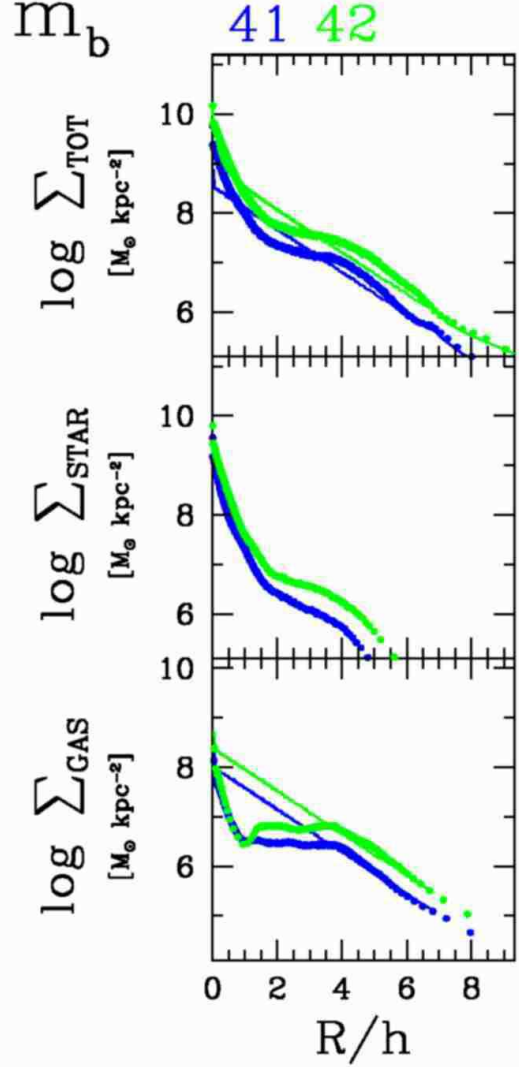}

\rm
\caption{Surface mass densities vs. $R/h$ for no bulge case (blue) or a bulge with 20\% of the disk 
mass (green) after 5 Gyr of evolution.  The other model parameters were held constant at 
$V_{200}$=160, $c$=5, $\lambda$=0.03 and $m_{d}$=0.025. Other lines and labels are as in 
Fig.~\ref{profvel}.}
\label{profmb}
\end{figure}

\begin{figure}
\centering

     \epsfxsize=8cm\epsfbox{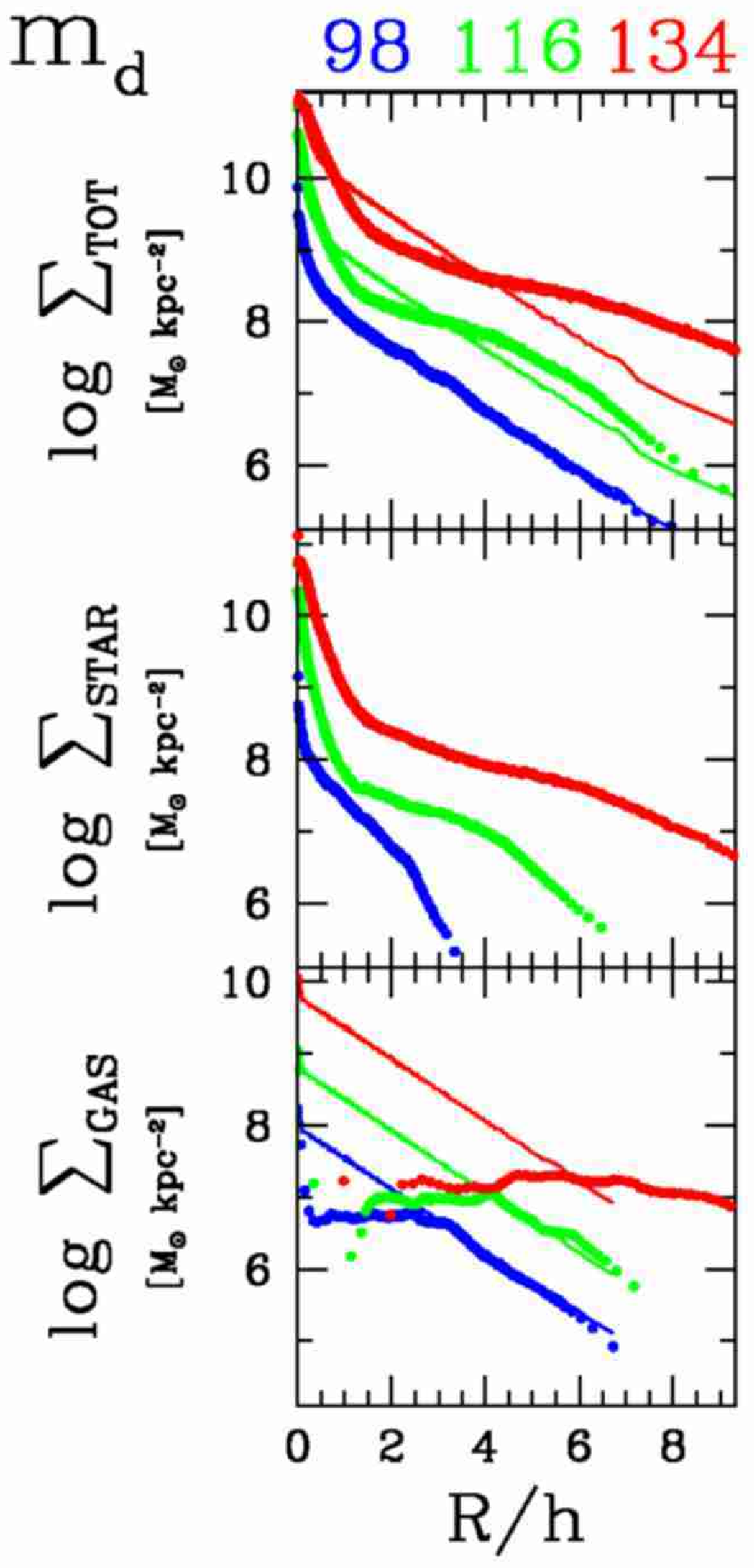}

\rm
\caption{Surface mass densities vs. $R/h$ for different values of $m_{d}$=0.025 (blue), 0.05 (green) and 0.1 (red) respectively after 5 Gyr of evolution.   The other model parameters were held constant at $V_{200}$=160, $c$=5, $\lambda$=0.03 and $m_{b}$=0. Other lines and labels are as in Fig.~\ref{profvel}.}
\label{profmd}
\end{figure}

The azimuthally-averaged mass density profiles for the three components -
gas, newly formed stars and the total baryons - are shown in Figs.~\ref{profspin}-~\ref{profmd}.  
The initial profiles at $t=0$ and the 5 Gyr profiles are depicted by
straight lines and filled circles, respectively.  The mass density profiles
are plotted in terms of the {\it initial} disk scale length in order to
provide a scaled comparison for all galaxies.  A first inspection shows
that the mass density profiles deviate significantly from the initial
profiles as $\lambda$ is decreased and $m_{d}$ is increased (Figs.~\ref{profspin} and ~\ref{profmd}).  The effects of $c$, $V_{200}$ and $m_{b}$ are less significant.
%%%%%%%%%%%%%%%%%%%%%%%%%%%%%%%%%%%%%%%%%%%%%%%%%%%%%%%%%%%%%%%%%%%
%%%%%%%%%%%%%%%%%%%%%%%%%%%%%%%%%%%%%%%%%%%%%%%%%%%%%%%%%%%%%%%%%
%%%%%%%%%%%%%%%%%%%%%%%%%%%%%%%%%%%%%%%%%%%%%%%%%%%%%%%%%%%%%%%%%%%
%%%%%%%%%%%%%%%%%%%%%%%%%%%%%%%%%%%%%%%%%%%%%%%%%%%%%%%%%%%%%%%%%
%%%%%%%%%%%%%%%%%%%%%%%%%%%%%%%%%%%%%%%%%%%%%%%%%%%%%%%%%%%%%%%%%%%
%%%%%%%%%%%%%%%%%%%%%%%%%%%%%%%%%%%%%%%%%%%%%%%%%%%%%%%%%%%%%%%%%
%%%%%%%%%%%%%%%%%%%%%%%%%%%%%%%%%%%%%%%%%%%%%%%%%%%%%%%%%%%%%%%%%%%
%%%%%%%%%%%%%%%%%%%%%%%%%%%%%%%%%%%%%%%%%%%%%%%%%%%%%%%%%%%%%%%%%
%%%%%%%%%%%%%%%%%%%%%%%%%%%%%%%%%%%%%%%%%%%%%%%%%%%%%%%%%%%%%%%%%%%

We can quantify the matter redistribution in the galaxy disk via any change
in the galaxy density concentration, $C_{28}\equiv5 \log(r_{80}/r_{20})$, where
$r_{20}$ and $r_{80}$ are the radii enclosing 20\% and 80\% of the total
baryonic mass.  A high value of $C_{28}$ may be due to a bar or large
bulge.  A straight exponential profile has $C_{28}$=2.8, while 
our galaxy
models with a bulge have an initial $C_{28} \approx 3.5$.

Under the action of angular momentum redistribution, matter in the inner parts of the disk (typically within a bar radius) will be driven to the nuclear regions.  This yields a flattening of the inner density profile.  The extension of this inner plateau to the outer disk produces an outer disk break feature at the location of $r_{out}/h$.  The $C_{28}$ value and the location of $r_{out}/h$ can thus be used as diagnostics for the extent of matter redistribution in the disk.  The development of a break feature and an increase of $r_{out}/h$ imply that the length of the inner plateau is growing and that matter redistribution is effective across the whole inner disk.

\begin{figure}
\centering

     \epsfxsize=8cm\epsfbox{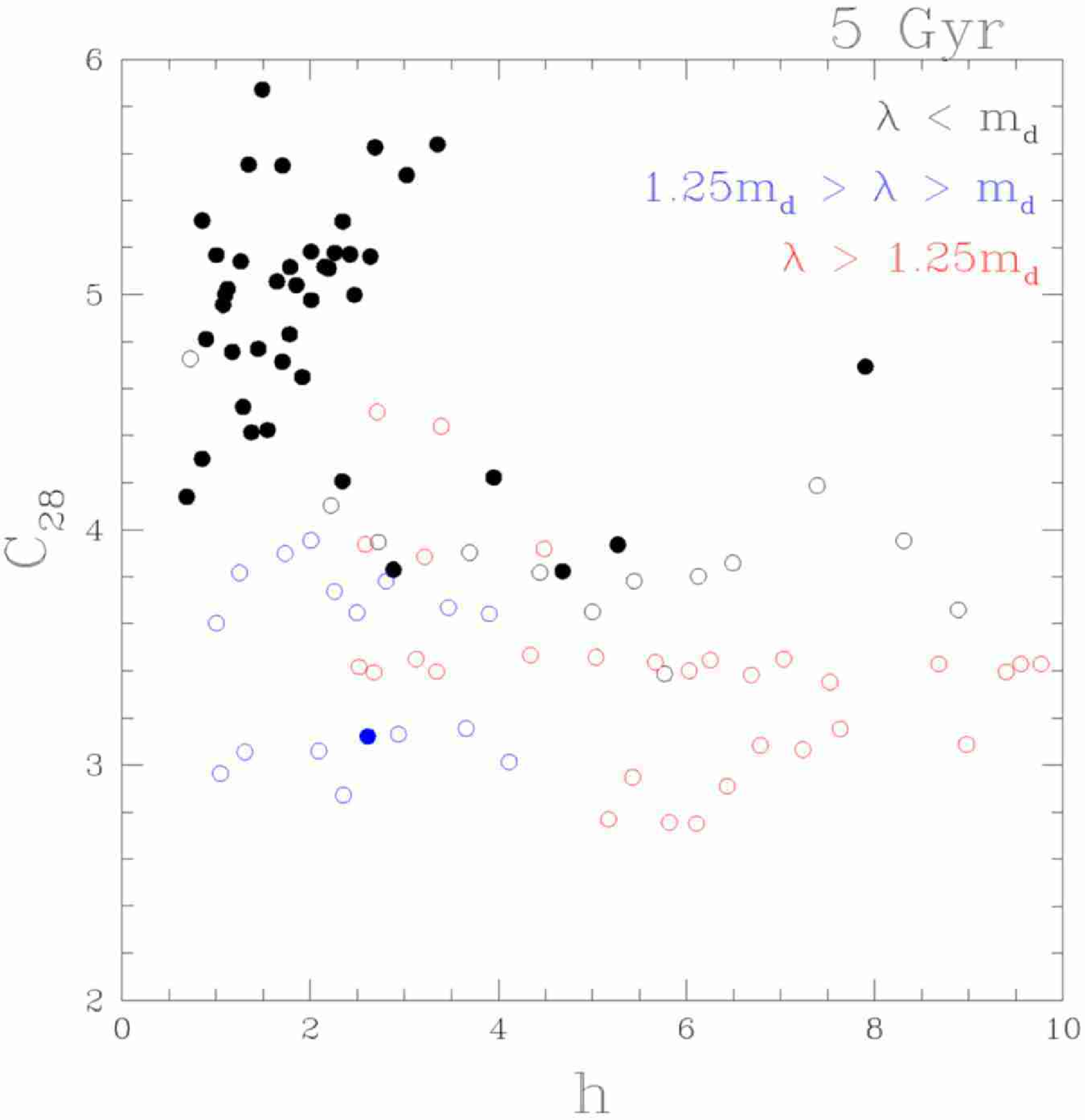}
     \epsfxsize=8cm\epsfbox{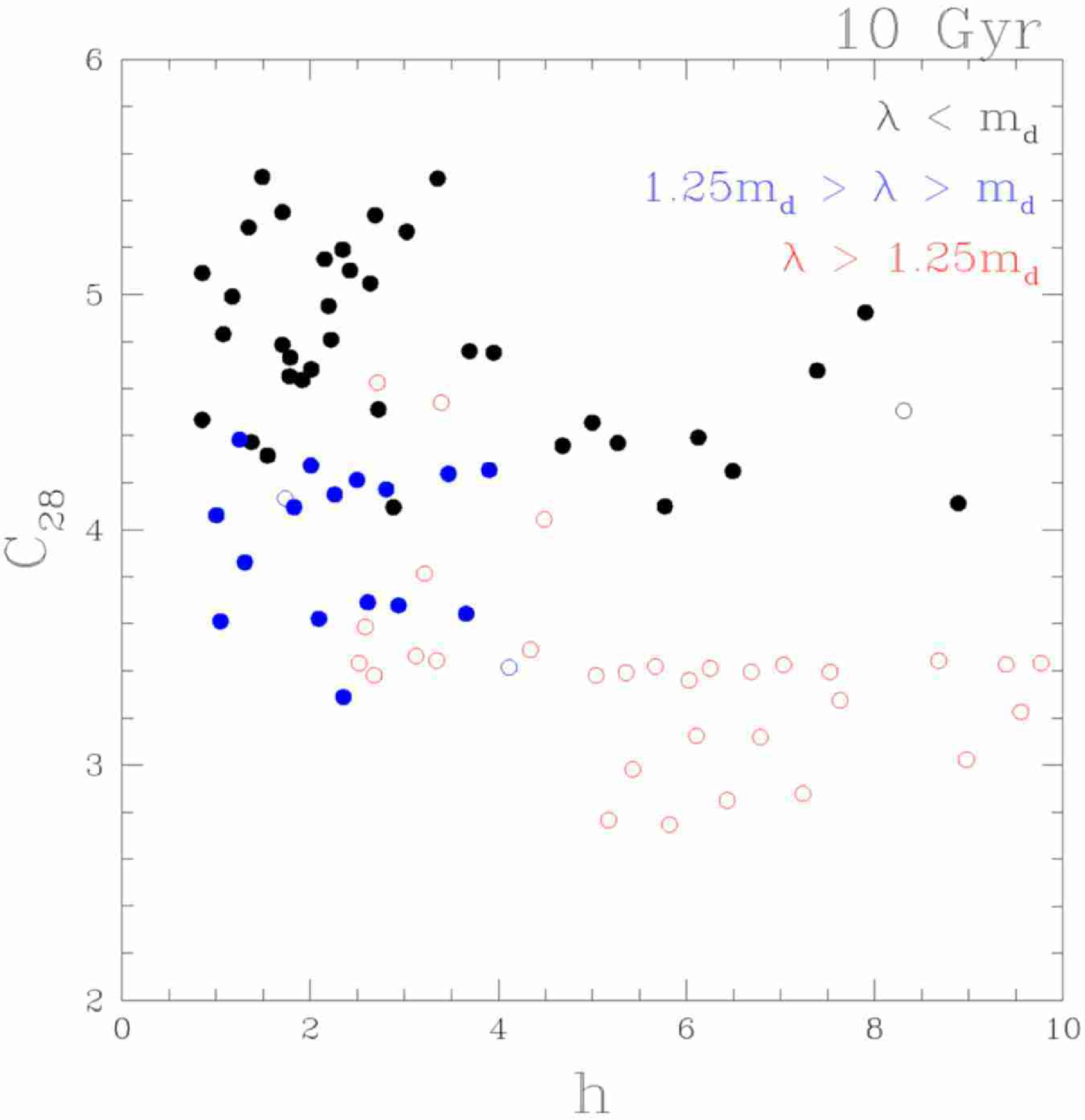}

\rm
\caption{Galaxy concentration, $C_{28}$, versus initial disk scale length for galaxies which develop breaks (filled circles) or not (open circles) in the total profiles evolved over 5 and 10 Gyr.}
\label{bkc28}
\end{figure}

 Given any stable galaxy model, we wish to determine if it will develop any
profile break and if so, what parameters control such a break?  Roughly
half of our galaxy models developed breaks in the total profile. Fig.~\ref{bkc28}
highlights those galaxies with a break (filled circles) in the
total profile versus those that did not develop a break (open circles) in terms of the initial
disk scale length, $h$, and the galaxy concentration, $C_{28}$.  Galaxies
that exhibit total profile breaks tend to have small initial disk scale
lengths and thus high concentrations.  This could be verified with existing galaxy samples that cover a broad range of surface brightnesses with identified break radii.  Such samples are currently being assembled.

Fig.~\ref{bkc28} shows the distribution of galaxies with (filled circles) and
without (open circles) breaks after 5 and 10 Gyr.  Regardless of time,
galaxies with $\lambda > 1.25m_{d}$ \footnote{Even though these galaxies may exhibit a minimum in their initial and final Q distribution.} never produce breaks.  However, in the
intermediate range with $1.25 m_{d}> \lambda > m_{d}$, breaks are seen
after 10 Gyr, even though most showed no breaks after only 5 Gyr. Indeed,
this comparison between the 5 and 10 Gyr snapshots suggests that galaxies
with a high ratio of $m_{d}/\lambda$ form breaks quickly.  As
$m_{d}/\lambda$ decreases, the physical time for break
formation decreases (due to the decreased dynamical timescale) and once
$m_{d}/\lambda<1$ breaks cease to form. It is worth noting that while the 
simulated
galaxies are compared at the same physical times, their orbital or
dynamical times (based on the number of rotations) will differ depending on
their density, with dense systems having shorter timescales.

\subsection{$m_{d}/\lambda$ Ratio}
The fact that $m_{d}$ and $\lambda$ control the mass density profile evolution to such a large extent can be understood in terms of disk stability arguments (see Eq. 2).  If $\lambda$ is less than $m_{d}$, the galaxy will be bar unstable.  Bar formation also plays a significant role in redistributing mass in the disk plane (D06), as material funneled by the bar produces a higher central density.

The direct correlation between break formation and the ratio $m_{d}/\lambda$ supports the notion that total profile breaks are associated with angular momentum redistribution.  Furthermore, all of the galaxies exhibiting a two-component profile also develop bars thus suggesting that profile breaks likely result from dynamical resonances induced by the central bar. 

\begin{figure}
\centering

     \epsfxsize=8cm\epsfbox{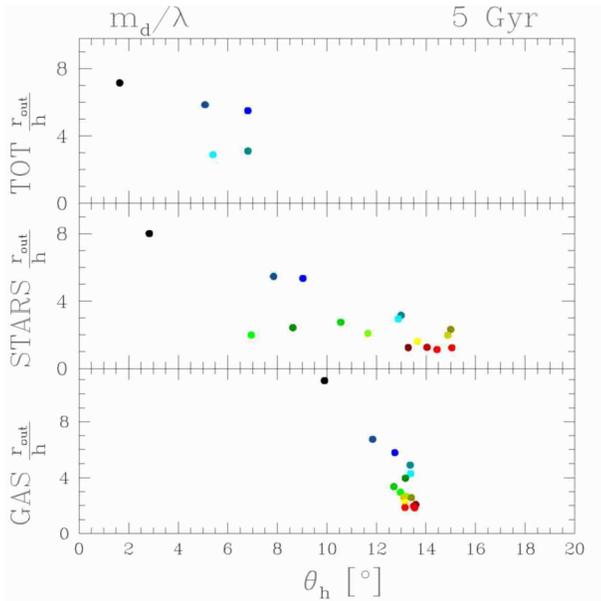}

\rm
\caption{Comparison of $r_{out}/h$ with $\theta_{h}$ for galaxies evolved up to 5 Gyr with different values of $\lambda$ ($\lambda$=0.015-0.1 from blue to red). The disk mass fraction is held constant at $m_{d}$=0.05.}
\label{spin3win}
\end{figure}

In order to closely examine properties of $m_{d}/\lambda$, we ran a new suite of additional simulations where we varied the spin parameter in increments of 0.005 from $\lambda=0.015$ to $0.1$.  All the other galaxy parameters were held constant ($c=10$, $V_{200}=160$, $m_{d}=0.05$, $m_{b}=0.2m_{d}$). The simulations were evolved for 5 Gyr.  Fig.~\ref{spin3win} shows the variations of $r_{out}/h$ and the strength of the break, $\theta_{h}$, for the gas, stars and total profile as $\lambda$ is varied from high (red) to low (blue) values. 

We see in Fig.~\ref{spin3win} that $r_{out}$ grows to greater radii in the gas and stars as $\lambda$ is reduced.  In the total profile, the development of a break only occurs for $\lambda \leq 0.04$.  Recall that our disk mass fraction was held constant at 0.05.  Indeed, $m_{d}/\lambda>1$, the disk will likely develop a two-component profile. It is also useful to calculate a bar strength, in addition to $C_{28}$, to examine any correlation with $m_{d}/\lambda$.  Galaxies with a high bar strength should also have a high galaxy concentration, but the latter does not necessarily imply the former if a galaxy has a large (classical) bulge formed via mergers.  The bar strength, $A_{\phi}$, is calculated as:
\begin{equation}
A_{\phi}=\frac{1}{N}|\Sigma_{j}e^{2i\phi_{j}}|
\end{equation}
where $\phi_{j}$ is the two dimensional cylindrical polar angle coordinate of particle $j$ (Sanders 1977; Athanassoula \& Misiriotis 2002; D06).

\begin{figure}
\centering

     \epsfxsize=8cm\epsfbox{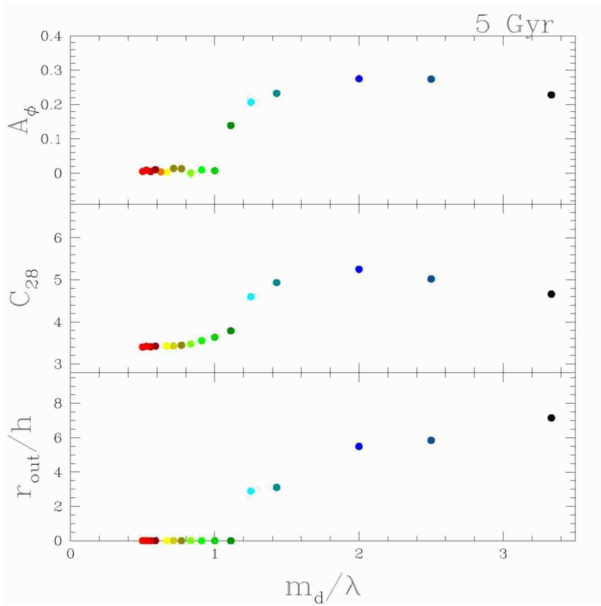}

\rm
\caption{$C_{28}$ and $A_{\phi}$ for galaxies with different values of $\lambda$ (see Fig.~\ref{spin3win}).  The disk mass fraction is held constant at $m_{d}$=0.05.}
\label{spinc28}
\end{figure}

Fig.~\ref{spinc28} shows the effect of $\lambda$ on  $C_{28}$ and $A_{\phi}$. The development of a break in the total profile is clearly correlated with a sudden increase in $C_{28}$ and the strength of the bar.  These results reaffirm the findings of D06 that the development of a bar ultimately drives the total profile breaks.  However, beyond $m_{d}/\lambda>2$, $A_{\phi}$ remains roughly constant between 0.2-0.3 and the galaxy concentration decreases; $r_{out}/h$, on the other hand, increases linearly with $m_{d}/\lambda$.  For $m_{d}/\lambda=1.4$-3.3, $r_{out}/h$ doubles.  Thus, while the development of a break clearly correlates with the development of a bar, the location of $r_{out}/h$ is not strictly tied to the strength of the bar and the galaxy concentration.  As $m_{d}/\lambda$ increases, so does $r_{out}/h$, regardless of $A_{\phi}$ and $C_{28}$.

\begin{figure}
\centering

     \epsfxsize=8cm\epsfbox{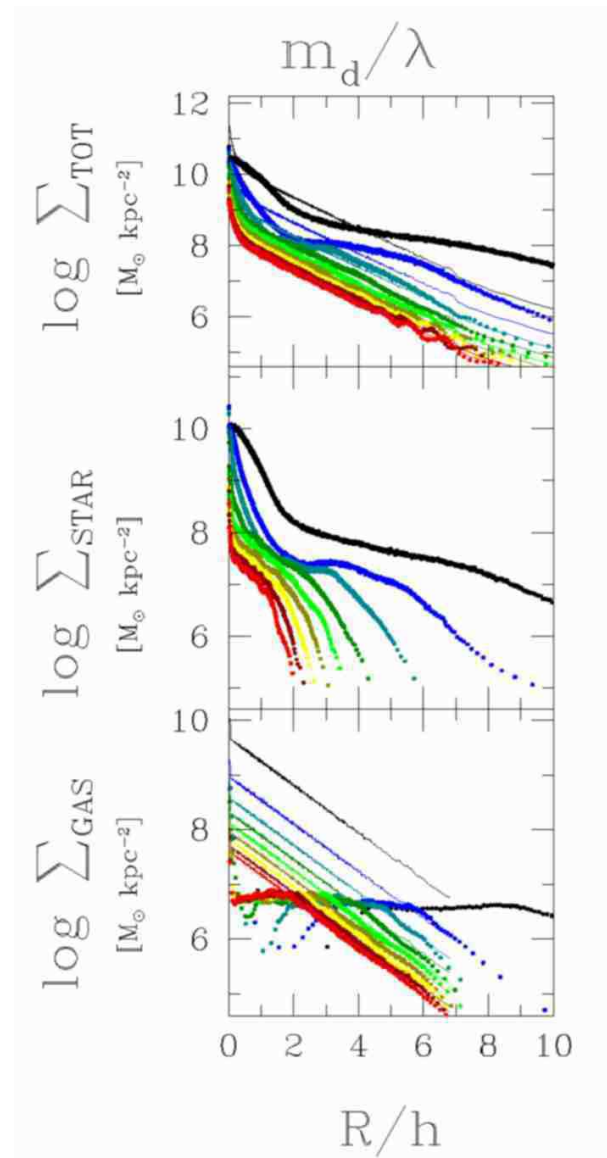}

\rm

\caption{Radial mass density profiles for the gas, stars and total baryonic components.  $\lambda$ 
values increase from blue to red ($\lambda$=0.015-0.1).  The disk mass fraction is held constant at 
$m_{d}$=0.05.}

\label{spinprof}
\end{figure}

Fig.~\ref{spinprof} shows a comparison of the radial profiles for the three galaxy
components - gas, stars and total baryons.  As $m_{d}/\lambda$
increases, the profiles deviate significantly from their initial state,
confirming that this ratio controls the extent to which matter is
redistributed into the disk.  For the smallest value of $\lambda=0.015$, the two-component profile extends unusually far into the disk, with the inner plateau making up most of the plotted profile. However, this low value of $\lambda$ is somewhat unrealistic and considered here for illustrative purposes only.  In the gaseous profiles, the interior regions
flatten to a common density as expected in the context of a star formation
threshold.  Once the gas density is reduced to a low value, star formation
ceases and gas depletion is halted.  Given that the high
$m_{d}/\lambda$ galaxies have an overall higher initial gas
density, we expect these galaxies to form significantly more stars.  
Indeed the middle panel of Fig.~\ref{spinprof} shows that stellar profiles are much
more extended and have higher densities when $m_{d}/\lambda$ is high.  
While galaxy breaks correlate with the development of a bar and a high
galaxy concentration, the bar strength does not correlate trivially with the location of $r_{out}/h$.  Recall also that these 
breaks are
not associated with a star formation threshold as the total baryonic
component is dominated by stars that were distributed exponentially in the
initial galaxy models.

%%%%%%%%%%%%%%%%%%%%%%%%%%%%%%%%%%%%%%%%%%%%%%%%%%%%%%%%%%%%%%%%%
%%%%%%%%%%%%%%%%%%%%%%%%%%%%%%%%%%%%%%%%%%%%%%%%%%%%%%%%%%%%%%%%%%%
%%%%%%%%%%%%%%%%%%%%%%%%%%%%%%%%%%%%%%%%%%%%%%%%%%%%%%%%%%%%%%%%%

%%%%%%%%%%%%%%%%%%%%%%%%%%%%%%%%%%%%%%%%%%%%%%%%%%%%%%%%%%%%%%%%%%%
%%%%%%%%%%%%%%%%%%%%%%%%%%%%%%%%%%%%%%%%%%%%%%%%%%%%%%%%%%%%%%%%%

The parameters $m_{d}$ and $\lambda$ are known to regulate many fundamental disk galaxy properties, including the central surface brightness and the disk scale length (DSS97; MMW98; HS06; Dutton et al. 2007).    The volume density of the disk scales as $(z_{o}h^{2})^{-1}$, where $z_{o}$ is the scale height of the disk and $z_{o} \propto \lambda^{-1}$, whereas $h \propto \lambda$.  The volume density of the disk scales as $\lambda^{-1}$ so clearly the density of the disk will increase linearly with $m_{d}/\lambda$.  An increase in $\lambda$ means a decrease in the potential well of the disk.  Low $\lambda$ galaxies are more compact and have enhanced self-gravity producing a higher rotational velocity.  The increased efficiency of low $\lambda$ galaxies to redistribute their angular momentum explains the development of breaks in these cases.  Due to the rapid rotation, the dynamical timescale of a galaxy also increases with decreasing $\lambda$.  This may explain why lowest $\lambda$ galaxies appear to have declining bar strengths and total concentrations.  Due to the increased dynamical time, the bar dissolves earlier on.  However, bar dissolution does not remove the break.  Thus, while the processes of bar and break formation are closely tied, one does not necessarily imply the other. 

We now examine whether the ratio of $m_{d}/\lambda$ also controls the time of 
break formation.  Since lower $\lambda$ galaxies have higher rotational velocities and shorter dynamical times, we expect breaks to develop sooner in compact galaxies given the enhanced efficiency of angular momentum redistribution.  Thus, breaks should develop quickly for high values of $m_{d}/\lambda$.

\begin{figure}
\centering

     \epsfxsize=8cm\epsfbox{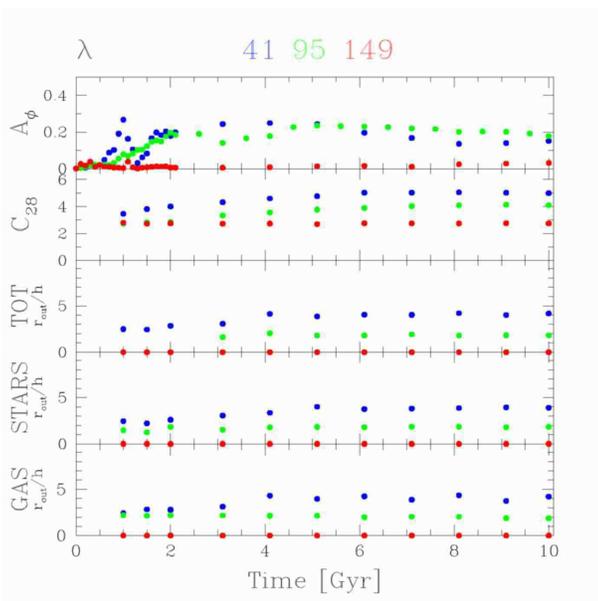}

\rm
\caption{Time evolution of $r_{out}/h$, $C_{28}$ and $A_{\phi}$ as a function of $\lambda$ for three randomly selected galaxies.  Blue, green and red correspond to  $\lambda=$ 0.02, 0.03 and 0.08, respectively.  The coloured numbers at the top are the model numbers.  The other model parameters were held constant at $m_{d}$=0.025, $c$=5, $V_{200}$=80 and $m_{b}$=0.}
\label{spintimebar}
\end{figure}

\begin{figure}
\centering

     \epsfxsize=8cm\epsfbox{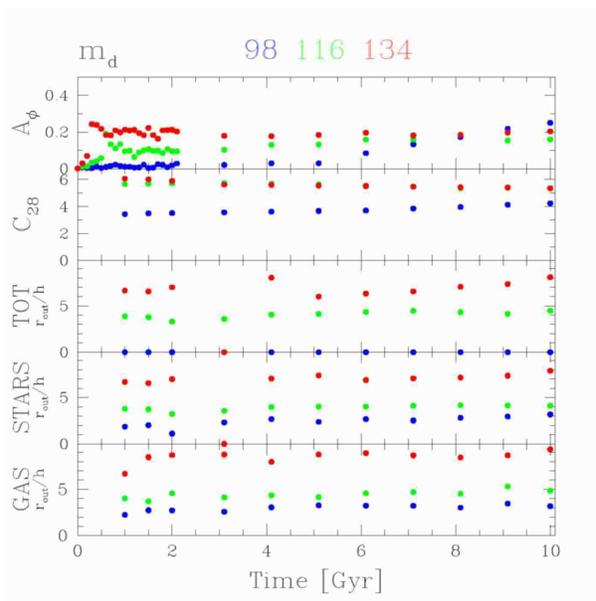}

\rm
\caption{Time evolution of $r_{out}/h$, $C_{28}$ and $A_{\phi}$ as a function of $m_{d}$ for three randomly selected galaxies selected at random.  Blue, green and red correspond to $m_{d}$ of 0.025, 0.05 and 0.1, respectively.  The coloured numbers at the are to the model numbers.  The other model parameters were held constant at $\lambda$=0.03, $c$=5, $V_{200}$=160 and $m_{b}$=0.}
\label{mdtimebar}
\end{figure}

Fig.~\ref{spintimebar} shows the time evolution of $r_{out}/h$ for the gas, stars and total baryons for three model galaxies where $\lambda$ is varied from 0.02 (model 41), 0.03 (model 95), and 0.08 (model 149).  Model 149, with the highest $\lambda$, shows no break features in any of the three profiles; while we always expect breaks in the stars and gas due to the star formation threshold in GADGET-2, this galaxy represents an exception since no stars ever formed over the course of its evolution.  The galaxy concentration of model 149 remains constant at $C_{28}=$2.8 as expected for a pure exponential disk.  Furthermore, model 149 shows no evidence of bar formation.  On the other hand model 41 with the lowest $\lambda$ develops breaks in all three profiles early on.  The galaxy concentration increases rapidly as does the bar strength.  While in the first few Gyr, there is a slight evolution of the break towards higher values of $r_{out}/h$, we see that beyond 4 Gyr the break location remains constant in all three profiles.  The galaxy concentration also remains constant, whereas the bar strength declines consistently beyond 5 Gyr.  Model 95, with its intermediate $\lambda$, develops a break early on in the gas and stars; however, a break develops in the total profile only after 3 Gyr.  Model 95 displays a rapid growth in disk concentration and bar strength at early times.  Thus, we see that despite the early presence of a bar, the break only develops later.  Once formed the break location remain fixed, but like model 41, model 95 displays a  bar strength strength beyond 5 Gyr. At all times, the location of $r_{out}/h$ is greater for model 41 than it is for model 95 due to the decreased value of $m_{d}/\lambda$.  Fig.~\ref{mdtimebar} depicts the time evolution of three galaxies with varying $m_{d}$ 
($m_{d}$=0.025, 0.05 and 0.1).  Once again, increasing $m_{d}$ moves $r_{out}/h$ further out into the disk.  Model 98 ($m_{d}$=0.025) never developed a break in the total profile, whereas model 116 ($m_{d}$=0.05) developed a break at $\sim$4.5$h$ and model 134 ($m_{d}$=0.1) developed a break at $\sim$6.5$h$.  The break locations remain fairly constant with time, although model 134 displays some variations in the first 5 Gyr. Initially model 98 showed no evidence of a bar or break; however, a bar develops after 5 Gyr, but there is no evidence of the subsequent development of a break.  Thus, we have seen that all of our galaxies with breaks in the total density profiles show evidence of bars.  However, some of our galaxies (\eg model 98) developed bars but did not develop breaks.  Those galaxies are characterized by  $m_{d}/\lambda < 0.9$. 

%%%%%%%%%%%%%%%%%%%%%%%%%%%%%%%%%%%%%%%%%%%%%%%%%%%%%%%%%%%%%%%%%%%
%%%%%%%%%%%%%%%%%%%%%%%%%%%%%%%%%%%%%%%%%%%%%%%%%%%%%%%%%%%%%%%%%

\begin{figure}
\centering

     \epsfxsize=8cm\epsfbox{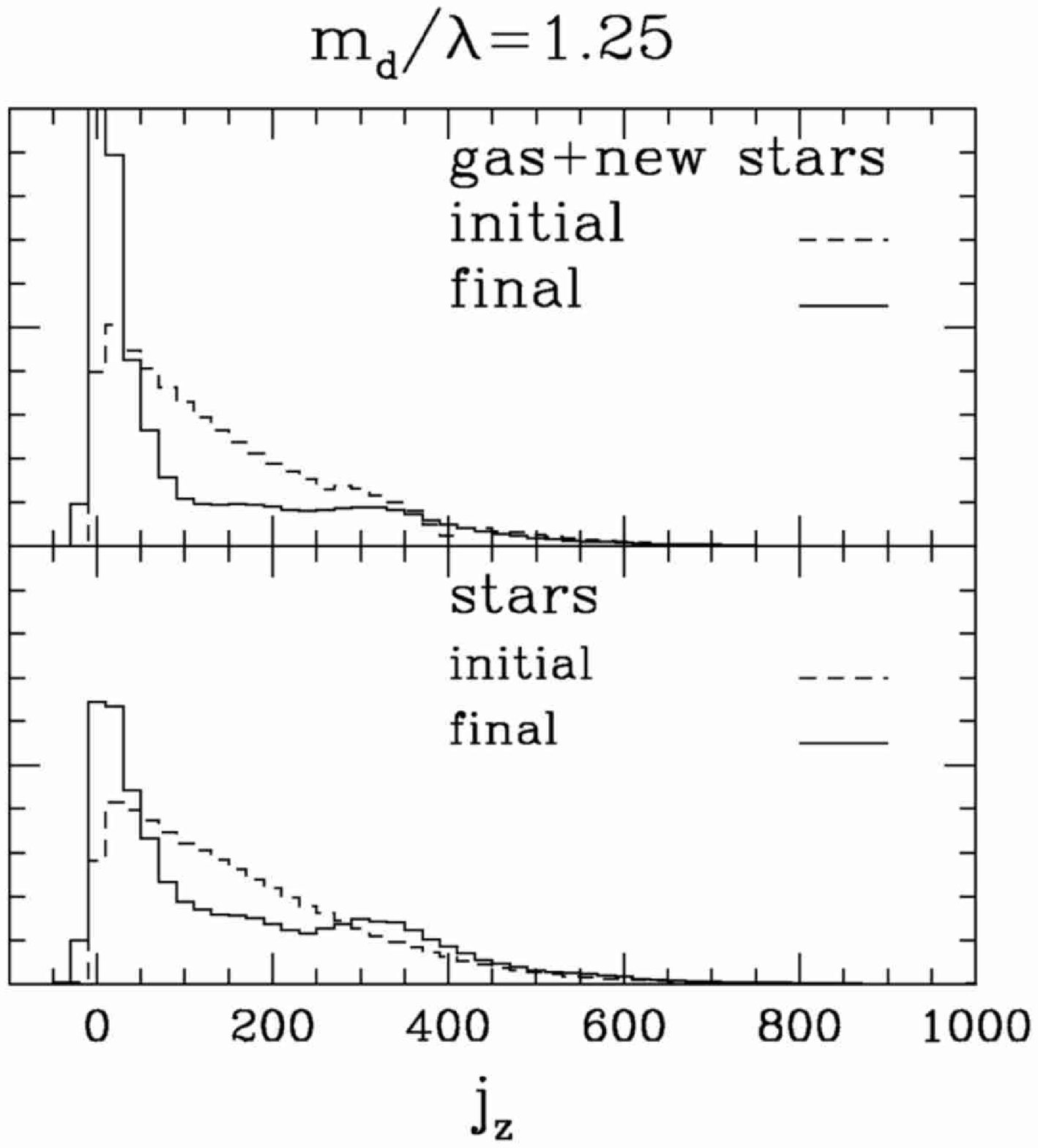}
     \epsfxsize=8cm\epsfbox{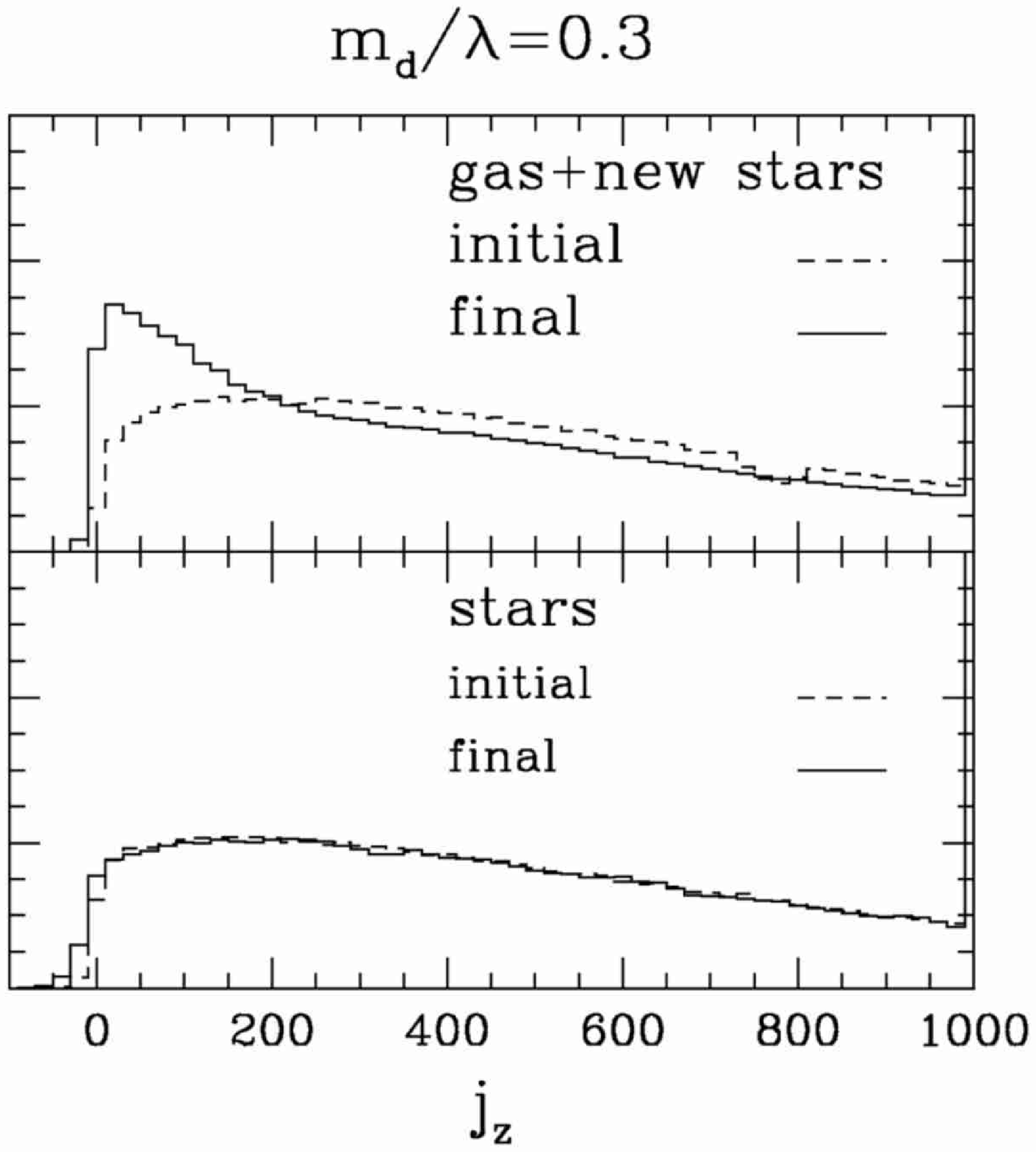}

\rm
\caption{(Left) Distribution of the specific angular momentum along the z-component, $j_{z}$, for the gas (upper panel) and the stars (lower panel) initially (dashed) and after 10 Gyr of evolution (solid) for a model galaxy with $m_{d}/\lambda=1.25$.  (Right) Same as left but for a model galaxy with $m_{d}/\lambda=0.3$.  Angular momentum redistribution is more effective for models with high $m_{d}/\lambda$.}
\label{anghist}
\end{figure}

The most significant parameters in the long term evolution of $r_{out}/h$ remain $m_{d}$ and $\lambda$.  The other three parameters ($c$, $V_{200}$, $m_{b}$) show no appreciable effect on the break location and time evolution of the density profile.  Fig.~\ref{anghist} shows variations of the gas and stars of the $z$-component of the specific angular momentum for two galaxies.  The gas histogram also includes the newly formed stars so as to account for the angular momentum carried away by these particles.  The model with the highest value of $m_{d}/\lambda$ shows significant angular momentum redistribution for both the gas and stars, while the model with lowest value of $m_{d}/\lambda$ shows nearly none.  

The time evolution of the bar and the break
are also not strictly coupled.  Once formed, a break persists, whereas the bar
strength may decline.  Furthermore, model 98 is an interesting case that
demonstrates that not all barred galaxies develop breaks.  With model 95, we verify that breaks may develop later in time despite
the early presence of a bar. 

\subsection{$m_{d}/\lambda$ versus $Q$} 

D06 linked the development of a bar and the subsequent development of a break with a small value of the 
{\it initial} value of the Toomre Q parameter.  We calculated the radial Toomre Q parameter, both initial 
and final, for all of our galaxies and compared the minimum of the Q-curve with the location of the 
break.  The Toomre Q parameter is computed from Eq.~\ref{Wang} for the case of a pure gas disk (i.e. the 
stellar component is not considered).

\begin{figure}
\centering

     \epsfxsize=8cm\epsfbox{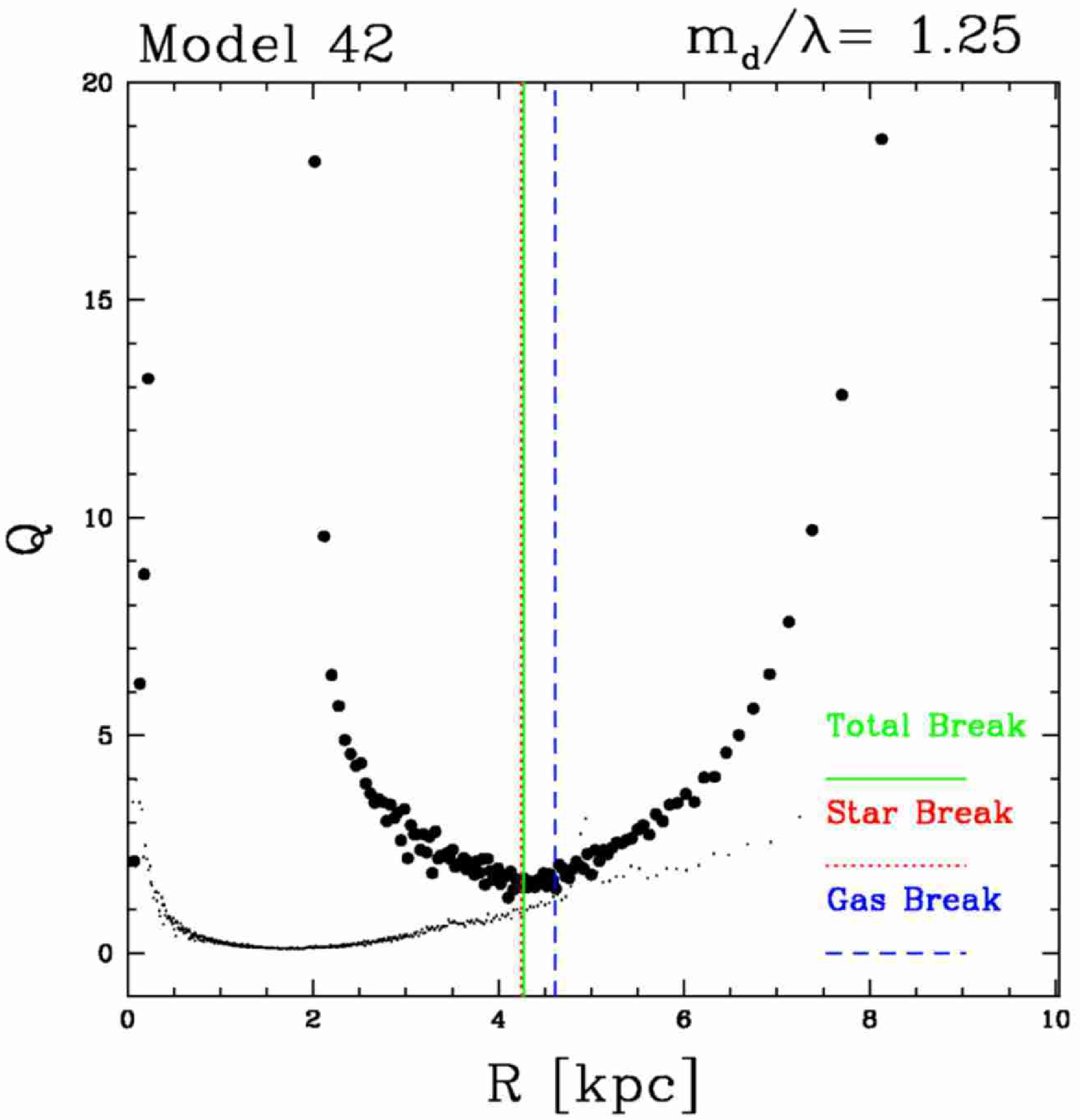}
     \epsfxsize=8cm\epsfbox{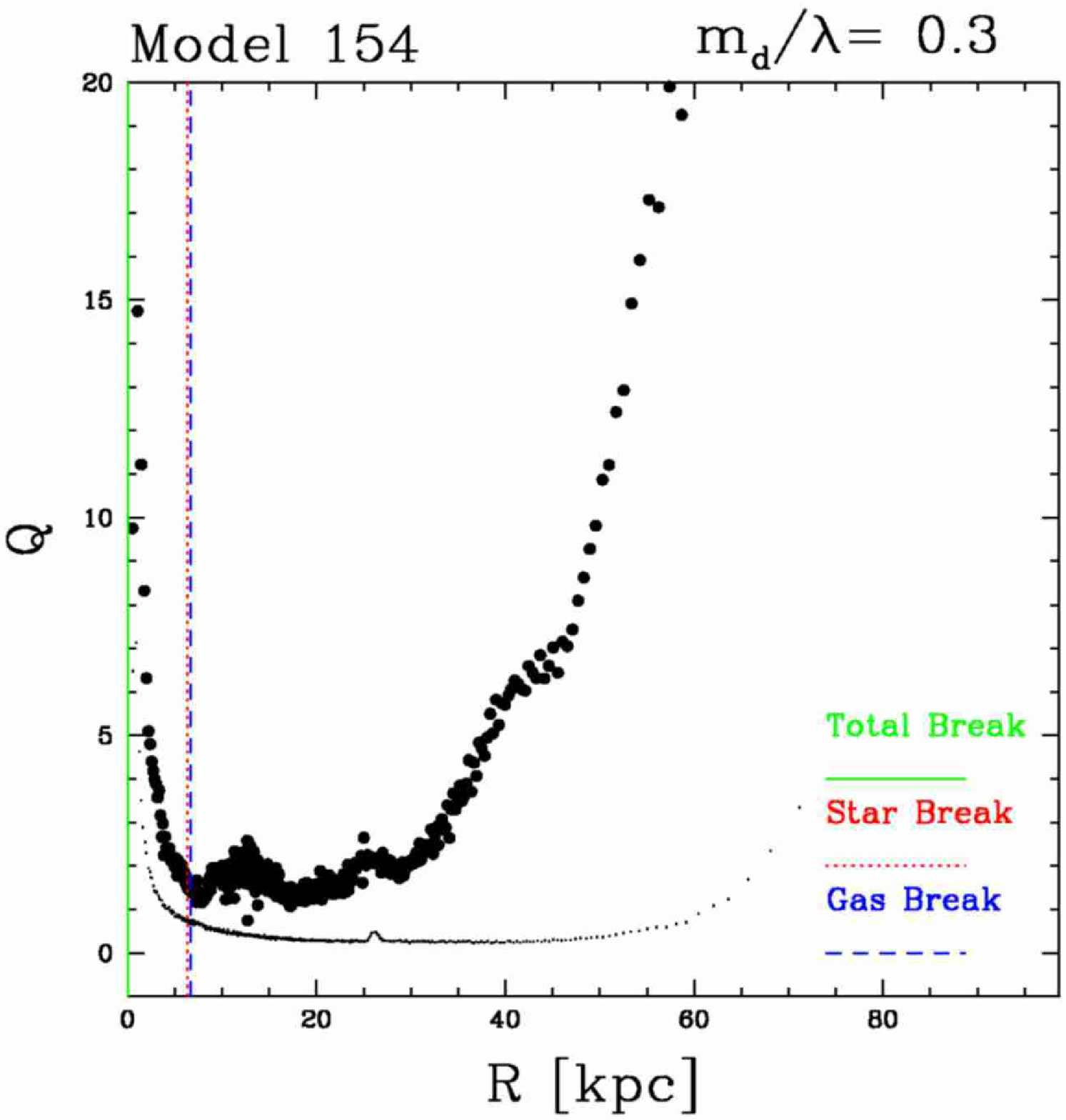}

\rm
\caption{Azimuthally averaged Toomre Q parameter versus radius for the initial snapshot (points) and the final snapshot (filled circles).  The location of the gas (blue dashed line), stars (red dotted line) and total (green solid line) break radius is shown.}
\label{Q}
\end{figure}

We compared the Toomre profiles for galaxies with and without breaks.  Fig.~\ref{Q} shows the radial profile of the Toomre Q parameter for two galaxies with different ratios of $m_{d}/\lambda$.  The open and filled circles represent the initial Q and the Q-parameter after 10 Gyr, respectively.  Model 42 has a high ratio ($m_{d}/\lambda=1.25$)  and subsequently developed a break in the total profile.  We see that the location of the break is roughly at the location of the minimum of the Toomre Q parameter.  Model 154 ($m_{d}/\lambda=0.3$) has a low value of  $m_{d}/\lambda$ and thus no break develops in the total profile, yet both Q curves are very similar for models 42 \& 154.  Indeed all of the galaxies, regardless of the ratio of $m_{d}/\lambda$ show a minimum in the Q-curve.  Thus, while the minimum of the Toomre Q parameter may dictate the location of the break, the Q-curves alone are not sufficient to determine the possibility of a break.

\subsection{$m_{d}/\lambda \approx 1$}

In the parameter space close to the stability threshold, the other parameters
$V_{200}$, $m_{b}$ and $c$ control whether or not a galaxy will develop a
break in the total profile.  In that same region of the parameter space some
galaxies may develop bars without breaks.

\begin{figure}
\centering

     \epsfxsize=8cm\epsfbox{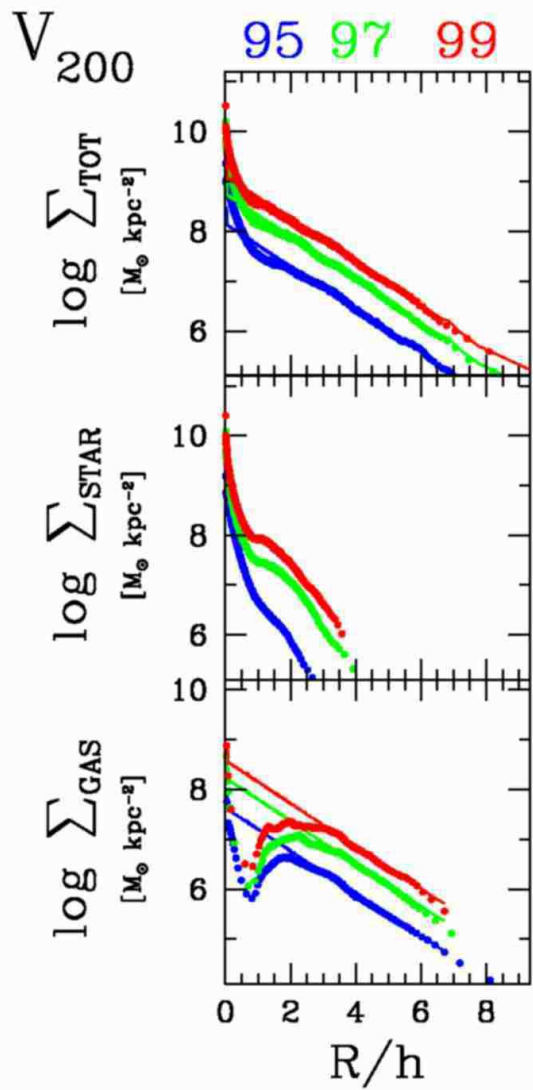}

\rm

\caption{Mass surface densities vs. $R/h$ for different values of $V_{200}$=80 (blue), 160 (green) and 
180 km s$^{-1}$ (red) respectively.  The coloured numbers at the top are the model numbers.  The other 
model parameters were held constant at $m_{d}$= 0.025, $c$=5, $\lambda$=0.03 and $m_{b}$=0. The 
straight lines show the initial profiles.}

\label{cvel1}
\end{figure}

Fig.~\ref{cvel1} shows a comparison of three galaxies with $m_{d}/\lambda\approx 1$
and all other parameters held constant except for $V_{200}$ which is varied
form 80 (model 95) to 160 (model 97) to $180$ km s$^{-1}$ (model 99).  We 
see that only model 95 
develops a
significant inner depression after 10 Gyr.  Models 97 and 99 remain largely
unchanged.  Thus, if $m_{d}$ and $\lambda$ are comparable in value, a low
$V_{200}$ value favors an increase in angular momentum redistribution.  
Thus, a baryon-dominated galaxy is more likely to redistribute
angular momentum efficiently.

\begin{figure}
\centering

     \epsfxsize=8cm\epsfbox{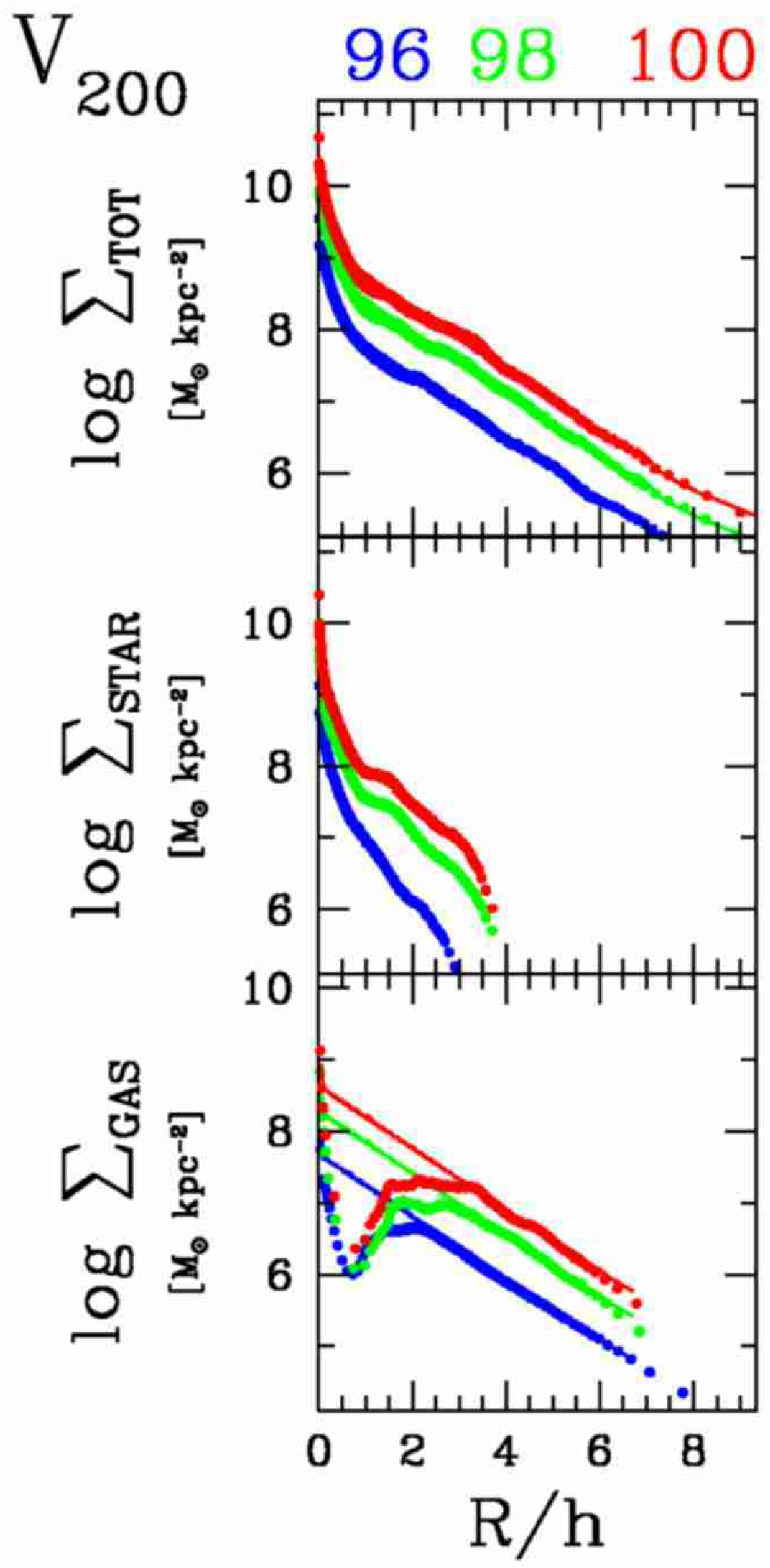}

\rm
\caption{Mass surface densities vs. $R/h$ for different values of $V_{200}$=80 (blue), 160 (green) and 180 km s$^{-1}$ (red) respectively.    The other model parameters were held constant at $m_{d}$=0.025, $c=$5, $\lambda$=0.03 and $m_{b}$=0.2$m_{d}$. Other lines and labels are as in Fig.~\ref{cvel1}.}
\label{cvel2}
\end{figure}

Comparison of Figs.~\ref{cvel1} and ~\ref{cvel2} highlights the impact of the presence, 
or lack there of, of a bulge component.
In the
bulge models, matter redistribution has been significantly impeded.  A
comparison of models 95 and 96 shows that any central depression in the
profile has been eliminated by the presence of the bulge.  Thus, if
$m_{d}/\lambda \approx 1$, a bulge can impede the development of a break.  
vdB01 also found for LSB systems that a bulge can prevent the disk from
becoming too centrally concentrated, thus preserving the initial pure
exponential profile.

\begin{figure}
\centering

     \epsfxsize=8cm\epsfbox{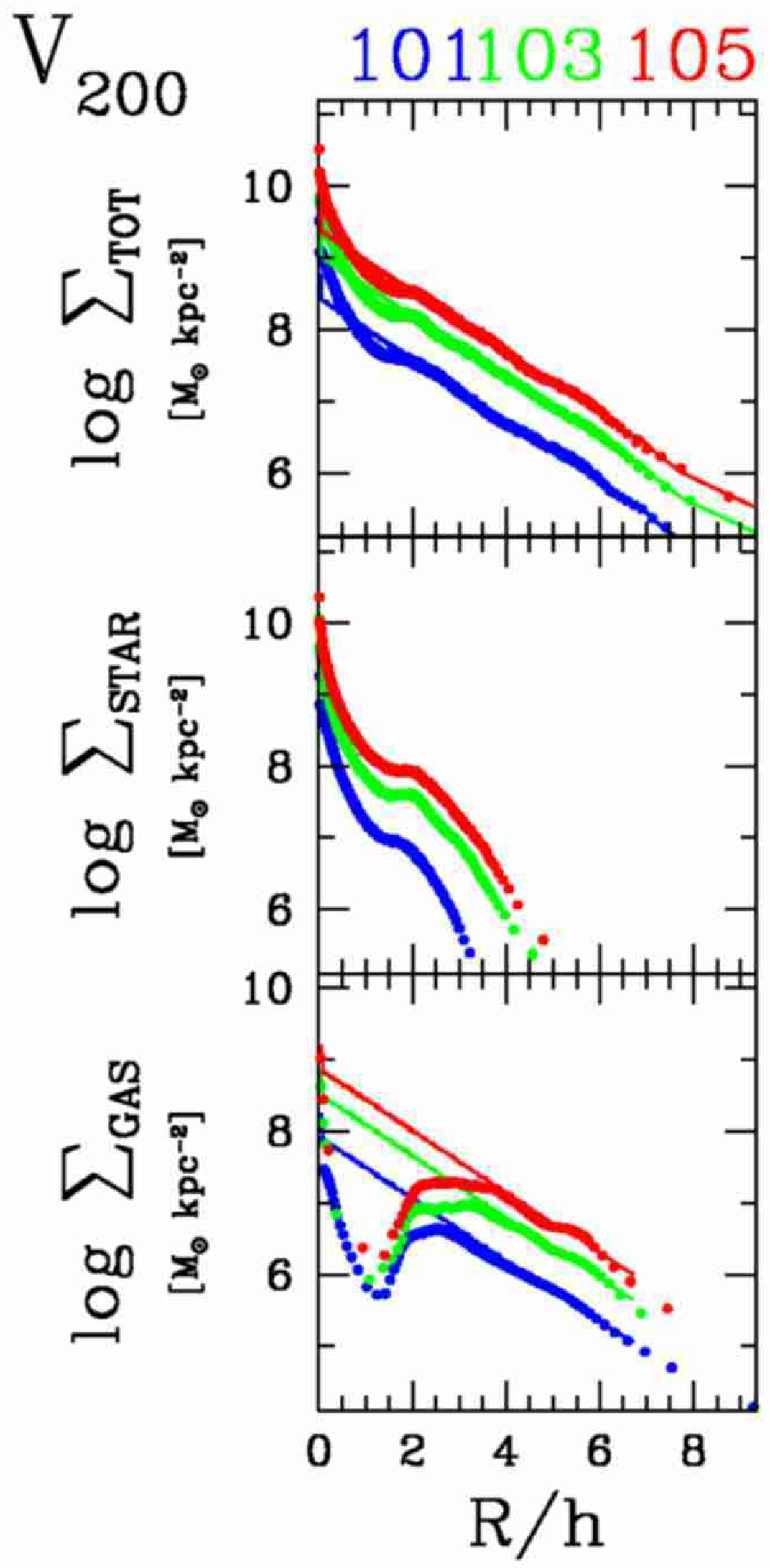}

\rm
\caption{Mass surface densities vs. $R/h$ for different values of $V_{200}$=80 (blue), 160 (green) and 180 km s$^{-1}$ (red) respectively.   The other model parameters were held constant at $m_{d}$=0.025, $c$=10, $\lambda$=0.03 and $m_{b}$=0.  Other lines and labels are as in Fig.~\ref{cvel1}.}
\label{cvel3}
\end{figure}

\begin{figure}
\centering

     \epsfxsize=8cm\epsfbox{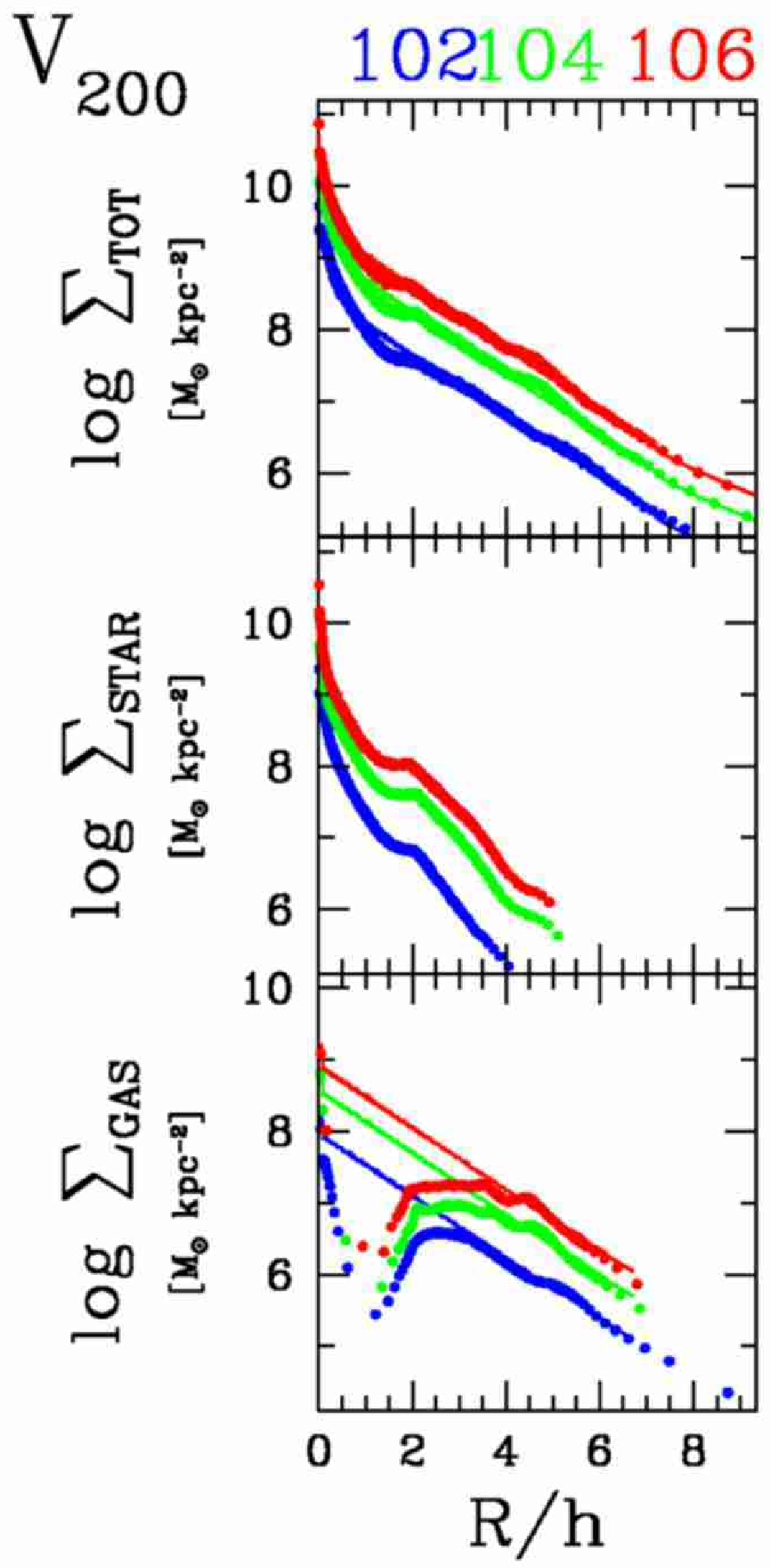}

\rm
\caption{Mass surface densities vs. $R/h$ for different values of $V_{200}$=80 (blue), 160 (green) and 180 km s$^{-1}$ (red) respectively.     The other model parameters were held constant at $m_{d}$=0.025, $c$=10, $\lambda$=0.03 and $m_{b}$=0.2$m_{d}$.  Other lines and labels are as in Fig.~\ref{cvel1}.}
\label{cvel4}
\end{figure}

\begin{figure}
\centering

     \epsfxsize=8cm\epsfbox{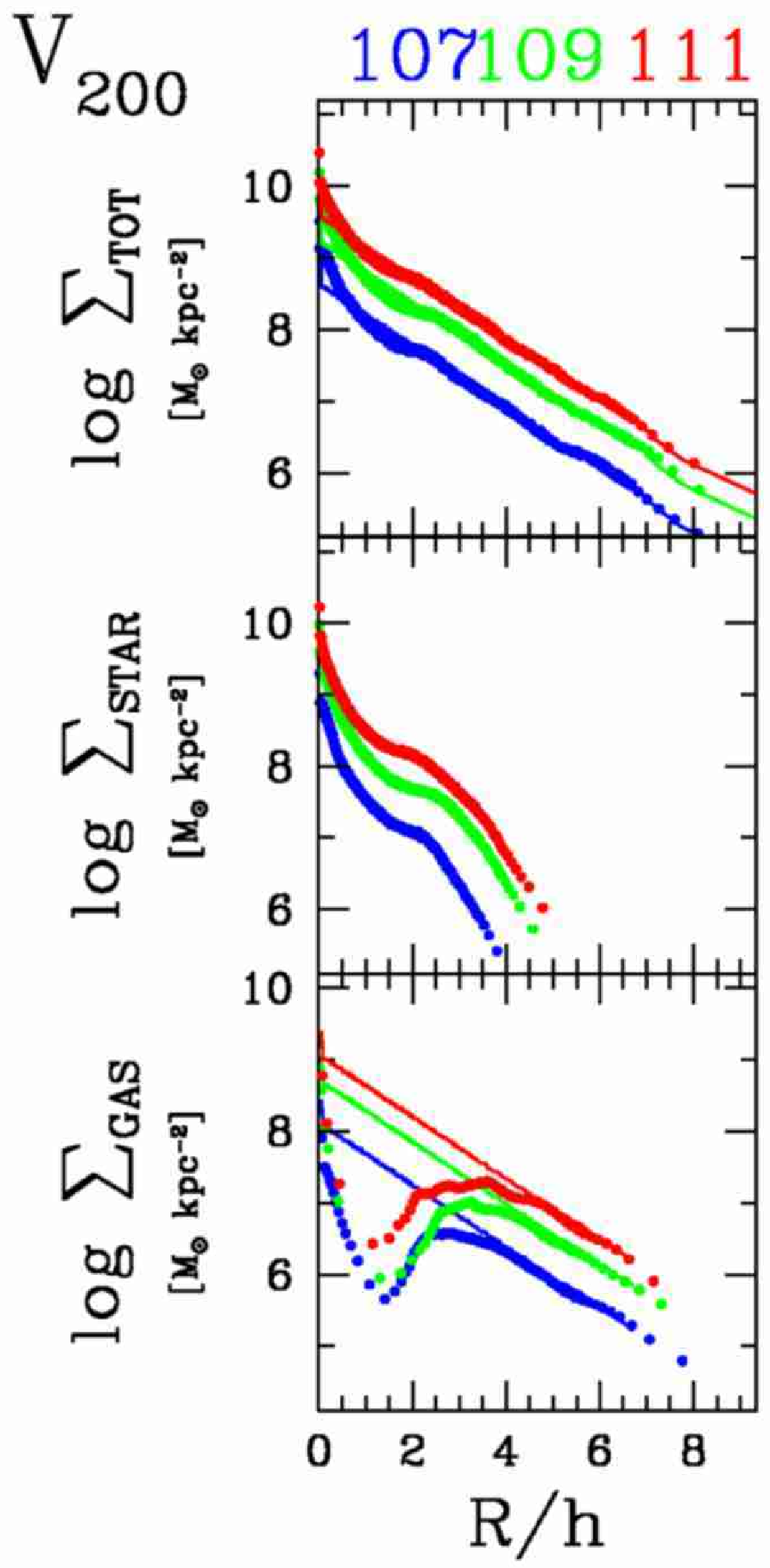}

\rm

\caption{Mass surface densities vs. $R/h$ for different values of $V_{200}=$ 80 (blue), 160 (green) 
and 180 km s$^{-1}$ (red) respectively.  The other model parameters were held constant at $m_{d}$= 
0.025, $c$=15, $\lambda$=0.03 and $m_{b}$=0.  Other lines and labels are as in Fig.~\ref{cvel1}.}

\label{cvel5}
\end{figure}

We now consider the effect of $c$ on density profiles.  Fig.~\ref{cvel3} displays
model galaxies with the same configuration of parameters as in Fig.~\ref{cvel1}, only
the halo concentration has been increased to $c$=10.  Fig.~\ref{cvel4} shows these
same parameter configurations but now with a bulge.  In Fig.~\ref{cvel1} 
we saw
that only the model with the lowest $V_{200}$ showed any evidence of
breaks.  Figs.~\ref{cvel3} and ~\ref{cvel5} show that an increased halo concentration
yields an inner depression via significant matter redistribution in the disk, and favors the
formation of a break.  However, reality is more complex.  Fig.~\ref{cvel5} shows the same
models with an even higher concentration ($c$=15).  Here, the amount of
matter redistribution has decreased.  Thus, the role of $c$ is quite
subtle.  If $m_{d}/\lambda \approx 1$, a low and a high $c$ value can
impede matter redistribution and the formation of break due to a depression
in the inner regions.  This is likely due to the relationship between the
bar and the halo, although a deeper investigation of this relationship is
needed.

 Thus, the conditions that favor the formation of a break, or the redistribution of matter within the disk, are a low $V_{200}$ value and the absence of a stabilizing bulge.  The concentration of the halo may foster a break formation provided it is neither too high nor too low.

%%%%%%%%%%%%%%%%%%%%%%%%%%%%%%%%%%%%%%%%%%%%%%%%%%%%%%%%%%%%%%%%%
%%%%%%%%%%%%%%%%%%%%%%%%%%%%%%%%%%%%%%%%%%%%%%%%%%%%%%%%%%%%%%%%%%%
%%%%%%%%%%%%%%%%%%%%%%%%%%%%%%%%%%%%%%%%%%%%%%%%%%%%%%%%%%%%%%%%%
%%%%%%%%%%%%

%%%%%%%%%%%%%%%%%%%%%%%%%%%%%%%%%%%%%%%%%%%%%%%%%%%%%%%
%%%%%%%%%%%%%%%%%%%%%%%%%%%%%%%%%%%%%%%%%%%%%%%%%%%%%%%%%%%%%%%%%
%%%%%%%%%%%%%%%%%%%%%%%%%%%%%%%%%%%%%%%%%%%%%%%%%%%%%%%%%%%%%%%%%%%
%%%%%%%%%%%%%%%%%%%%%%%%%%%%%%%%%%%%%%%%%%%%%%%%%%%%%%%%%%%%%%%%%
%%%%%%%%%%%%%%%%%%%%%%%%%%%%%%%%%%%%%%%%%%%%%%%%%%%%%%%%%%%%%%%%%%%

\subsection{Mass Surface Density of Profile Breaks}
\begin{figure}
\centering

     \epsfxsize=8cm\epsfbox{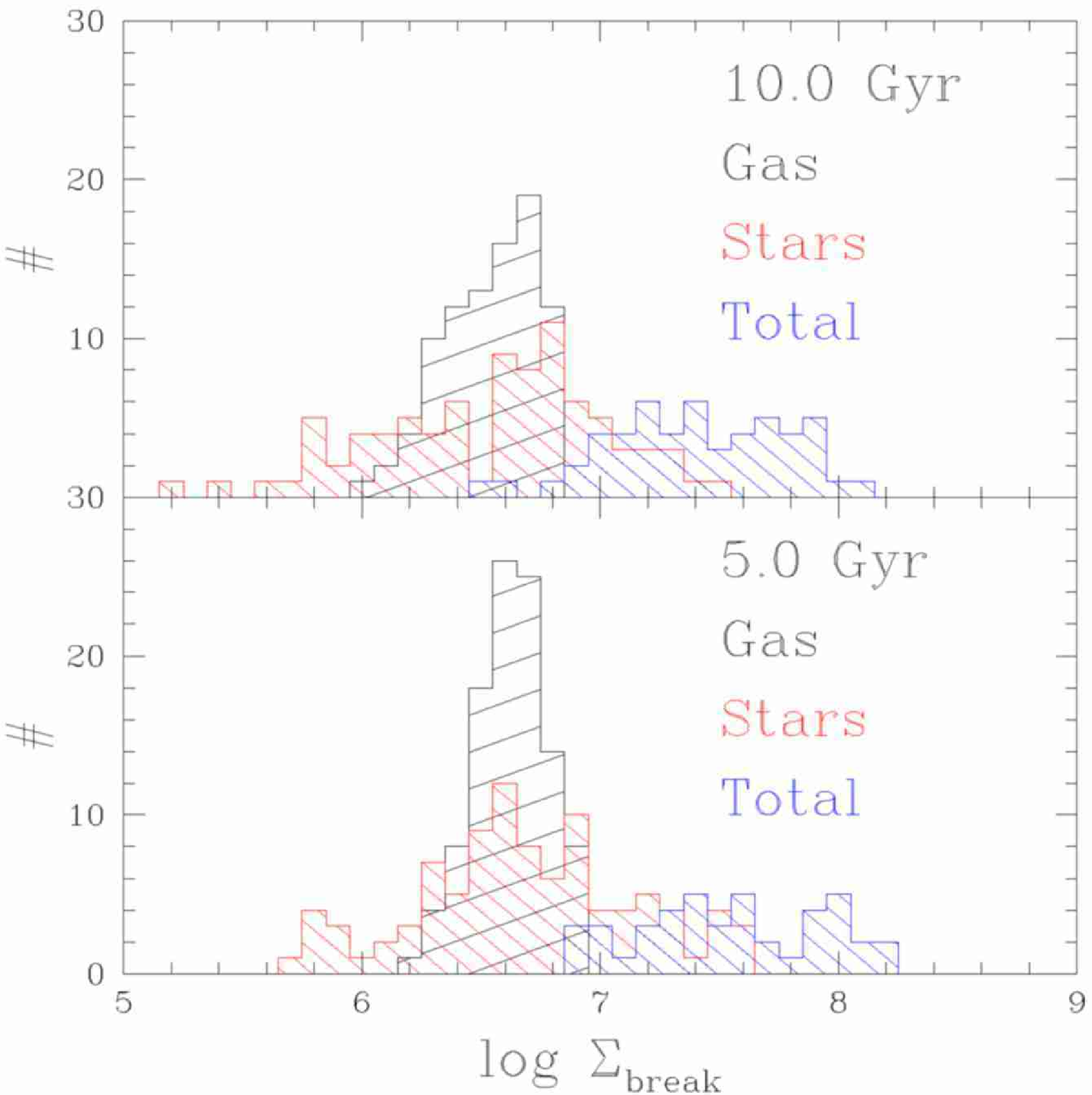}

\rm
\caption{Distribution of the mass surface density at outer break location for the gas (black), newly formed stars (red) and total baryonic (blue) components after 5 and 10 Gyr.}
\label{hbkhist}
\end{figure}

While the break radius depends on the ratio $m_{d}/\lambda$, we now investigate if breaks occur at a specific mass surface density.  Fig.~\ref{hbkhist} shows two histograms of the break locations in mass surface density for the gas (black), newly formed stars (red) and total baryonic components (blue) after 5 and 10 Gyr.  We see that the gas breaks have roughly the same mass surface density with $\log \Sigma_{break}\approx6.6 \pm 0.4$.  Breaks in the gas profiles do not exceed $\log \Sigma_{break}$=7, as a result of the built-in star formation threshold at 10 M$_{\odot}$ pc$^{-2}$.  The break radii in the newly formed stars have a fairly broad distribution ($\log \Sigma_{break}=$5.8-7.8); while the stellar profile will truncate based on the density of the gas, the surface density of this break should vary.  The range for the mass surface density of the total profile breaks is slightly narrower ($\log \Sigma_{break}$=6.8-8.2).  In order to determine which parameters control the amplitude of the break in terms of surface density, Table 3 lists the average mass surface density of the total profiles binned according to the model parameters.  We find the greatest variations in the mass surface density with $V_{200}$ and $c$;  as these values increase, the mass surface density increases too.
\begin{table}
\caption{Comparison of Break Mass Density with Model Parameters after 5 Gyr}
\begin{tabular}{cc}
\hline\hline
$c$ & Total $\log \Sigma_{break}$\\
\hline
5 &  $7.2 \pm 0.2$ \\
10 & $7.7 \pm 0.3$ \\
15 &  $7.9 \pm 0.4$ \\
\hline\hline
$V_{200}$ &  \\
\hline
80 &  $7.2 \pm 0.4$ \\
160 & $7.8 \pm 0.5$ \\
180 &  $7.6 \pm 0.4$ \\
\hline\hline
$\lambda$ & \\
\hline
0.02 &  $7.6 \pm 0.3$ \\
0.03 & $7.6 \pm 0.4$ \\
0.08 &  $7.2 \pm 0.4$ \\
\hline\hline
$m_{b}$ &\\
\hline
0 &  $7.5 \pm 0.3$ \\
0.2$m_{d}$ & $7.6 \pm 0.3$ \\
\hline\hline
$m_{d}$ & \\
\hline
0.025 &  $7.6 \pm 0.6$ \\
0.05 & $7.6 \pm 0.4$ \\
0.1 &  $7.5 \pm 0.5$ \\
\hline\hline
\end{tabular}
\end{table}
Whereas the break location was determined by $m_{d}/\lambda$, the mass surface density of the break is largely independent of it.  This is due to the opposite dependences of the mass density profile on $m_{d}$ and $\lambda$.  The central mass surface density of a galaxy is $\Sigma_{o} \propto F M_{TOT}^{1/3} \lambda^{-2+6F}$, where $F$ is the baryonic fraction and $M_{TOT}$ is the total mass of the galaxy (DSS97).  Thus, the total disk mass is given by the baryonic fraction times the total galaxy mass.  If $\lambda$ is held constant and the disk mass fraction increases, the central surface density and the entire profile amplitude will increase.  However, an increase in the disk mass also pushes the break further out.  These two opposing effects lead to the break occurring at roughly the same mass surface density.

A similar opposing effect is seen when $\lambda$ is varied.  If $m_{d}$ is held constant and $\lambda$ increases, the mass surface density will decrease but the break will also move inward, thus the mass surface density of the break remains largely constant.  Conversely, the parameters $V_{200}$ and $c$ have little effect on the break location but do affect the mass density (amplitude) of the break.  An increase in $c$ and $V_{200}$ leads to an increase in the disk scale length.  Thus, the mass surface density of the break will increase as these parameters increase if the break location is held constant.

\subsection{Disk Scale Length}
Given a two-component density profile, it is unclear which of the inner and
outer disk scale lengths should be most representative of the disk.  
Traditionally, observations have been expressed in terms of the inner disk
scale length (\eg PT06).  This is a sensible approach if the formation of
a break in the total profile is caused by a star formation threshold which
leads to a rapid decline in the outer parts of the surface brightness
profile.  However, if a two-component profile is caused by
dynamical instabilities, it is unclear which of the inner or outer scale
lengths is the most representative structural parameter.

\begin{figure}
\centering

     \epsfxsize=8cm\epsfbox{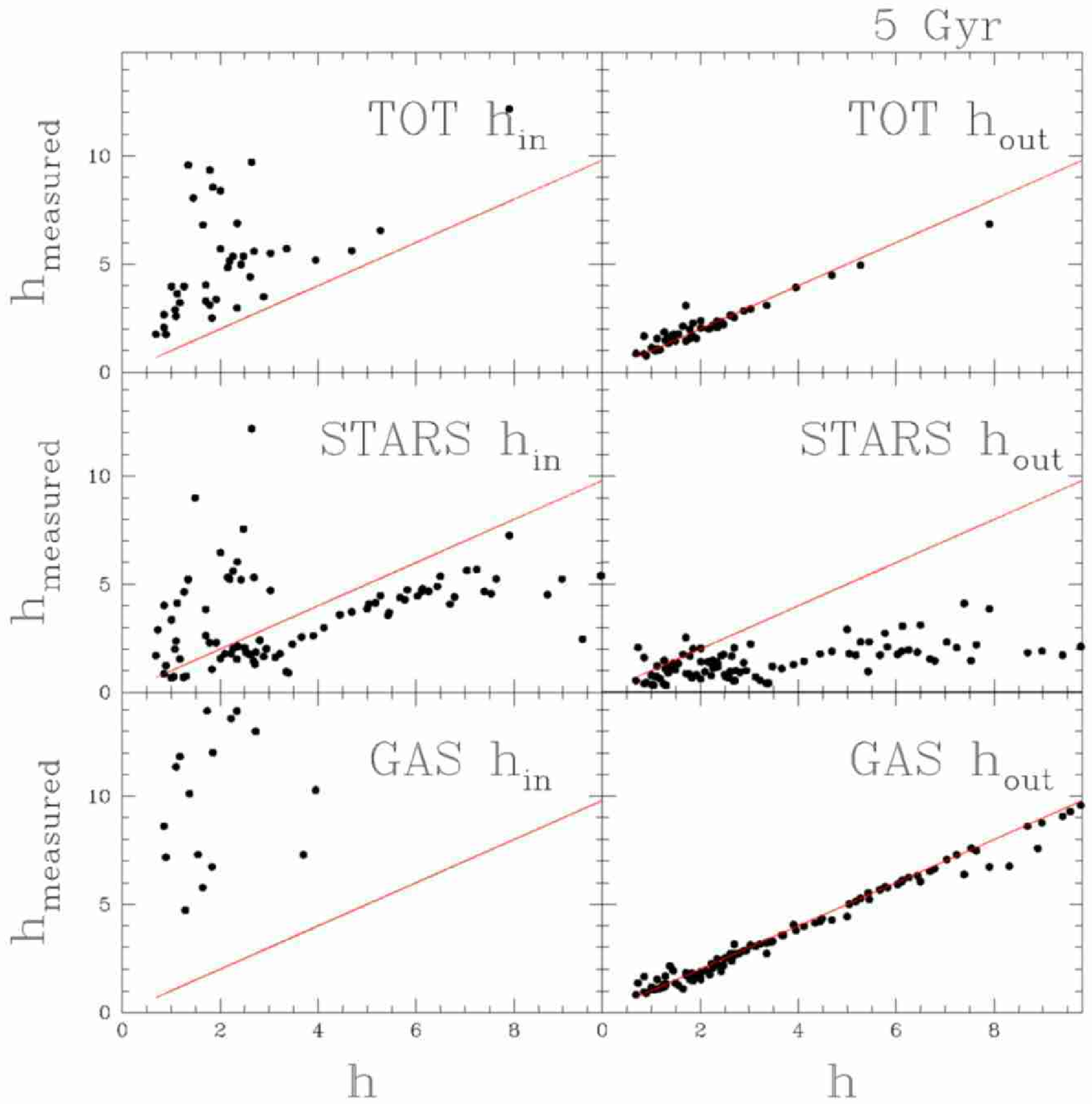}
     \epsfxsize=8cm\epsfbox{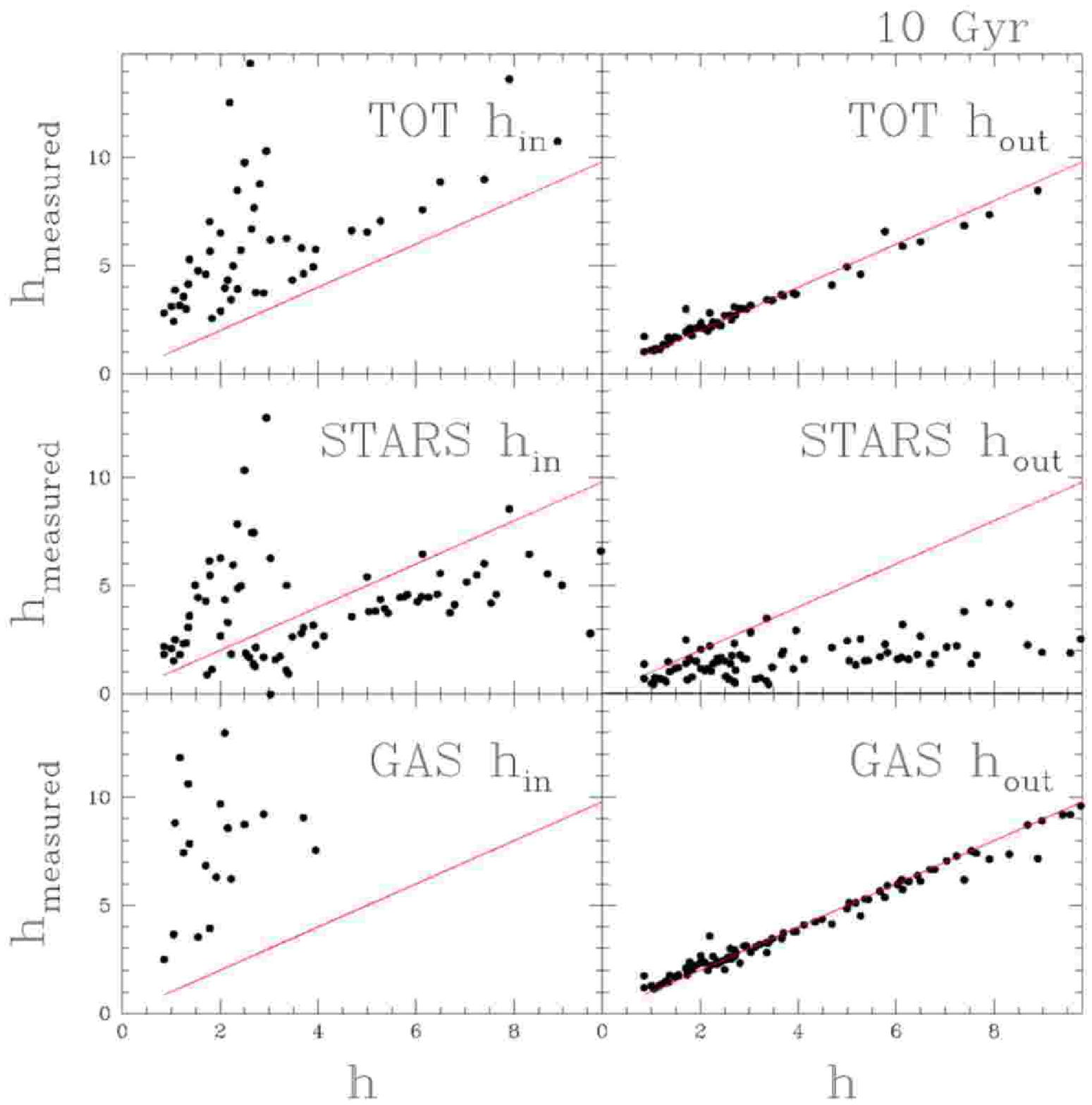}

\rm
\caption{The left panels show the inner versus the initial disk scale lengths for three profile types after 5 and 10 Gyr.  Likewise on the right panels, but for the outer disk scale lengths.}
\label{scalecomp}
\end{figure}

Fig.~\ref{scalecomp} shows the variations of the inner and outer disk scale lengths versus the initial disk scale length after 5 and 10 Gyr.  We see that the outer disk scale length is in much closer agreement with the initial disk scale length for both the total and the gas density profiles.  However, the inner disk scale length remains closer to the initial disk scale length for the case of the newly formed star profiles due to the sharp cut-off caused by the SF threshold.  If the development of a two-component profile is associated with dynamical properties, the outer disk scale length should better describe the evolutionary properties of the disk.  However, if a two-component profile is tied to a star formation threshold, the inner disk scale length should be preferred. 

\begin{figure}
\centering

     \epsfxsize=8cm\epsfbox{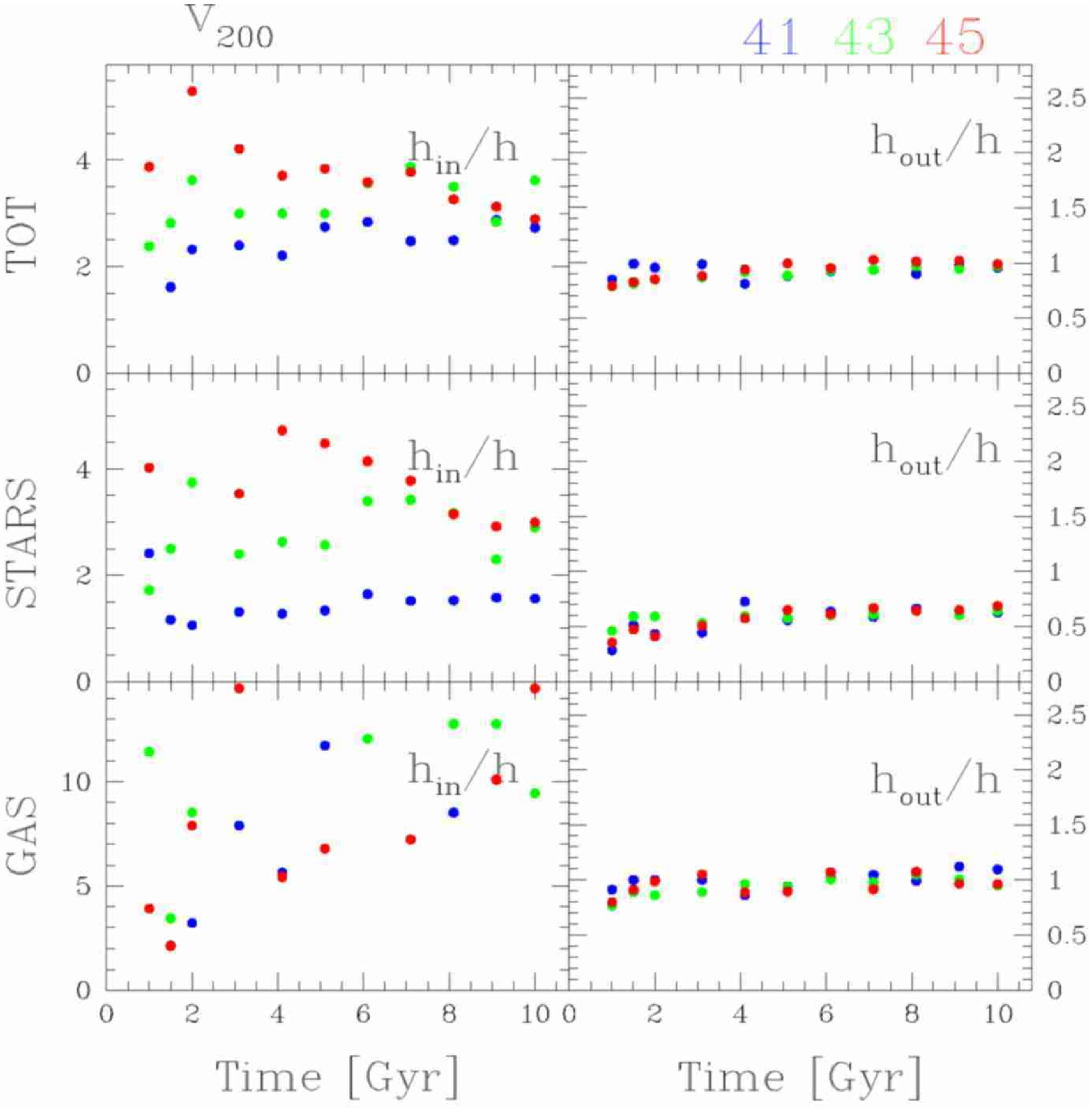}

\rm
\caption{Time evolution of $h_{in}/h$ and $h_{in}/h$ as a function of $V_{200}$ for three galaxies selected randomly.  Blue, green and red correspond to $V_{200}$ of 80, 160 and 180 km s$^{-1}$ respectively.  The numbers at the top are the model numbers.   The other model parameters were held constant at $m_{d}$=0.025, $c$=5, $\lambda$=0.02 and $m_{b}$=0.}
\label{scaletime}
\end{figure}

We now examine the evolution of the inner and outer disk scale length with time.  We consider three randomly selected models and examine how the inner and outer scale lengths evolve with time. Fig.~\ref{scaletime} compares the time evolution of the inner and outer disk scale length for variations of $V_{200}$ (80, 160 and 180 km s$^{-1}$).  The outer disk scale length not only remains close to the initial disk scale length, it is also more stable with time.  For the stars, while the outer disk scale length is lower than the initial disk scale length, the former proves to be more stable over time than the latter.  The fact that $h_{out}/h$ remains close to 1, whereas $h_{in}/h$ evolves to much higher values, makes sense in the context of matter redistribution in the disk.  We have seen that the break feature in the central regions forms largely due to matter funneled towards the nucleus.  Thus, during the evolution of the disk, the inner regions will continue to grow, further depleting the inner disk and enhancing the bulge (secular evolution).  The outer regions of the disk remain largely unaffected.  These conclusions will be altered in the context of gas accretion scenarios.  For instance, cosmological simulations with gas infall suggest that the break radius grows with time (Roskar et al. 2007).  Once gas accretion settles, the position of the break must however stay fixed but its amplitude may still grow via internal dynamical processes (\S 4.5).


\section{Comparisons with Observations}

\subsection{Break Positions}
PT06 compared the surface brightness profiles of 90 face-on galaxies from the SDSS.  Among the Type II galaxies in their sample, they noted a correlation between the break radius and the inner disk scale length. While this correlation is of value, the outer disk scale length correlates more tightly with the dynamical parameters of the disk including angular momentum, velocity and mass.  The inner disk scale length is of lesser interest since it deviates significantly from the initial disk scale length and evolves considerably over time.  However, in the presence of accretion, the final outer scale length may also deviate from the initial outer scale length.

\begin{figure}
\centering

     \epsfxsize=8cm\epsfbox{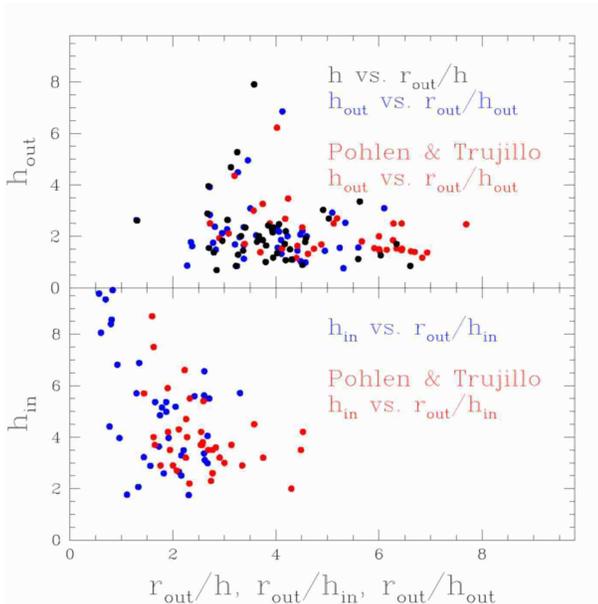}

\rm
\caption{(Upper panel) Comparison of the outer disk scale length with the break radius in terms of the outer disk scale length.  PT06's Type II galaxies are shown in red and our model galaxies in blue.  The black points refer to our initial disk scale lengths versus the break radius in terms of the initial disk scale length.  (Lower panel) Comparison of PT06's data in red, with our values for the inner disk scale lengths versus the break radius in terms of the inner disk scale length (blue points).  Both panels use model data evolved up to 5 Gyr.}
\label{pohlencomp}
\end{figure}

In Fig.~\ref{pohlencomp}, we plot as red points the data of PT06 in terms of the inner and outer disk scale lengths; our data are shown as blue and black points corresponding to the outer and initial disk scale lengths, respectively.  Here, we use break radii for the total baryonic profile only in order to correspond to observed values.  We see that the distributions for $h_{in}$ and $h_{out}$ in terms of $r_{out}/h$ (where $h$ may be the initial, inner or outer scale length) are similar for the data and models.  Considering only the inner disk scale length, our values of $h_{in}/h$ are lower than those observed.  However, the fact that the inner disk scale length evolves considerably over time may explain the disagreement;  we have seen that the inner disk scale length shifts to larger values over time and the sample of PT06 may simply be more dynamically evolved than ours.  We note the close agreement between the blue and black points in the upper panel of Fig.~\ref{pohlencomp}, suggesting again that the outer disk scale length has not evolved significantly with time.  
\subsubsection{Broken Exponentials: SF Thresholds versus Angular Momentum Redistribution?}
\begin{figure}
\centering

     \epsfxsize=4cm\epsfbox{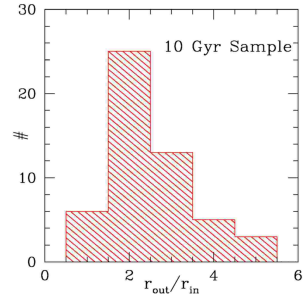}

\rm
\caption{Histogram of the ratio of the outer to the inner break radii determined by-eye. $r_{in}$ can be loosely taken as the bar length.   Outer profile breaks do preferentially occur at twice the bar length, corresponding to a galaxy's OLR.}
\label{OLRhist}
\end{figure}

We have focused thus far on two-component profiles in the total stellar profile.  However, if breaks are due at least in part to a star formation threshold, those seen in the newly formed stars may possibly better match the observed breaks.  PT06 distinguished between different types of Type II galaxies.  ``Type II-OLR'' galaxies, which make up only 15\% of their sample, are galaxies with breaks associated with the Outer Lindblad Resonance and thus linked to the presence of a bar.  Breaks associated with bars are typically located at roughly twice the bar radius, the location of the OLR.  While we did not specifically measure bar lengths in our simulations\footnote{We investigated the methods outlined by Michel-Dansac and Wozniak (2006) to determine bar lengths. Measurements for the latter are difficult due to the many different ways to calculate it observationally.  We were unable to settle on a robust measure of bar length for our simulations.}, we eye-balled the location in the disk that  marks roughly the transition between the outer bulge and inner disk.  For barred model galaxies, the inner break radius, $r_{in}$, corresponds roughly to the bar length.  We see in Fig.~\ref{OLRhist} that most breaks do indeed occur at twice this inner radius.  The closeness of these two values further supports our interpretation that breaks in the total profile are likely coupled with bar formation.  
PT06 also found a large number of ``classically truncated'' galaxies, or ``Type II-CTs'', which make up 32 \% of their sample.  These galaxies have breaks that are seemingly not associated with bars.  The breaks in these systems could potentially be described by a SF threshold.  Our stellar density profile breaks, which form due to the star formation threshold in GADGET-2, might better match this group.  

\begin{figure}
\centering

     \epsfxsize=8cm\epsfbox{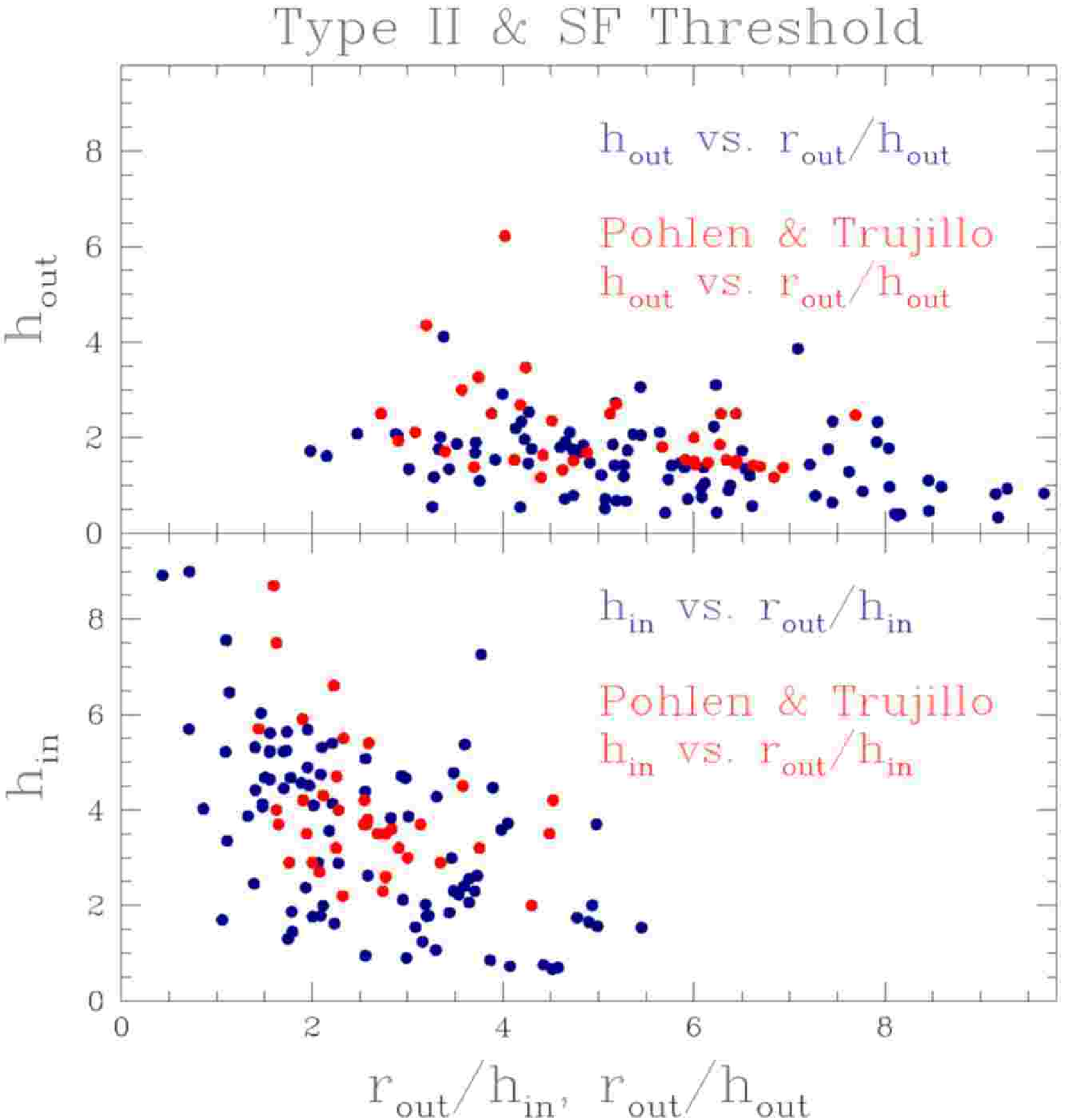}
    \epsfxsize=8cm\epsfbox{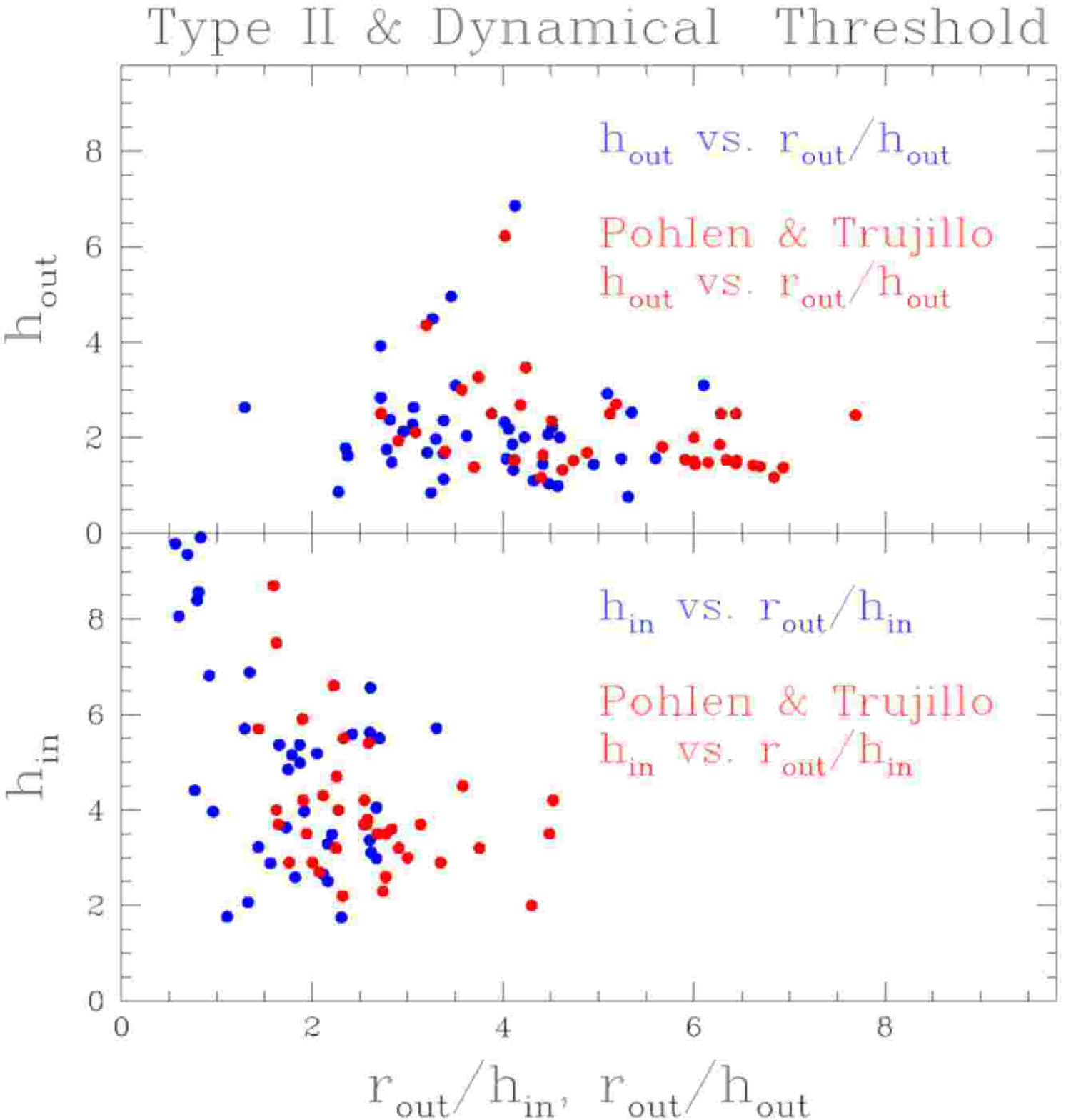}

\rm
\caption{(Upper panels) Comparison of the outer disk scale length with the break radius in terms of the outer disk scale length.  PT06's Type II galaxies are shown in red.  Our model galaxies shown in blue have breaks in the newly formed star density profiles confirming that the star formation threshold model does match well with observed breaks.  (Lower panels) Comparison of the inner disk scale lengths with the break radius in terms of the inner disk scale length.  Our model galaxies shown in blue have breaks in the total density profiles confirming that breaks may be produced by dynamical effects in the disk.  Our data correspond to models evolved up to 5 Gyr.}
\label{pohlencomp2}
\end{figure}

We can now compare our results for stars and total profiles with PT06's Type
II-CT and Type II-OLR galaxies.  Their Type II-OLR galaxies have a mean
break radius in terms of the inner disk scale length of 1.7 $h_{in}$
compared  to a mean break radius of 1.6 $\pm$ 0.8
$h_{in}$ for our models.  Their Type II-CT galaxies have a mean break
radius of 2.5 $\pm$ 0.6 $h_{in}$, compared to 2.4 $\pm$ 1.3 $h_{in}$ for
our stellar density profiles, again in close agreement.  We show in Fig.~\ref{pohlencomp2}
the distribution of our total and stellar profile breaks in comparison with PT06's Type IIs.  Once more, both distributions are in fairly
close agreement.  However, we note one important point about the notion
that breaks are solely regulated by a star formation threshold; all
of our model galaxies produced breaks in the stellar profiles by design.  Thus, a pure SF threshold explanation could not account
simultaneously for the observed detection of Type I and Type II galaxies.  
Elmegreen \& Hunter (2006) invoked turbulent compression in order to
produce stars in the outer parts of the galaxies and thus enable the
production of Type I galaxies with a SF threshold.  This hypothesis ought to be tested soon with enhanced simulations.

\subsection{Break Evolution}
We have found in \S4 that once a two-component profile is established, the break remains at the same physical location over time.  The time for break development depends on the ratio  $m_{d}/\lambda$.  If $m_{d}/\lambda$ is high a break develops quickly;  if $m_{d}/\lambda$ is low a break develops later, if at all.  The location of the break is also controlled by this ratio (but less so than Q; \S 4.3).  The higher $m_{d}/\lambda$, the further out in to the disk will the break occur.  In light of these findings we would expect fewer galaxies with breaks at high redshift and that these breaks would be seen further out into the disk in terms of the outer disk scale length.

P\'{e}rez (2004) and Trujillo \& Pohlen (2005) have shown that observed face-on galaxies at high redshift
($z \approx 1$) have breaks at {\it smaller} galactocentric radii relative to
nearby galaxies, \ie in contradiction with our expectations from pure 
secular 
evolution.  Their
findings are consistent with a disk size increase of 25\% over the last 8
Gyr.  This is in close agreement with Somerville et al.
(2008), who found an evolution of the disk size of 15-20\% between a nearby sample of galaxies from the SDSS and a
high-redshift sample of galaxies from the GEMS (Galaxy Evolution from
Morphology and SEDs) survey. 

Despite these apparent contradictions, note that the analysis
of Trujillo \& Pohlen (2005) reported  break radii for 36 distant nearly face-on
galaxies in physical units rather than versus disk scale length.  They find
break radii in distant galaxies at smaller physical locations than seen
locally.  Based on our findings, galaxies with a low $h$ should develop
breaks early.  The results of Trujillo \& Pohlen (2005) may be significantly different in terms
of disk scale length.

P\'{e}rez observed 16 unbarred galaxies at high redshift ($z \sim$1) and reported breaks at $r_{out}/h_{in} < 3.5$, but with six galaxies having $r_{out}/h_{in} < 1.8$.   Our models have $r_{out}/h_{in} < 3.6$ after 5 Gyr of evolution and $r_{out}/h_{in} < 3.4$ after 10 Gyr of evolution.  The mean break radii is $r_{out}/h_{in}=1.7$ for the total baryonic for both the 5 and 10 Gyr evolutionary stages.  The lack of evolution in the break position is likely due to the fact that these models did not include gas accretion.  

Finally, the high dust content of distant galaxies can hinder the simple
detection of a bar before any considerations of bias 
in the
measurement of a break radius (unless distant galaxy disks can be assumed
to be transparent).

\section{Discussion}

Based on high resolution simulations, D06 showed that bar formation
plays a significant role in the generation of double exponential density
profiles.  They demonstrated that profile breaks correspond to locations 
where
angular momentum is shed by the bar and carried away by resonantly-coupled
spiral arms (such as the OLR).  D06 also showed that the presence of a bar
does not necessarily imply that a break will develop; the latter is in fact
correlated with the Toomre-Q parameter.  If Q is small, the inner disk must
shed a lot of angular momentum in order to form a bar.  This leads to a
higher central density and ultimately to the formation of a break by
creating an inner depression in the density profile with respect to the outer
regions of the profile.  If Q is high, the density change is smaller across
the disk and neither a bar nor a break may form.  D06 also showed that the
development of a break will naturally alter the inner disk scale length
sometimes by as much as a factor of 2. As a direct result, D06 stressed 
that the specific
angular momentum cannot be directly related to the distribution of inner
disk scale lengths, which has important implications for predictions of
disk galaxy structure.

Our work has not specifically focused on the connection between bars and breaks, but rather on the dynamical parameters that lead to the formation of a break in the total profile which is nonetheless, as we have seen, closely linked the formation of a bar.  However, some of our galaxies with bars did not form breaks and not all galaxies formed breaks at the same absolute time.  We find that the ratio of $m_{d}/\lambda$ is central in determining the development and location of breaks whereas Q prescribes the break location.  We have determined that if $m_{d}/\lambda > 1$, a break will develop.  If $m_{d}/\lambda \approx 1$ a break may or may not develop depending on the values of other structural parameters, such as the halo concentration, $c$ and $V_{200}$.  $m_{d}/\lambda$ also determines when and where a break will develop.  If $m_{d}/\lambda$ is high the break will develop quickly and be located further out into the disk.  If the disk is highly unstable, \ie $m_{d}/\lambda \gg 1$, a significant amount of angular momentum would be shed in order to form a bar leading to subsequent matter redistribution in the disk.  As $m_{d}/\lambda$ decreases, the disk stability increases and, while a bar may still form, the amount of angular momentum redistribution will be significantly reduced.  This will impede the development of a break, or if a break still forms, it will not develop as far out into the disk.  Therefore the requirement that broken exponentials form when $m_{d}/\lambda > 1$ suggests that pure exponential disks ought to have low surface brightnesses.

Like D06, we have shown that the inner disk scale length evolves 
considerably over time, as a result of star formation, and cannot be naively assumed to be representative of the initial profile disk scale length.  However, we have also shown that the outer disk scale length, not only remains stable over the disk's evolution, but is a much closer match to the initial disk scale length.  Thus, while D06 have argued that the inner disk scale length cannot be trivially related to the distribution of the dark matter halo specific angular momenta, we have shown that the outer disk scale length is a more stable and reliable reference, at least is the context of isolated evolution.

Our models have not included gas accretion and have used pre-assembled disks.  This means, particularly in the case of time evolution of the two-component profile, that our results may not accurately represent observed galaxies. Rather, our study will be used to separate the effects of pure dynamical evolution from those of  cosmological evolution with gas accretion and merging.

vdB01 took a semi-analytical approach to study the surface 
brightness profiles of galaxies and, in particular, addressed possible solutions to the problem of overly concentrated disks in simulated galaxies.  His models make use of star formation, bulge formation and feedback as possible solutions to the aforementioned problem.  The inclusion of a bulge component alleviates the problem of overly concentrated disks in HSB systems as it prevents systems with strongly declining rotation curves.  Bulge formation depletes the cold central gas to values below the critical level.  This, in turn, suppresses star formation resulting in an exponential stellar disk.  However, in LSB systems, which typically have low bulge-to-disk ratios, the model disks that formed even with a realistic bulge were still overly concentrated.  vdB01 reported that his break radii in stellar profiles were in close agreement with observations using a Kennicutt star formation threshold.  His stellar profile breaks ranged from 2.5 to 4.5 disk scale lengths.  However, vdB01 does not report whether he could model non-truncated galaxies (Type I).  Indeed, the main problem for the star formation threshold as an explanation for profile breaks is that it fails to explain  purely exponential galaxies out to large radii (Schaye 2004).    

While a star formation threshold may prove necessary to explain observed profiles, semi-analytic models cannot address the secular evolution of the disk and the effects induced by angular momentum redistribution within the disk.  Thus, vdB01's study could not address the formation of breaks coupled with bar formation.  vdB01 also assumed that the distribution of observed (inner) disk scale lengths would be coupled to the specific angular momenta of the halo.  If, however, the observed disk scale lengths evolve over time, as our simulations have shown, this assumption may be ill-advised.

Our work has enabled the demonstration that dynamics and secular
evolution can produce galaxy breaks and may significantly reshape the
mass density profiles of galaxies, particularly regarding HSB galaxies.
While the star formation threshold implemented in GADGET-2 produced
star formation breaks which matched the stellar breaks understood by
PT06 as Type II-CT breaks, we stress that a star formation threshold
alone is insufficient to explain Type I light profiles.  We showed
that the ratio $m_{d}/\lambda$ is the key factor in determining if,
when and where a galaxy will develop a two-component threshold.  We
stress that these results are based on the physics of an isolated disk
alone.  This work can be extended to galaxies subjected to merging and
gas accretion in order to assess their contribution to the properties
of density profile breaks.  A comprehensive study of the observed
light profiles of low surface brightness galaxies would strengthen
our predictions for the variations of profile break properties as
a function of surface brightness.

\section{Conclusion}
We have studied the behavior of profile breaks in 162 model galaxies with a 
range of structural
parameters for the halo $\lambda$, $c$, and $V_{200}$, a disk mass 
parameter $m_d$ and the
presence of a bulge component.  Our principal conclusions are:
\renewcommand{\labelenumi}{\roman{enumi}}
\begin{enumerate}
\item Given the use of a star formation threshold, breaks are always seen in the gas and newly formed stars for all models (but not necessarily in the total profiles): breaks may develop in the total baryonic profile regardless of a star formation threshold depending on the amount of angular momentum redistribution;
\item The development of a break is coupled to the development of a bar and an increased central mass concentration (but the converse is not necessarily true);
\item  The development of breaks with time is controlled by the ratio $m_{d}/\lambda$: if $m_{d}/\lambda$ is high, breaks will develop quickly: If $m_{d}/\lambda \approx 1$ a break may or may not develop depending on the combination of the other model parameters:
\begin{itemize}
\item a low $V_{200}$ value in this situation will favor break development
\item a bulge in this situation will impede break development
\item a high and low halo concentration will impede break development
\end{itemize}
\item The radial position of the profile break is constant with time and  depends on the ratio of $m_{d}/\lambda$ for a given SF threshold; if this ratio is high the break radius will be larger.  A minimum in the final Q curve predicts the location of the break, but the Q curve alone is insufficient to predict the development of a break.  The latter is determined by $m_{d}/\lambda$ (see iii).;
\item Increasing the halo concentration may decrease the break radius but not as efficiently as $m_{d}/\lambda$;
\item The presence of a bulge may increase the break radius but not as efficiently as $m_{d}/\lambda$;
\item The mass density of a two-component profile break is controlled by $V_{200}$ and $c$, and is independent of $m_{d}/\lambda$;
\item The outer disk scale length of the gas and total baryonic profile is in close agreement with the initial disk scale length of the isolated galaxy model;
\item The inner disk scale length of the newly formed stars is in close agreement with the initial disk scale length of the isolated galaxy model;
\item The inner disk scale length evolves considerably over time due to star formation.

\end{enumerate}

We acknowledge valuable discussions with Volker Springel, Josh Barnes and
James Wadsley.  We are grateful to Volker Springel for providing us with a
private version of GADGET-2, without which this work would have never been
possible.  We thank the SHARCNET allocations committee, led by Hugh
Couchman, for the allocation of all of our requested computing time to
carry out such an exhaustive study.  Finally, we thank the referee, Fabio Governato, for a comprehensive review and useful comments.  S.C. and R.J.T. acknowledge the 
support of NSERC through respective Discovery Grants. R.J.T. also 
acknowledges funding from the Canada Research Chairs program and Canada 
Foundation for Innovation.

%\bigskip
%\bibliographystyle{apj}

\section{Appendix A - Model Parameters}

We list the model parameters and disk scale lengths of the simulated galaxies in Table 
\ref{parameters}.  Models 
in bold failed 
to progress beyond 2 Gyr.

\begin{table}
\caption{Model Parameters}
\begin{tabular}{ccccccc}
\hline\hline
{Model} & {\bf $\lambda$} & {\bf $m_{d}$} & {\bf $m_{b}$} & {\bf $V_{200}$} & {\bf $c$} & {\bf $h$}\\
\hline\hline
41 & 0.02 & 0.025 & 0.000 & 80.0 & 5 & 1.17 \\ 
42 & 0.02 & 0.025 & 0.005 & 80.0 & 5 & 1.08 \\ 
43 & 0.02 & 0.025 & 0.000 & 160.0 & 5 & 2.34 \\ 
44 & 0.02 & 0.025 & 0.005 & 160.0 & 5 & 2.15 \\ 
45 & 0.02 & 0.025 & 0.000 & 180.0 & 5 & 2.64 \\ 
46 & 0.02 & 0.025 & 0.005 & 180.0 & 5 & 2.42 \\ 
47 & 0.02 & 0.025 & 0.000 & 80.0 & 10 & 0.85 \\ 
48 & 0.02 & 0.025 & 0.005 & 80.0 & 10 & 0.79 \\ 
49 & 0.02 & 0.025 & 0.000 & 160.0 & 10 & 1.70 \\ 
50 & 0.02 & 0.025 & 0.005 & 160.0 & 10 & 1.58 \\ 
51 & 0.02 & 0.025 & 0.000 & 180.0 & 10 & 1.92 \\ 
52 & 0.02 & 0.025 & 0.005 & 180.0 & 10 & 1.78 \\ 
53 & 0.02 & 0.025 & 0.000 & 80.0 & 15 & 0.69 \\ 
54 & 0.02 & 0.025 & 0.005 & 80.0 & 15 & 0.64 \\ 
55 & 0.02 & 0.025 & 0.000 & 160.0 & 15 & 1.37 \\ 
56 & 0.02 & 0.025 & 0.005 & 160.0 & 15 & 1.29 \\ 
57 & 0.02 & 0.025 & 0.000 & 180.0 & 15 & 1.55 \\ 
58 & 0.02 & 0.025 & 0.005 & 180.0 & 15 & 1.45 \\ 
{\bf 59} & 0.02 & 0.050 & 0.000 & 80.0 & 5 & 0.84 \\ 
60 & 0.02 & 0.050 & 0.010 & 80.0 & 5 & 0.72 \\
{\bf 61} & 0.02 & 0.050 & 0.000 & 160.0 & 5 & 1.69 \\ 
62 & 0.02 & 0.050 & 0.010 & 160.0 & 5 & 1.45 \\ 
{\bf 63} & 0.02 & 0.050 & 0.000 & 180.0 & 5 & 1.90 \\ 
64 & 0.02 & 0.050 & 0.010 & 180.0 & 5 & 1.63 \\ 
{\bf 65} & 0.02 & 0.050 & 0.000 & 80.0 & 10 & 0.64 \\ 
66 & 0.02 & 0.050 & 0.010 & 80.0 & 10 & 0.56 \\ 
{\bf 67} & 0.02 & 0.050 & 0.000 & 160.0 & 10 & 1.28 \\ 
68 & 0.02 & 0.050 & 0.010 & 160.0 & 10 & 1.12 \\ 
{\bf 69} & 0.02 & 0.050 & 0.000 & 180.0 & 10 & 1.44 \\ 
70 & 0.02 & 0.050 & 0.010 & 180.0 & 10 & 1.26 \\ 
{\bf 71} & 0.02 & 0.050 & 0.000 & 80.0 & 15 & 0.53 \\ 
{\bf 72} & 0.02 & 0.050 & 0.010 & 80.0 & 15 & 0.47 \\ 
{\bf 73} & 0.02 & 0.050 & 0.000 & 160.0 & 15 & 1.06 \\ 
74 & 0.02 & 0.050 & 0.010 & 160.0 & 15 & 0.94 \\ 
{\bf 75} & 0.02 & 0.050 & 0.000 & 180.0 & 15 & 1.20 \\ 
76 & 0.02 & 0.050 & 0.010 & 180.0 & 15 & 1.05 \\ 
{\bf 77} & 0.02 & 0.100 & 0.000 & 80.0 & 5 & 0.49 \\ 
{\bf 78} & 0.02 & 0.100 & 0.020 & 80.0 & 5 & 0.40 \\ 
{\bf 79} & 0.02 & 0.100 & 0.000 & 160.0 & 5 & 0.98 \\
80 & 0.02 & 0.100 & 0.020 & 160.0 & 5 & 0.80 \\ 
{\bf 81} & 0.02 & 0.100 & 0.000 & 180.0 & 5 & 1.11 \\ 
82 & 0.02 & 0.100 & 0.020 & 180.0 & 5 & 0.90 \\ 
{\bf 83} & 0.02 & 0.100 & 0.000 & 80.0 & 10 & 0.39 \\ 
84 & 0.02 & 0.100 & 0.020 & 80.0 & 10 & 0.32 \\ 
{\bf 85} & 0.02 & 0.100 & 0.000 & 160.0 & 10 & 0.78 \\ 
86 & 0.02 & 0.100 & 0.020 & 160.0 & 10 & 0.65 \\ 
{\bf 87} & 0.02 & 0.100 & 0.000 & 180.0 & 10 & 0.88 \\ 
88 & 0.02 & 0.100 & 0.020 & 180.0 & 10 & 0.73 \\ 
{\bf 89} & 0.02 & 0.100 & 0.000 & 80.0 & 15 & 0.33 \\ 
{\bf 90} & 0.02 & 0.100 & 0.020 & 80.0 & 15 & 0.28 \\ 
{\bf 91} & 0.02 & 0.100 & 0.000 & 160.0 & 15 & 0.67 \\ 
92 & 0.02 & 0.100 & 0.020 & 160.0 & 15 & 0.55 \\ 
{\bf 93} & 0.02 & 0.100 & 0.000 & 180.0 & 15 & 0.75 \\ 
94 & 0.02 & 0.100 & 0.020 & 180.0 & 15 & 0.62 \\ 
95 & 0.03 & 0.025 & 0.000 & 80.0 & 5 & 1.83 \\ 
96 & 0.03 & 0.025 & 0.005 & 80.0 & 5 & 1.73 \\ 
97 & 0.03 & 0.025 & 0.000 & 160.0 & 5 & 3.66 \\ 
98 & 0.03 & 0.025 & 0.005 & 160.0 & 5 & 3.47 \\ 
99 & 0.03 & 0.025 & 0.000 & 180.0 & 5 & 4.11 \\ 
100 & 0.03 & 0.025 & 0.005 & 180.0 & 5 & 3.90 \\ 
101 & 0.03 & 0.025 & 0.000 & 80.0 & 10 & 1.31 \\ 
102 & 0.03 & 0.025 & 0.005 & 80.0 & 10 & 1.25 \\ 
103 & 0.03 & 0.025 & 0.000 & 160.0 & 10 & 2.61 \\ 
\hline
\label{parameters}
\end{tabular}
\end{table}

\begin{table}
\contcaption{}
\begin{tabular}{ccccccc}
\hline\hline
{Model} & {\bf $\lambda$} & {\bf $m_{d}$} & {\bf $m_{b}$} & {\bf $V_{200}$} & {\bf $c$} & {\bf $h$}\\
\hline\hline
104 & 0.03 & 0.025 & 0.005 & 160.0 & 10 & 2.50 \\ 
105 & 0.03 & 0.025 & 0.000 & 180.0 & 10 & 2.94 \\ 
106 & 0.03 & 0.025 & 0.005 & 180.0 & 10 & 2.81 \\ 
107 & 0.03 & 0.025 & 0.000 & 80.0 & 15 & 1.05 \\ 
108 & 0.03 & 0.025 & 0.005 & 80.0 & 15 & 1.00 \\
109 & 0.03 & 0.025 & 0.000 & 160.0 & 15 & 2.09 \\ 
110 & 0.03 & 0.025 & 0.005 & 160.0 & 15 & 2.01 \\ 
111 & 0.03 & 0.025 & 0.000 & 180.0 & 15 & 2.35 \\ 
112 & 0.03 & 0.025 & 0.005 & 180.0 & 15 & 2.26 \\ 
113 & 0.03 & 0.050 & 0.000 & 80.0 & 5 & 1.49 \\ 
114 & 0.03 & 0.050 & 0.010 & 80.0 & 5 & 1.35 \\ 
115 & 0.03 & 0.050 & 0.000 & 160.0 & 5 & 2.98 \\ 
116 & 0.03 & 0.050 & 0.010 & 160.0 & 5 & 2.69 \\ 
117 & 0.03 & 0.050 & 0.000 & 180.0 & 5 & 3.35 \\ 
118 & 0.03 & 0.050 & 0.010 & 180.0 & 5 & 3.03 \\ 
119 & 0.03 & 0.050 & 0.000 & 80.0 & 10 & 1.10 \\
120 & 0.03 & 0.050 & 0.010 & 80.0 & 10 & 1.00 \\ 
121 & 0.03 & 0.050 & 0.000 & 160.0 & 10 & 2.20 \\ 
122 & 0.03 & 0.050 & 0.010 & 160.0 & 10 & 2.01 \\ 
123 & 0.03 & 0.050 & 0.000 & 180.0 & 10 & 2.47 \\ 
124 & 0.03 & 0.050 & 0.010 & 180.0 & 10 & 2.26 \\ 
125 & 0.03 & 0.050 & 0.000 & 80.0 & 15 & 0.89 \\ 
126 & 0.03 & 0.050 & 0.010 & 80.0 & 15 & 0.82 \\ 
127 & 0.03 & 0.050 & 0.000 & 160.0 & 15 & 1.79 \\ 
128 & 0.03 & 0.050 & 0.010 & 160.0 & 15 & 1.65 \\ 
129 & 0.03 & 0.050 & 0.000 & 180.0 & 15 & 2.01 \\ 
130 & 0.03 & 0.050 & 0.010 & 180.0 & 15 & 1.85 \\ 
{\bf 131} & 0.03 & 0.100 & 0.000 & 80.0 & 5 & 1.01 \\ 
132 & 0.03 & 0.100 & 0.020 & 80.0 & 5 & 0.85 \\ 
{\bf 133} & 0.03 & 0.100 & 0.000 & 160.0 & 5 & 2.02 \\ 
134 & 0.03 & 0.100 & 0.020 & 160.0 & 5 & 1.70 \\ 
{\bf 135} & 0.03 & 0.100 & 0.000 & 180.0 & 5 & 2.27 \\ 
136 & 0.03 & 0.100 & 0.020 & 180.0 & 5 & 1.92 \\ 
{\bf 137} & 0.03 & 0.100 & 0.000 & 80.0 & 10 & 0.78 \\ 
{\bf 138} & 0.03 & 0.100 & 0.020 & 80.0 & 10 & 0.67 \\ 
{\bf 139} & 0.03 & 0.100 & 0.000 & 160.0 & 10 & 1.56 \\ 
140 & 0.03 & 0.100 & 0.020 & 160.0 & 10 & 1.34 \\ 
{\bf 141} & 0.03 & 0.100 & 0.000 & 180.0 & 10 & 1.76 \\ 
142 & 0.03 & 0.100 & 0.020 & 180.0 & 10 & 1.51 \\ 
{\bf 143} & 0.03 & 0.100 & 0.000 & 80.0 & 15 & 0.65 \\ 
{\bf 144} & 0.03 & 0.100 & 0.020 & 80.0 & 15 & 0.56 \\ 
{\bf 145} & 0.03 & 0.100 & 0.000 & 160.0 & 15 & 1.31 \\ 
146 & 0.03 & 0.100 & 0.020 & 160.0 & 15 & 1.13 \\ 
{\bf 147} & 0.03 & 0.100 & 0.000 & 180.0 & 15 & 1.47 \\ 
{\bf 148} & 0.03 & 0.100 & 0.020 & 180.0 & 15 & 1.27 \\ 
149 & 0.08 & 0.025 & 0.000 & 80.0 & 5 & 4.78 \\ 
150 & 0.08 & 0.025 & 0.005 & 80.0 & 5 & 4.70 \\ 
151 & 0.08 & 0.025 & 0.000 & 160.0 & 5 & 9.55 \\ 
152 & 0.08 & 0.025 & 0.005 & 160.0 & 5 & 9.40 \\ 
153 & 0.08 & 0.025 & 0.000 & 180.0 & 5 & 10.75 \\ 
154 & 0.08 & 0.025 & 0.005 & 180.0 & 5 & 10.57 \\ 
155 & 0.08 & 0.025 & 0.000 & 80.0 & 10 & 3.39 \\ 
156 & 0.08 & 0.025 & 0.005 & 80.0 & 10 & 3.34 \\ 
157 & 0.08 & 0.025 & 0.000 & 160.0 & 10 & 6.78 \\ 
158 & 0.08 & 0.025 & 0.005 & 160.0 & 10 & 6.69 \\ 
159 & 0.08 & 0.025 & 0.000 & 180.0 & 10 & 7.63 \\ 
160 & 0.08 & 0.025 & 0.005 & 180.0 & 10 & 7.52 \\ 
161 & 0.08 & 0.025 & 0.000 & 80.0 & 15 & 2.71 \\ 
162 & 0.08 & 0.025 & 0.005 & 80.0 & 15 & 2.68 \\
163 & 0.08 & 0.025 & 0.000 & 160.0 & 15 & 5.43 \\ 
164 & 0.08 & 0.025 & 0.005 & 160.0 & 15 & 5.36 \\ 
165 & 0.08 & 0.025 & 0.000 & 180.0 & 15 & 6.11 \\ 
166 & 0.08 & 0.025 & 0.005 & 180.0 & 15 & 6.03 \\ 
\hline
\end{tabular}
\end{table}

\begin{table}
\contcaption{}
\begin{tabular}{ccccccc}
\hline\hline
{Model} & {\bf $\lambda$} & {\bf $m_{d}$} & {\bf $m_{b}$} & {\bf $V_{200}$} & {\bf $c$} & {\bf $h$}\\
\hline\hline
167 & 0.08 & 0.050 & 0.000 & 80.0 & 5 & 4.49 \\ 
168 & 0.08 & 0.050 & 0.010 & 80.0 & 5 & 4.34 \\ 
169 & 0.08 & 0.050 & 0.000 & 160.0 & 5 & 8.98 \\ 
170 & 0.08 & 0.050 & 0.010 & 160.0 & 5 & 8.68 \\ 
171 & 0.08 & 0.050 & 0.000 & 180.0 & 5 & 10.10 \\ 
172 & 0.08 & 0.050 & 0.010 & 180.0 & 5 & 9.77 \\ 
173 & 0.08 & 0.050 & 0.000 & 80.0 & 10 & 3.22 \\ 
174 & 0.08 & 0.050 & 0.010 & 80.0 & 10 & 3.13 \\ 
175 & 0.08 & 0.050 & 0.000 & 160.0 & 10 & 6.43 \\ 
176 & 0.08 & 0.050 & 0.010 & 160.0 & 10 & 6.25 \\ 
177 & 0.08 & 0.050 & 0.000 & 180.0 & 10 & 7.24 \\ 
178 & 0.08 & 0.050 & 0.010 & 180.0 & 10 & 7.03 \\ 
179 & 0.08 & 0.050 & 0.000 & 80.0 & 15 & 2.59 \\ 
180 & 0.08 & 0.050 & 0.010 & 80.0 & 15 & 2.52 \\ 
181 & 0.08 & 0.050 & 0.000 & 160.0 & 15 & 5.17 \\ 
182 & 0.08 & 0.050 & 0.010 & 160.0 & 15 & 5.04 \\ 
183 & 0.08 & 0.050 & 0.000 & 180.0 & 15 & 5.82 \\ 
184 & 0.08 & 0.050 & 0.010 & 180.0 & 15 & 5.67 \\ 
185 & 0.08 & 0.100 & 0.000 & 80.0 & 5 & 3.95 \\ 
186 & 0.08 & 0.100 & 0.020 & 80.0 & 5 & 3.69 \\ 
187 & 0.08 & 0.100 & 0.000 & 160.0 & 5 & 7.90 \\ 
188 & 0.08 & 0.100 & 0.020 & 160.0 & 5 & 7.39 \\ 
189 & 0.08 & 0.100 & 0.000 & 180.0 & 5 & 8.89 \\ 
190 & 0.08 & 0.100 & 0.020 & 180.0 & 5 & 8.31 \\ 
191 & 0.08 & 0.100 & 0.000 & 80.0 & 10 & 2.88 \\ 
192 & 0.08 & 0.100 & 0.020 & 80.0 & 10 & 2.72 \\ 
193 & 0.08 & 0.100 & 0.000 & 160.0 & 10 & 5.77 \\ 
194 & 0.08 & 0.100 & 0.020 & 160.0 & 10 & 5.45 \\ 
195 & 0.08 & 0.100 & 0.000 & 180.0 & 10 & 6.49 \\ 
196 & 0.08 & 0.100 & 0.020 & 180.0 & 10 & 6.13 \\ 
197 & 0.08 & 0.100 & 0.000 & 80.0 & 15 & 2.34 \\ 
198 & 0.08 & 0.100 & 0.020 & 80.0 & 15 & 2.22 \\ 
199 & 0.08 & 0.100 & 0.000 & 160.0 & 15 & 4.68 \\ 
200 & 0.08 & 0.100 & 0.020 & 160.0 & 15 & 4.44 \\ 
201 & 0.08 & 0.100 & 0.000 & 180.0 & 15 & 5.27 \\ 
202 & 0.08 & 0.100 & 0.020 & 180.0 & 15 & 5.00 \\ 
\hline
\end{tabular}
\end{table}

\section{Appendix B - Break Radii Determination}
We discuss the three methods used to determine the break radii of our simulated galaxies. 
\subsection{Break Radius from the Profile Derivatives}

Considering a smooth galaxy density profile, the location of the break at the transition between the inner and outer disks may be found in principle at the peak of the second derivative of this profile.  However, simulated (and observed) profiles contain many local minima and maxima and the process of finding the peak in the second 
profile derivative is non-trivial.
Hence, we developed a search method for the second 
derivative which proceeds from the inner regions of a 
surface density profile outwards to locate the second derivative maximum while  
avoiding maxima associated with raw (unsmoothed) and transient features. 

\subsection{Break Radius from the Deviation from the Initial Mass 
Density Profile}

Our second method involves comparing any evolved density profile to its 
initial profile.  This method is used only for the total and gas 
profiles, as the new stars have no initial profiles.  This technique can 
also not be applied to observationally derived density profiles since 
initial profiles are unknown. Our break identification involves 
starting 
from the outermost profile point and stepping inwards towards smaller radii and tagging the location where the current profile deviates steadily (by more than 1.5 \%) from the initial profile.

\subsection{Break Radius Determined by Eye}

A further test for the validity of these two methods is the comparison 
of the aforementioned break identifications with those estimated by-eye. 
The inner break radius is roughly the transition from the 
outer bulge, if any, to the inner disk; the outer break radius is the transition between the inner and outer disk 
components (see Fig.~\ref{mark}).  The slopes (inverse scale lengths) of the 
inner and outer profile components can be determined and the relative 
sharpness of the 
break is computed as the angle, $\theta$, between the two fitted 
exponential lines.  Fig.~\ref{mark} shows a total (stars $+$ gas) mass density profile 
with the relevant parameters.

\label{lastpage}
\subsection{Comparison of the Three Methods}

\begin{figure}
\centering

     \epsfxsize=8cm\epsfbox{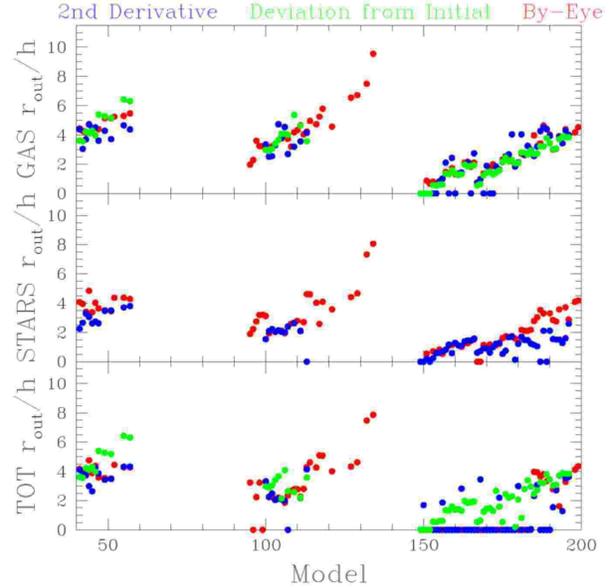}

\rm
\caption{Comparison of the two numerical break-finding methods for the gas, newly formed stars and total baryonic profiles.  The Y-axis is the ratio of the numerically determined outer break radius, $r_{out}$, (see Fig.~\ref{mark}) to the initial disk scale length, $h$.  The different panels represent values for the gas, stars and total (gas $+$ stars) density profiles from bottom to top.  Galaxy model numbers are shown on the X-axis (see Appendix A for the model parameters).  The by-eye method is shown in red while the numerical methods are shown in green and blue.  $r_{out}/h=0$ where breaks could not be identified.}
\label{progcomp}
\end{figure}

While the two numerical methods predicted break points reasonably well, 
their locations did not always closely match the estimates by-eye.  Fig.~\ref{progcomp} shows a comparison of the two numerical methods against eye-ball markings for the 10 Gyr sample.  
The discrepancies are especially acute for snapshots earlier than 3 Gyr where 
transient features persist.  The comparison is significantly better 
for the 10 Gyr 
models.  While close agreement is 
seen with gas break estimates, we find a greater dispersion for the 
total profile measurements.  The numerical techniques are clearly 
promising and worthy of further investigation.  However, for simplicity, 
we will rely on eye-ball estimates which, for most simulations, are 
in close agreement with the numerical predictions.

%%%%%%%%%%%%%%%%%%%%%%%%%%%%%%%%%%%%%%%%%%%%%%%%%%%%%%%%%%%%%%%%%%%
%%%%%%%%%%%%%%%%%%%%%%%%%%%%%%%%%%%%%%%%%%%%%%%%%%%%%%%%%%%%%%%%%

\end{document}